\documentclass[a4paper,11pt]{article}
\usepackage[T1]{fontenc}
\usepackage[utf8]{inputenc}
\usepackage{amsfonts,amsmath,amssymb}
\usepackage[left=25mm,right=20mm,bottom=25mm,top=25mm,headheight=10mm]{geometry}
\usepackage{lastpage}
\usepackage{bm}
\usepackage{xcolor}
\usepackage{cite}
\usepackage[nottoc]{tocbibind}
\usepackage{titletoc}
\usepackage{fancyhdr}
\usepackage{booktabs}
\usepackage{graphicx}
\usepackage{hyperref}
\usepackage{enumitem}
\usepackage[font=small]{caption}
\usepackage{fontawesome}
\usepackage{authblk}
\usepackage{tikz}
\usetikzlibrary{calc, intersections, arrows}
\dottedcontents{section}[2.5em]{}{1.5em}{0.5pc}
\dottedcontents{subsection}[5.5em]{}{2em}{0.5pc}
\numberwithin{equation}{section}

\hypersetup{linktoc = all}

\makeatletter
\newcommand{\customlabel}[2]{%
   \protected@write \@auxout {}{\string \newlabel {#1}{{#2}{\thepage}{#2}{#1}{}} }%
   \hypertarget{#1}{#2}
}
\makeatother

\definecolor{lime}{HTML}{A6CE39}
\DeclareRobustCommand{\orcidicon}{
   \begin{tikzpicture}
   \draw[lime, fill=lime] (0,0) 
   circle [radius=0.16] 
   node[white] {{\fontfamily{qag}\selectfont \tiny ID}};
   \draw[white, fill=white] (-0.0625,0.095) 
   circle [radius=0.007];
   \end{tikzpicture}
   \hspace{-2mm}
}

\foreach \x in {A, ..., Z}{\expandafter\xdef\csname orcid\x\endcsname{\noexpand\href{https://orcid.org/\csname orcidauthor\x\endcsname}
         {\noexpand\orcidicon}}
}

\definecolor{bluish}{HTML}{228CEE}
\definecolor{greenish}{HTML}{419127}
\definecolor{orangish}{HTML}{FFB347}
\hypersetup{%
  colorlinks=true,
  linkcolor=bluish,
  anchorcolor=green,
  citecolor=orangish,
  filecolor=cyan,
  menucolor=magenta,
  runcolor=yellow,
  urlcolor=greenish
}

\title{\bf\textsc{Scalar induced gravitational waves review}}
\author[$\ddagger$\orcidA{}]{Guillem Dom{\`e}nech\footnote{\faEnvelope: \href{mailto:domenech@pd.infn.it}{domenech@pd.infn.it}\quad;\quad \faHome: \href{https://domenechcosmo.netlify.app}{domenechcosmo.netlify.app}}}
\affil[$\ddagger$]{INFN Sezione di Padova, I-35131 Padova, Italy}

\begin{document}

\clearpage\maketitle
\thispagestyle{empty}

\hrule
\section*{Abstract}
We provide a review on the state-of-the-art of gravitational waves induced by primordial fluctuations, so-called induced gravitational waves. We present the intuitive physics behind induced gravitational waves and we revisit and unify the general analytical formulation. We then present general formulas in a compact form, ready to be applied. This review places emphasis on the open possibility that the primordial universe experienced a different expansion history than the often assumed radiation dominated cosmology. We hope that anyone interested in the topic will become aware of current advances in the cosmology of induced gravitational waves, as well as becoming familiar with the calculations behind.\\

\noindent\textbf{Keywords:} Induced Gravitational Waves; Primordial Black Holes; Early universe; Inflation.\\

\hrule

\vspace*{5mm}

\tableofcontents

\section{Introduction \label{sec:intro}}

\begin{figure}
\begin{minipage}{0.5\columnwidth}
\includegraphics[width=\columnwidth]{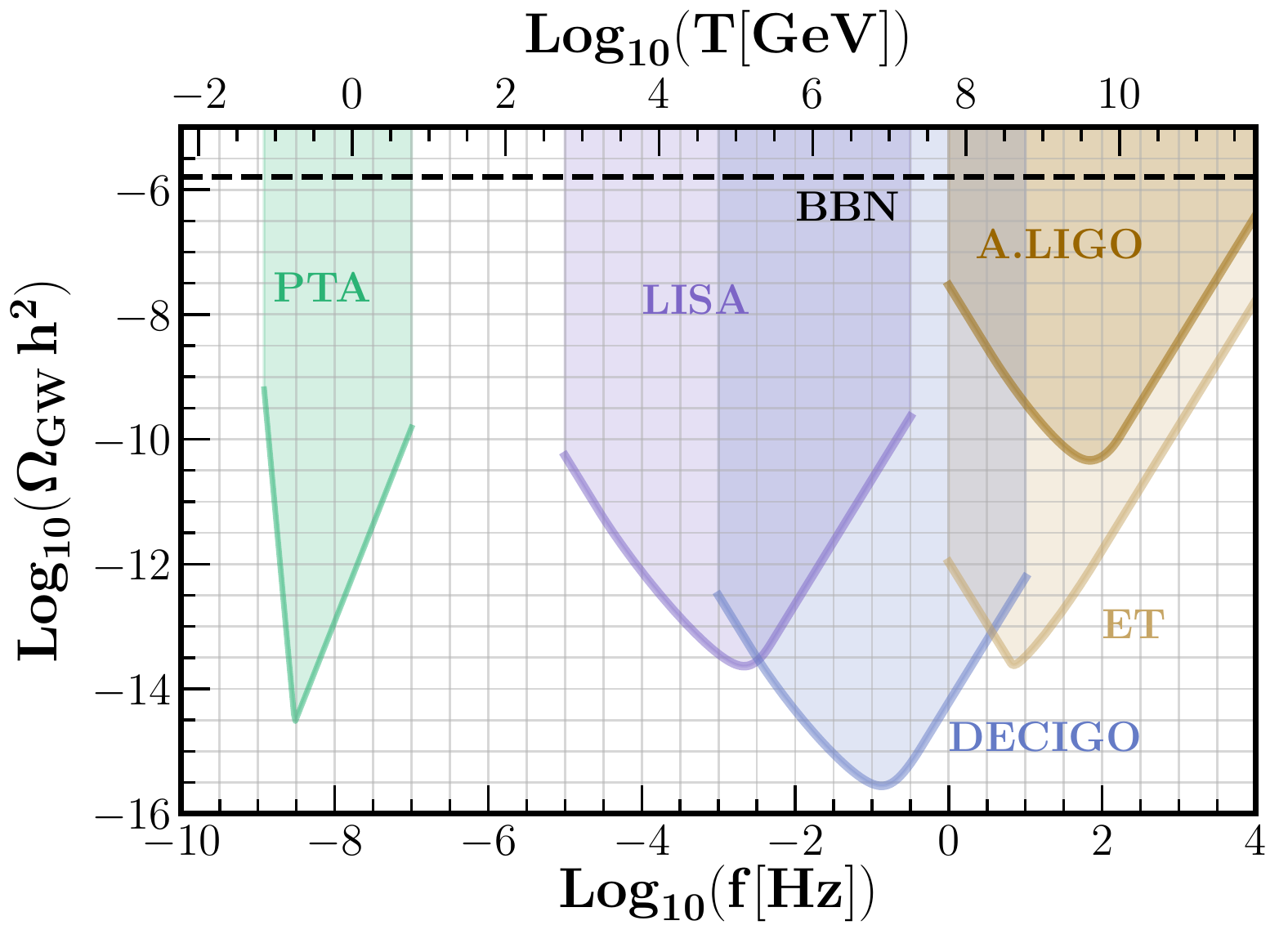}
\end{minipage}
\begin{minipage}{0.5\columnwidth}
\includegraphics[width=\columnwidth]{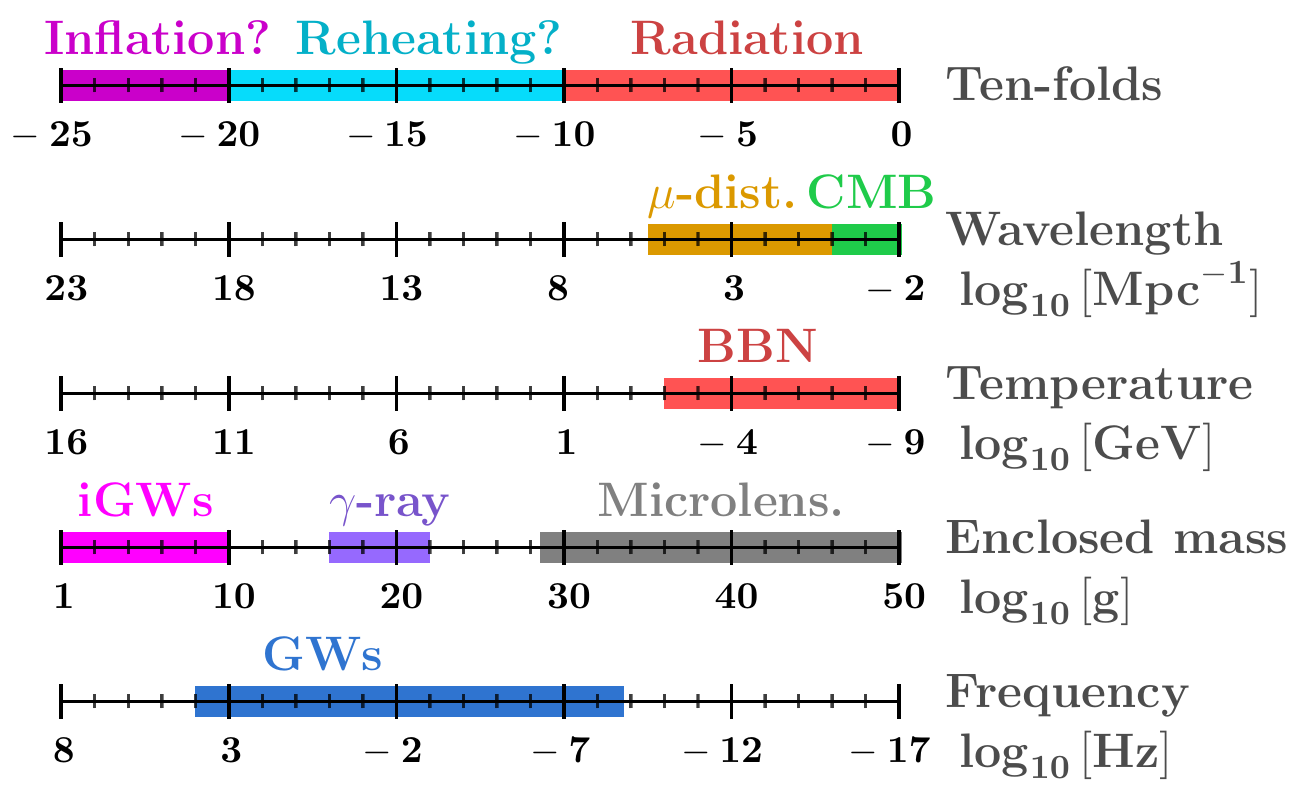}
\end{minipage}
\caption{On the left we show the power-law integrated sensitivity curves \cite{Thrane:2013oya} in terms of the GW spectral density for some GW detectors. In the lower and upper horizontal axis, we respectively show the GW frequency and the associated temperature of the universe at the time of GW generation, assuming radiation domination. On the right, we show the typical quantities associated to the cosmological horizon at a particular time, given in terms of ten-folds from matter-radiation equality. We also show the ranges covered by CMB anisotropies and $\mu$ spectral distortions, BBN constraints and GWs. If PBH form, they can be probed by induced GWs (iGW), emission of $\gamma$ rays and microlensing. \label{fig:ruler}}
\end{figure}

Cosmology with Gravitational Waves (GWs) is a new and unique doorway to the primordial universe. The further in time we can observe with electromagnetic waves is around neutrino decoupling, roughly one second after the Big Bang and some time before Big Bang Nucleosynthesis (BBN). Prior to that, we have scarce evidence of the evolution of the primordial universe. See Ref.~\cite{Allahverdi:2020bys} for a review on possible expansion histories of the early universe. Nevertheless, we learned a lot from studies of the Cosmic Microwave Background\footnote{Observations of the first photons that decoupled from the thermal plasma after neutral hydrogen formed} (CMB). For instance, we know that there are Gaussian primordial fluctuations with an almost scale invariant spectrum \cite{Aghanim:2018eyx,Akrami:2018odb}, which by causality must have been generated before the Big Bang. This fact provides strong evidence that there was a period of accelerated expansion in the primordial universe, so-called inflation \cite{Brout:1977ix,Starobinsky:1979ty,Guth:1980zm,Sato:1980yn}. During inflation, quantum vacuum fluctuations are stretched out to the largest scales and become primordial fluctuations \cite{Mukhanov:1981xt,Linde:1981mu,Albrecht:1982wi,Sasaki:1983kd,Kodama:1985bj}, which we later see as CMB anisotropies and galaxies. Thus, through the observation of primordial fluctuations we have access to inflation. However, the CMB only probes the largest scales and, therefore, only a small fraction of inflation. Any information on much smaller scales,\footnote{To have a quantitative idea, the comoving wavenumber corresponding to the size of Hubble horizon at the time of neutrino decoupling is roughly $k\sim 10^{4}\,{\rm Mpc}^{-1}$. CMB observations constraint the primordial fluctuations roughly on scales $k\sim 10^{-4}-5\times 10^{-1}{\rm Mpc}^{-1}$. Any information on much smaller scales, that is for $k\gg10^{4}\,{\rm Mpc}^{-1}$, is erased by complicated astrophysical processes. On the largest scales we are limited by cosmic variance. CMB spectral distortions might probe down to $k\sim10^{5}\,{\rm Mpc}^{-1}$.} and the last stages of inflation, is basically erased by complicated astrophysical processes. In contrast, GWs barely interact with intervening matter. This means that with GWs we can explore the universe before neutrino decoupling and even before the Big Bang towards the last stages of inflation. GWs provide a unique opportunity to complete our picture of the very early universe. A quantitative picture of the potential of cosmology with GWs can be found in Fig.~\ref{fig:ruler}.

GWs generated by a cosmological process in the early universe will appear today randomly distributed in all directions and with a very large number of unresolved sources. For example, take one source per Hubble patch and count how many Hubble patches at a given redshift fit the current universe. The number actually grows as $(1+z)^3$ and we are talking about GWs generated at $z>10^{10}$. On top of that, the angular resolution of GW detectors is not enough to pinpoint cosmic GWs generated in tiny Hubble patches. The superposition of all incoming unresolved GWs leads to the Stochastic GW Background (SGWB). Encouragingly, the typical frequency of such cosmic GWs falls right into the frequency range of future GWs detectors. For example, GWs generated when the universe had a temperature of around $1\,{\rm GeV}$ and $10^{10}\,{\rm GeV}$, respectively have a typical (peak) frequency roughly around $10^{-8}\,{\rm Hz}$ and $100\,{\rm Hz}$. Pulsar Timing Arrays (PTA) \cite{Lentati:2015qwp,Shannon:2015ect,Arzoumanian:2015liz,Qin:2018yhy,Aggarwal:2018mgp,Arzoumanian:2020vkk}, cover around $10^{-9}\sim10^{-7}\,{\rm Hz}$,  LIGO, VIRGO, KAGRA and ET \cite{Maggiore:2019uih}, are sensitive roughly for $10\sim10^3\,{\rm Hz}$. In between, we will have LISA \cite{Audley:2017drz,Barausse:2020rsu}, DECIGO \cite{Seto:2001qf,Yagi:2011wg,Kawamura:2020pcg}, AION/MAGIS \cite{Badurina:2019hst}, Taiji \cite{Guo:2018npi} and Tainqin \cite{Luo:2015ght}. For an illustration see Fig.~\ref{fig:ruler}.

Typical sources of cosmological GWs in the very early universe include (for more details see \cite{Caprini:2018mtu}): phase transitions, which may lead to collisions of bubbles or a universe filled with cosmic strings, resonances during reheating, quantum (gravity) fluctuations during inflation (so-called primordial GWs) and GWs induced from large primordial fluctuations. Such large primordial fluctuations may also collapse to form primordial black holes (PBHs). Out of these cosmological sources, GWs induced by primordial fluctuations, and to some extend PBHs, are our direct window to the latest stages of inflation. The information we could gain about inflation by the so-called induced GWs potentially covers scales around $k\sim 10^7-10^{18}\,{\rm Mpc}^{-1}$, otherwise inaccessible by any other probe. Even the absence of induced GWs will place new constraints on the primordial spectrum in unexplored regimes, potentially down to ${\cal P}_{\cal R}\sim 10^{-4}-10^{-5}$ \cite{Gow:2020bzo}. On top of these very promising prospects, induced GWs are generated when primordial fluctuations re-enter the horizon sometime between the end of inflation and BBN. Since we have no evidence of the content and expansion history of the universe around that time, induced GWs not only provide access to the last stages of inflation but they also contain information on the content of the primordial universe. In this review, we will be open to the possibility that the universe much before BBN was not dominated by radiation. In the future, information on the primordial spectrum obtained by induced GWs will complement those from other probes such as spectral distortions \cite{Chluba:2019kpb,Kite:2020uix} in the multimessenger cosmology era \cite{Unal:2020mts}.

\subsection{Induced GWs history \label{sec:history}}

The possibility of having GWs induced by density fluctuations was first noticed, as far as the author is aware, by K.~Tomita in 1967 \cite{Tomita}. They were later rediscovered in the 90's by Matarrese, Pantano and Saez \cite{Matarrese:1992rp,Matarrese:1993zf} when studying second order cosmological perturbations in a dust dominated universe. Quite interestingly, in 1997 Matarrese, Mollerach and Bruni \cite{Matarrese:1997ay} noticed that induced GWs in a dust dominated universe suffer from gauge ambiguities. They proceeded to argue that only the oscillating part of the induced tensor modes were identified as \textit{true} gravitational waves. We will discuss more about the gauge ambiguities in Sec.~\ref{sec:gauge}. However, Refs.~\cite{Matarrese:1992rp,Matarrese:1993zf,Matarrese:1997ay} concluded that induced GWs were too small to be practically observed. 

It was not until 20 years later that Ananda, Clarkson and Wands \cite{Ananda:2006af} started to uncover the potential of induced GWs. In Ref.~\cite{Ananda:2006af} they proposed to use induced GWs, generated in a radiation dominated universe, to constrain the spectral tilt of primordial fluctuations. A blue tilted primordial spectrum even with the CMB normalization might end up yielding large enough induced GWs. Some months later, Baumann, Steinhardt, Takahashi and Ichiki \cite{Baumann:2007zm} looked into more detail at the transfer function of induced GWs taking into account the radiation-matter equality at around $z\sim 3400$ and the anisotropic stress due to neutrinos (see also \cite{Mangilli:2008bw} on the latter). They found that assuming the CMB normalisation of an almost scale invariant spectrum, modes with $k\sim k_{\rm eq}\sim 0.01{\rm Mpc}^{-1}$ were enhanced due to the matter dominated stage. Notably, the peak in the induced GW background can be larger than the primordial one if the tensor to scalar ratio is $r<0.1$. However, the frequency of the peak of such GWs is extremely low, around $f_{\rm eq}\sim 10^{-17}{\rm Hz}$. This is only observable perhaps by CMB B-mode polarization experiments or cosmic shear \cite{Sarkar:2008ii}. Nevertheless, this was an indication that an early epoch of dust domination might yield interesting results. In the same period, Martineau and Brandenberger \cite{Martineau:2007dj} derived a lower bound to induced GWs for various inflationary models and alternatives, although they referred to it as lower bound from ``backreaction'' of scalar fluctuations. Another application of GWs induced by primordial fluctuations in the curvaton scenario, sourced between the end of inflation and the curvaton decay, was studied by Bartolo, Matarrese, Riotto and V{\"a}ihk{\"o}nnen \cite{Bartolo:2007vp}. An action formalism for a scalar field acting as a perfect fluid and its induced GWs was studied by Boubekeur, Creminelli, Nore{\~n}a and Vernizzi \cite{Boubekeur:2008kn}.
 
An important realization was done by Saito and Yokoyama in 2008 \cite{Saito:2008jc,Saito:2009jt}: if the primordial spectrum of fluctuations is large enough on small scales, it does not only induce GWs but might also collapse to form PBHs. This means that the induced GW spectrum can be used in the future to place bounds on the PBH abundance. Even more, if PBHs were found, one should also find the induced GW counterpart. These ideas were further pursued by Bugaev and Klimai \cite{Bugaev:2009zh,Bugaev:2009kq,Bugaev:2010bb}. Extending the work of Saito and Yokoyama a bit further, Assadullahi and Wands proposed induced GWs as probes of the primordial spectrum \cite{Assadullahi:2009jc}. They also considered the early dust dominated case \cite{Assadullahi:2009nf} which confirmed the large enhancement of induced GWs. The same group, now including Arroja, Assadullahi, Koyama and Wands \cite{Arroja:2009sh}, studied the matching conditions on superhorizon scales for such GWs induced at second order. Later, in 2012 the induced GWs from particular models of inflation were studied by Alabidi, Kohri, Sasaki and Sendouda in radiation \cite{Alabidi:2012ex} and dust domination \cite{Alabidi:2013lya}. Around the same time, a curvaton scenario with a blue tilted primordial spectrum yielding large induced GWs was proposed by Kawasaki, Kitajima and Yokoyama \cite{Kawasaki:2013xsa}. More investigations on the upper bounds of induced GWs by overproduction of PBHs were done by Nakama and Suyama \cite{Nakama:2015nea,Nakama:2016enz}. Other works related to induced GWs include the decay of two curvatons by Suyama and Yokoyama \cite{Suyama:2011pu}, the impact of anisotropic stress due to free streaming particles by Saga, Ichiki and Sugiyama \cite{Saga:2014jca} and the B-modes due to induced GWs in the CMB by Fidler et al.~\cite{Fidler:2014oda}.

In 2016 LIGO reported the detection of GWs from the merger of a binary black hole \cite{Abbott:2016nmj} and the amount of new works on induced GWs and their PBH counterpart exploded. The amount of works is so large that we will not attempt to go through all of them in chronological order but instead we will classify them into seven main directions below. We give a more detailed discussion in the corresponding sections.
\begin{itemize}
\item[--] \textit{General semi-analytical formulation:} Since induced GWs are a second order effect one needs to integrate in time and momenta over the linear evolution of primordial fluctuations. The analytical transfer functions for radiation domination are derived in Refs.~\cite{Espinosa:2018eve,Kohri:2018awv} and later generalized to constant equation of state parameter in Ref.~\cite{Domenech:2019quo}.
\item[--] \textit{Induced GWs for different expansion histories and different contents of the universe:} Induced GWs may have been generated in a non-radiation dominated universe. This leaves characteristic signatures in the induced GW spectrum. Studies in early matter era can be found in Refs.~\cite{Inomata:2019zqy,Inomata:2019ivs,Dalianis:2020gup}. The extension to an early PBH dominated epoch is investigated in Refs.~\cite{Inomata:2020lmk,Papanikolaou:2020qtd,Domenech:2020ssp,Domenech:2021wkk}. More general thermal histories are studied in Refs.~\cite{Hajkarim:2019nbx,Bhattacharya:2019bvk,Domenech:2019quo,Domenech:2020kqm,Dalianis:2020cla,Abe:2020sqb}. The impact of additional free streaming particles is studied in Ref.~\cite{Hook:2020phx}.
\item[--] \textit{Induced GW spectral features:} There are cases where the induced GW spectrum may be investigated semi-analytically. These are for example, the low frequency tail \cite{Cai:2019cdl,Yuan:2019wwo,Domenech:2020kqm}, the UV tail \cite{Liu:2020oqe,Atal:2021jyo} and the log-normal peak in the primordial spectrum \cite{Pi:2020otn}. Furthermore, the primordial spectrum may also present oscillatory features which are captured into the induced GW spectrum \cite{Cai:2019amo,Fumagalli:2020nvq,Braglia:2020eai,Braglia:2020taf,Fumagalli:2021cel}. On top of that, large primordial non-Gaussianities may have a non-trivial impact on the induced GW spectrum \cite{Garcia-Bellido:2017aan,Cai:2018dig,Unal:2018yaa,Yuan:2020iwf,Atal:2021jyo,Adshead:2021hnm,Ragavendra:2021qdu}. Other effects include: anisotropic non-gaussianities, which may be a source of superhorizon tensor modes \cite{Ota:2020vfn}, resonances that may occur during inflation, which enhance the induced GW spectrum \cite{Cai:2019jah,Cai:2019bmk,Zhou:2020kkf,Cai:2021yvq}, and non-Bunch Davies initial conditions in inflation\cite{Ragavendra:2020vud}, although the latter does not yield an observable signature.
\item[--] \textit{Explanations of current observations:} Induced GWs have been extensively used as counterpart of the PBH scenario as a totality or a fraction of dark matter. For example, the induced GWs from various inflationary models can be found in Refs.~\cite{Garcia-Bellido:2016dkw,Gong:2017qlj,Ando:2018nge,Byrnes:2018txb,Gao:2019kto,Xu:2019bdp,Mahbub:2019uhl,Mishra:2019pzq,Fu:2019vqc,Ozsoy:2019lyy,Ozsoy:2020kat,Gao:2020tsa,Bhaumik:2020dor,Lin:2020goi,Ragavendra:2020sop,Yi:2020cut,Gao:2021vxb,Gao:2021dfi,Solbi:2021wbo} and \cite{Espinosa:2018eve,Drees:2019xpp,Yi:2020kmq,Gao:2021vxb} in the context of Higgs inflation. In particular, the large induced GW counterpart to PBH as the totality of dark matter is studied in Refs.~\cite{Garcia-Bellido:2017aan,Kohri:2018qtx,Bartolo:2018evs,Bartolo:2018rku,Tada:2019amh,Ballesteros:2020qam,Ozsoy:2020kat}. Other possibilities include an explanation to the LIGO observations \cite{Inomata:2016rbd,Nakama:2016gzw,Ando:2017veq,Kohri:2018qtx} and the NANOGrav results \cite{Vaskonen:2020lbd,DeLuca:2020agl,Kohri:2020qqd,Domenech:2020ers,Bhattacharya:2020lhc,Inomata:2020xad,Yi:2021lxc}. 
\item[--] \textit{Current and future GW constraints:} It is important to place constraints on the current absence of induced GWs and also asses future capabilities to constraint/find different models. A study using current PTA and LIGO data on SGWBs can be found in Refs.~\cite{Orlofsky:2016vbd,Cai:2019elf,Chen:2019xse,Chen:2019irf} and an analysis of future GW prospects in Refs.~\cite{Inomata:2018epa,Clesse:2018ogk,Wang:2019kaf,Yuan:2019udt,Lu:2019sti,Unal:2020mts,Dalianis:2020cla,Zhang:2020ptw,Wang:2016tbj}.
\item[--] \textit{SGWB anisotropies:} In the far future we may not only observe the GW spectrum but also its anisotropies. The anisotropies due to induced GWs associated to PBH formation in the case of a monochromatic mass function are investigated in Refs.~\cite{Bartolo:2018qqn,Bartolo:2018evs,Bartolo:2018rku,Bartolo:2019zvb}.
\item[--] \textit{The gauge issue of induced GWs:} Tensor modes are subject to gauge ambiguities at second order due to mode mixing. Since the work of Hwang, Jeong and Noh \cite{Hwang:2017oxa} in 2017 there has been an extensive discussion on the gauge issue of induced GWs \cite{Domenech:2017ems,Gong:2019mui,Tomikawa:2019tvi,DeLuca:2019ufz,Inomata:2019yww,Yuan:2019fwv,Chang:2020iji,Chang:2020tji,Lu:2020diy,Ali:2020sfw,Chang:2020mky,Domenech:2020xin,Gurian:2021rfv}. The source of the problem and the applicability of predictions is by now well understood, although the gauge issue persists in the strictest sense.
\end{itemize}

\subsection{Structure and scope of the review\label{sec:structure}}

Induced GWs are the most promising probe of the primordial universe. In this review, we will revisit and unify the current analytical formulation and main predictions for induced GWs. Note that the scope of this review is basically limited to the author's works on induced GWs \cite{Domenech:2017ems,Domenech:2019quo,Domenech:2020kqm,Domenech:2020ssp,Domenech:2020ers,Domenech:2020xin,Atal:2021jyo,Domenech:2021wkk}, although we discuss and properly cite the relevant literature. As interesting as it is, we will not enter in the details of PBH formation nor on the GW data analysis. Nevertheless, we briefly discuss these relevant topics and give appropriate credit, when needed.

For convenience, we proceed to briefly explain the organization of the review. In this way, the reader interested in certain aspects may jump directly to the relevant section. In sec.~\ref{sec:estimates} we present an intuitive picture of the induced GW generation. We estimate their typical frequency, the spectral shape and we briefly comment on their PBH counterpart for different expansion histories in the primordial universe. The details of the calculations and derivation of analytical approximations are reviewed in Secs.~\ref{sec:general} and \ref{sec:analytical}. The analytical approximations for the induced GW spectrum for typical primordial curvature spectra are presented in \ref{sec:typical}. The special case of the dust dominated universe is discussed in some detail in Sec.~\ref{sec:dust}. We also include the case where PBH dominate the universe in Sec.~\ref{sec:PBHdom}. The gauge issue of induced GWs is explained in Sec.~\ref{sec:gauge}. We list other GW counterparts associated to PBHs in Sec.~\ref{sec:othersources} and discuss the future observational prospects in Sec.~\ref{sec:observations}. We also provide a summary of the main formulas in Sec.~\ref{sec:summary}. These formulas are ready for calculating the induced GW spectrum for general primordial spectra. Thus, the reader interested solely in the final formulas of the induced GW may jump directly to Sec.~\ref{sec:typical} and/or Sec.~\ref{sec:summary}.. We end with a discussion on future directions in the conclusions Sec.~\ref{sec:conclusions}. We present many details of the numerical factors, formulas and calculations used in the main text in the Appendices.

We work in natural units where $\hbar=c=1$. We keep the reduced Planck mass $M_{\rm pl}=(8\pi G)^{-1/2}$ in the equations, except in Secs.~\ref{sec:induced1} and \ref{sec:analytical} where for simplicity in the calculations we set it to $M_{\rm pl}=1$. At the beginning of each section we list the literature in which the section is based, except for Secs.~\ref{sec:othersources} and \ref{sec:observations} which are relatively short. We provide a list of useful related reviews below.\\

\noindent\textit{A list of other useful reviews.} The context of induced GWs is very broad. It covers cosmological perturbation theory up to second order, stochastic GW backgrounds, primordial black holes and primordial non-Gaussianity. Unfortunately, all the details cannot be covered in this review. Instead, we list below existing reviews on topics directly related to induced GWs, which have been useful in writing this manuscript. 
\begin{itemize}
\item\textit{Cosmological perturbation theory.} The classical reviews on cosmological perturbation theory at linear order are the one by Kodama and Sasaki \cite{Kodama:1985bj} and by Mukhanov, Feldman and Brandenberger \cite{Mukhanov:1990me}. A typical reference for second order cosmological perturbation theory is the review by Malik and Wands \cite{Malik:2008im}. Since then there have been many other reviews on cosmological perturbation theory. Some that we found particularly up to date and useful are Refs.~\cite{Durrer:2004fx,Langlois:2010xc,Baumann:2009ds,Piattella:2018hvi}. For a take on multi-field inflation we suggest Ref.~\cite{Gong:2016qmq}. For primordial features in the primordial spectrum we refer the reader to Ref.~\cite{Chen:2011zf}.
\item\textit{Stochastic GW backgrounds.} Induced GWs are not the only source of GW backgrounds. Reviews on the different cosmic and astrophysical sources can be found in the review by Caprini and Figueroa \cite{Caprini:2018mtu} and by Christensen \cite{Christensen:2018iqi}. A useful collection of cosmic GW spectra can be found in Kuroyanagi, Chiba and Takahashi \cite{Kuroyanagi:2018csn}, although it is technically not a review. A review focused on GWs from inflation is given by Guzzetti, Bartolo, Liguori and Matarrese \cite{Guzzetti:2016mkm}.
\item\textit{Primordial black holes.} The literature on primordial black holes is very vast and currently under refinement. A review that has been used in particular is the one by Sasaki, Suyama, Tanaka and Yokoyama \cite{Sasaki:2018dmp}. Other interesting reviews are Refs.~\cite{Khlopov:2008qy,Carr:2020gox,Carr:2020xqk,Green:2020jor}. A complementary review on induced GWs with more focus on the PBH counterpart is given by Yuan and Huang \cite{Yuan:2021qgz}.
\item\textit{Primordial non-Gaussianity.} Although quantum fluctuations during inflation are drawn from a Gaussian distribution, they can develop small departures from such Gaussian distribution due to gravitational or general interactions. The reader interested in primordial non-Gaussianities may check the reviews in the context of inflation and CMB observations, e.g., \cite{Komatsu:2010hc,Bartolo:2010qu,Byrnes:2010em,Koyama:2010xj}. 
\item \textit{Alternative expansion histories.} A recent review encompassing many of the new physics of a primordial universe which is not filled with radiation is given in Ref.~\cite{Allahverdi:2020bys}.
\end{itemize}


\section{Estimates and intuitive picture \label{sec:estimates}}

In this section we present the intuitive physical picture behind induced GWs. We do not intend to do a rigorous derivation but to lay out the basic assumptions behind the calculations. Before we go into the details of the induced GWs themselves, we first review in Sec.~\ref{sec:spectraldensity} what is considered to be the ``observable'' quantity for cosmic GWs. This will also be useful when dealing with the gauge issue in Sec.~\ref{sec:gauge}. We later in \ref{sec:induced1} derive intuitively, based on heuristic calculations and order of magnitude estimations, the main features of the induced GW spectrum in general cosmological backgrounds. Lastly in Sec.~\ref{sec:PBH}, we give the relation between the typical frequencies of the induced GWs with the corresponding PBH masses also in general cosmological backgrounds.\\

\noindent\faBook\,\,\textit{Main references:} Section \ref{sec:spectraldensity} is mainly based on Misner, Thorne and Wheeler's \cite{Misner:1974qy} and Maggiore's \cite{Maggiore:1900zz} books, while the original work is essentially due to Isaacson \cite{Isaacson:1967zz,Isaacson:1968zza}. We follow the notation of, e.g., Ref.~\cite{Cai:2019cdl,Domenech:2020xin}. Sec.~\ref{sec:induced1} essentially follows from \cite{Domenech:2020kqm} although it builds up from previous works in the literature, e.g., \cite{Baumann:2007zm,Kohri:2018awv,Cai:2019cdl,Domenech:2019quo}. Sec.~\ref{sec:PBH} is mainly extracted from \cite{Sasaki:2018dmp,Domenech:2019quo,Domenech:2020ers}. All relevant equations can be found in App.~\ref{app:generalgauge}.

\subsection{The spectral density of GWs in cosmology\label{sec:spectraldensity}}

A GW detector, such as an interferometer, measures the GW strain at a given frequency and time. In the case of the SGWB, one is measuring the time and/or ensemble average of the superposition of all incoming GWs. For convenience, cosmologists often use the energy density fraction of GWs as the ``observable''. However, it is not an easy task to theoretically associate an energy density to GWs in general situations. This is because in general relativity there is no local notion of energy of the gravitational field. By the equivalence principle, one may always go to a locally flat frame and erase any sign of gravity. Of course, we know that GWs carry momentum and energy (after all, they have been detected). Thus, indeed one can define the energy of GWs in some limits of interest. For instance, when the frequency of the GW is much higher than that of a slowly varying background \cite{Maggiore:1900zz}.

Relevant to the later discussion, let us briefly review the case of GWs propagating in a Ricci flat spacetime, i.e., a spacetime with no matter sources. We can then split the total metric as $g_{\mu\nu}=\bar g_{\mu\nu}+h_{\mu\nu}$ where $\bar g_{\mu\nu}$ is the background metric and $h_{\mu\nu}\ll1$ are the GWs, treated as a high frequency perturbation. Then, by the non-linear nature of gravity, these GW backreact onto the background metric. Mathematically speaking, if we expand the Einstein tensor up to second order we will find terms quadratic in $h_{\mu\nu}$ which can be thought of as a ``matter'' source. This means that, if we somehow ``integrate out'' the high frequency perturbation, we can write\footnote{We have that the linear terms in the perturbative expansion in $h_{\mu\nu}$ vanish after averaging.}
\begin{equation}
\bar G_{\mu\nu}[\bar g_{\mu\nu}]={M_{\rm pl}^{-2}}t_{\mu\nu}^{\rm GW}[h_{\mu\nu}]\,,
\end{equation}
where $G_{\mu\nu}$ is the Einstein tensor and $t_{\mu\nu}^{\rm GW}$ is the (pseudo)-energy momentum tensor of GWs. A way to ``integrate out'' the high frequency modes is by taking an average. For instance, we may focus on a small box and take the volume (and/or time) average. The box shall not be too small, so that several GW wavelengths fit in. In this case, we have that the (pseudo)-energy momentum tensor of GWs reads \cite{Isaacson:1967zz,Isaacson:1968zza}
\begin{equation}\label{eq:pseudotensor}
t_{\mu\nu}^{\rm GW}=\frac{M_{pl}^2}{4}\bigg\langle\partial_\mu h^{\alpha\beta}\partial_\nu h_{\alpha\beta}-\frac{1}{2}\bar g_{\mu\nu}\partial_\sigma h^{\alpha\beta}\partial^\sigma h_{\alpha\beta}\bigg\rangle\,,
\end{equation}
where all contractions of indices are done with the background metric.

In cosmology we face the extra difficulty of having a matter source to the Einstein equations which drives the expansion of the universe. Because of the expansion, we also have a cosmological horizon roughly given by $1/H$ where $H$ is the Hubble parameter and quantifies the expansion rate. For these reasons, the definition of GWs in a cosmological background is more subtle than in Ricci flat spacetimes. Nevertheless, with some tweaks we shall write in analogy
\begin{equation}\label{eq:Einsteincosmo}
G_{\mu\nu}={M_{\rm pl}^{-2}}\left(T_{\mu\nu}+t_{\mu\nu}^{\rm GW}\right)\,,
\end{equation}
where $T_{\mu\nu}$ is the matter energy-momentum tensor.
This time, however, we must be more careful on the averaging procedure. Furthermore, we must introduce a slightly different, but important, notation. First, in the helicity decomposition of the metric, the transverse-traceless component of the spatial metric is referred to as tensor mode. A tensor mode with a comoving wavenumber $k$ stays constant if its physical size is larger than the cosmological horizon, i.e., $k/a\ll H$. In the opposite limit, a tensor mode behaves as a GW if its physical size is much smaller than the cosmological horizon, that is $k/a\gg H$. Thus, Eq.~\eqref{eq:Einsteincosmo} only makes sense for the high frequency tensor modes deep inside the horizon. This also means that the small boxes we used to take the average have to be much smaller than the horizon size but still larger than several GW wavelengths. In addition to that, in cosmology we often deal with a SGWB and, by ergodicity, we may take an {ensemble average} over all small boxes. Under these assumptions we may use Eq.~\eqref{eq:pseudotensor} to compute the energy density of GWs in cosmology. Be aware though that these assumptions are strictly speaking not enough as we shall argue in Sec.~\ref{sec:gauge}.

Proceeding with the cosmological SGWB, we consider GWs propagating in a Friedmann-Lema{\^i}tre-Robertson-Walker (FLRW) metric, namely\footnote{Note that sometimes in the literature there is an additional factor $1/2$ in front of $h_{ij}$, e.g., in Refs.~\cite{Ananda:2006af,Kohri:2018awv}. This slightly changes the numerical factors in different steps of the calculations but, of course, yields the same results in the end.}
\begin{equation}\label{eq:FLRW}
 ds^2=-dt^2+a^2(t)(\delta_{ij}+h_{ij})dx^idx^j\,,
\end{equation} 
where $a(t)$ is the scale factor. The evolution of the scale factor is dictated by the Friedmann equations, which we provide in App.~\ref{app:generalgauge}. Using Eq.~\eqref{eq:pseudotensor} under the ensemble average procedure, we find that the energy density of GWs (the $00$ component of $t_{\mu\nu}^{\rm GW}$) in Fourier space (see App.~\ref{app:polarization} for notations) reads
\begin{equation}\label{eq:spectumdensity}
\rho_{\rm GW}=M_{\rm pl}^2\int d\ln k \frac{k^3}{16\pi^2}\left\{\sum_\lambda\langle \dot h_{\bm{k},\lambda}\dot h_{-\bm{k},\lambda}\rangle^{!}+ \frac{k^2}{a^2} \langle h_{\bm{k},\lambda}h_{-\bm{k},\lambda}\rangle^{!}\right\}\,,
\end{equation}
where $\dot\,\equiv d/dt$, $\lambda$ is the GW polarization and we already used that the coincident ensemble average of two tensor modes in a homogeneous and isotropic universe is proportional to a Dirac delta,\footnote{We are assuming that the tensor modes are drawn from a Gaussian distribution. This is essentially the case when they are generated by quantum fluctuations during inflation. Then, the expectation of a two-point correlation function of a random variable that follows a Gaussian distribution is proportional to the Dirac delta.} namely
\begin{equation}\label{eq:hh!}
\langle h_{\bm{k},\lambda}h_{\bm{k}',\lambda}\rangle= \langle h_{\bm{k},\lambda}h_{-\bm{k},\lambda}\rangle^{!}(2\pi)^3\delta^{(3)}(\bm{k}+\bm{k'})\,.
\end{equation}
The notation "!" in Eqs.~\eqref{eq:spectumdensity} and \eqref{eq:hh!} is to indicate that the Dirac delta has been factored out, so that $\langle h_{\bm{k},\lambda}h_{-\bm{k},\lambda}\rangle^{!}$ is directly related to the power spectrum, which by isotropy is a function of $k=|\bm{k}|$ only.
Eq.~\eqref{eq:spectumdensity} is the total energy density of the GWs filling the universe. It is often more convenient to use the spectral density fraction, i.e., the GW power per logarithm of a given wavenumber $k$ (or frequency) over the critical density, which is defined as
\begin{equation}\label{eq:spectraldensity}
\Omega_{\rm GW}(k)\equiv\frac{1}{3 M_{\rm pl}^2 H^2}\frac{d \rho_{\rm GW}}{d\ln k}=\frac{k^2}{12a^2H^2}\sum_\lambda{\cal P}_{h,\lambda}(k)\,,
\end{equation}
where $H=\dot a/a$.
This is often what is used to compare theoretical predictions with current constraints and future observational prospects. Note that in Eq.~\eqref{eq:spectraldensity} we used the definition of the dimensionless power spectrum, concretely
\begin{equation}\label{eq:hh}
\langle h_{\bm{k},\lambda}h_{-\bm{k},\lambda}\rangle^{!}= \frac{2\pi^2}{k^3}{\cal P}_{h,\lambda}(k)\,.
\end{equation}
We also used the approximation that for a freely propagating wave we have that 
\begin{align}
\dot h_{\bm{k},\lambda}\approx\frac{k}{a} h_{\bm{k},\lambda}\,.
\end{align}
With these ingredients we can use Eq.~\eqref{eq:spectraldensity} to estimate the amplitude and spectral shape of induced GWs.

\subsection{GWs induced by primordial fluctuations \label{sec:induced1}}

Let us derive some estimates by considering a rough, yet instructive, toy model for induced GWs. Take a tensor mode $h_{ij}$ in a FLRW background \eqref{eq:FLRW} sourced by scalar field fluctuations $\delta\varphi$. Its equations of motion, in terms of conformal time $d\tau=dt/a$, read
\begin{equation}\label{eq:eom}
h''_{ij}+2{\cal H}h'_{ij}-\partial_k\partial^k h_{ij}={\cal P}^{ab}\,_{ij}\left\{T_{ab}\right\}\approx {\cal P}^{ab}\,_{ij}\left\{\partial_a\delta\varphi\partial_b\delta\varphi\right\}\,,
\end{equation}
where $'\equiv d/d\tau$, ${\cal H}=H/a=a'/a$, $T_{ij}$ is the spatial components of the energy-momentum tensor of a scalar field (see Eq.~\eqref{eq:Tmunuphi}) and ${\cal P}^{ab}\,_{ij}$ is the transverse-traceless projector defined in App.~\ref{app:generalgauge}. Without understanding much about the evolution of these fluctuations, we can already see that 
\begin{equation}
{\cal P}_{h}\equiv\sum_\lambda{\cal P}_{h,\lambda}\propto {\cal P}^2_{\delta\varphi}\,.
\end{equation}
If the scalar field fluctuations were sourced during inflation, we can relate them to the curvature perturbation ${\cal R}$ so that ${\cal P}_{\delta\varphi}\sim {\cal P}_{\cal R}$. Furthermore, the spectral density of GWs in a radiation dominated universe remains constant, as GWs behave as radiation themselves. Thus, we can boldly estimate that the amplitude of induced GW spectral density measured today is given by
\begin{equation}\label{eq:spectraldensityestimate}
\Omega^{\rm induced}_{\rm GW}h^2\sim\frac{1}{12}\Omega_{r,0} h^2\times{\cal P}^2_{\cal R}\,,
\end{equation}
where $\Omega_{r,0}\equiv \rho_{r,0}/(3H_0^2M_{\rm pl}^2)$ is the density fraction of radiation, $H_0$ is the Hubble rate today and $h\equiv H_0/(100 km/Mpc/s)$. We have introduced $\Omega_{r,0}$ to take into account the dilution of the GWs (as radiation) as the universe expands and goes through the cold dark matter and dark energy dominated stages. We give more details on this factor in Sec.~\ref{sec:typical}. $h$ considers the uncertainty in the exact value of $H_0$. From the latest Planck 2018 results \cite{Aghanim:2018eyx} we have that $\Omega_{r,0} h^2\approx 4\times 10^{-5}$. With all the above simplifications, we arrive at
\begin{equation}\label{eq:spectraldensityestimate2}
\Omega^{\rm induced}_{\rm GW}h^2\sim 10^{-6}{\cal P}^2_{\cal R}\,.
\end{equation}

The estimate \eqref{eq:spectraldensityestimate2} can be now contrasted with future observational prospects. The sensitivity of future GW detectors after collecting data over several years might reach $\Omega^{\rm induced}_{\rm GW}h^2\sim 10^{-16}$, if we use the power-law sensitivity curves of Thrane and Romano \cite{Thrane:2013oya}. This optimistic value is for the best DECIGO sensitivity \cite{Kawamura:2011zz,Yagi:2011wg}. Then, in the absence of an induced GW detection we can use Eq.~\eqref{eq:spectraldensityestimate2} to place an optimistic upper bound to the primordial spectrum on the corresponding scales (roughly at $10^{16}$-$10^{12}$ ${\rm Mpc}^{-1}$ ) of about ${\cal P}_{\cal R}<10^{-5}$. This is already 3 orders of magnitude better than current constraints from the absence of PBHs (around ${\cal P}_{\cal R}<10^{-2}$ \cite{Sasaki:2018dmp}), although it is still 4 orders of magnitude above the CMB extrapolation which would be around ${\cal P}_{\cal R}\sim10^{-9}$ \cite{Akrami:2018odb}. The most interesting point though is that PBHs which formed by the collapse of large primordial fluctuations, which requires ${\cal P}_{\cal R}\gtrsim 10^{-2}$, must have an observable large induced GW counterpart. Thus, the absence of induced GWs completely rules out the PBH scenario from large primordial fluctuations \cite{Hawking:1974sw,Carr:1974nx}. Note that PBH could have formed by other mechanisms though. For example, PBH could form in first order phase transitions \cite{Crawford:1982yz,Kodama:1982sf}, tunnelling during inflation \cite{Garriga:2015fdk}, the collapse of Q-balls \cite{Cotner:2016cvr,Cotner:2019ykd} and might also be the result of long range interactions stronger than gravity \cite{Amendola:2017xhl,Savastano:2019zpr,Flores:2020drq}.

In the above estimations we have completely neglected the evolution of the scalar fluctuations since the end of inflation onwards, which is by no means irrelevant. As shown in Refs.~\cite{Assadullahi:2009nf,Alabidi:2013lya,Inomata:2019ivs,Inomata:2019zqy}, a period of dust domination might substantially enhance the amplitude of induced GWs. The amplification is such that GWs induced by tiny primordial fluctuations might observable \cite{Alabidi:2013lya,Inomata:2019ivs}. This case exemplifies the importance of the expansion history of the primordial universe. At this point, we would like to emphasise that contrary to studies of the CMB we have scarce evidence of the content of the primordial universe. Therefore, when deriving theoretical predictions, and eventually contrasting them with the data, it would be a good idea to be open and acknowledge our ignorance by allowing a free equation of state parameter $w\equiv \rho/p$ and a free propagation speed of fluctuations $c_s$, which describe the unknown content of the primordial universe. A nice review on the implications of various expansion histories can be found in Ref.~\cite{Allahverdi:2020bys}. For these reasons, let us estimate the general impact of the equation of state $w$ and sound speed $c_s$ onto the induced GW spectrum. To do that, we continue with the toy model of the scalar field. This time, however, we will consider the evolution of $\delta\varphi$. Fluctuations of a $K$-essence massless scalar field roughly follow the Klein-Gordon equation
\begin{equation}\label{eq:deltaphi}
\delta\varphi_k''+2{\cal H}  \delta\varphi_k'+c_s^2k^2\delta\varphi_k=0\,,
\end{equation}
where for simplicity we considered a constant $w$ and $c_s$. The former implies constant expansion rate, i.e.,
\begin{equation}\label{eq:epsilon}
\epsilon\equiv-\frac{\dot H}{H^2}=\frac{3}{2}(1+w)={\rm constant}\,.
\end{equation}
Such constant expansion rate implies a scale factor which goes as a power-law of conformal time, explicitly from the Eqs. in App.~\ref{app:generalgauge} we have
\begin{equation}\label{eq:b}
a\propto\tau^{1+b}\quad{\rm where}\quad b=\frac{1-3w}{1+3w}\,.
\end{equation}
We introduced the parameter $b$ for later convenience. $b=0$ corresponds to a universe filled with radiation ($w=1/3$). Then, $b<0$ and $b>0$ respectively correspond to a stiffer and softer fluid. Note that if we allow $w\gg1$ the range of $b$ goes as $-1<b<\infty$ for $\infty>w>-1/3$
.

For the sake of simplicity, we solve the Klein-Gordon equation \eqref{eq:deltaphi} in the super-(sound)horizon, $c_sk\ll{\cal H}$, and sub-(sound)horizon, $c_sk\gg{\cal H}$, limits. This leads us to
\begin{equation}\label{eq:deltaphics}
\delta\varphi_k\approx\delta\varphi_i\left\{
\begin{aligned}
&{\rm constant} & c_sk\tau\ll1 \\
&a^{-1}{e^{ic_sk\tau}}& c_sk\tau\gg 1 
\end{aligned}
\right.\,,
\end{equation}
where $\delta\varphi_i$ is an initial value.
The above solutions are all we need to solve for the induced GWs. For analytical viability, let us take that there are fluctuations of the scalar field only in a single scale $k_p$. Namely the scalar fluctuations are Dirac-delta like in Fourier space, i.e.,
\begin{equation}\label{eq:deltap}
\delta\varphi_i\sim \delta\varphi_p\times\delta(\ln(k/k_p))\,,
\end{equation}
where $\delta\varphi_p$ is the primordial amplitude, e.g., set by inflation.
All the above simplifications lead to a simple differential equation for the tensor modes given by
\begin{equation}\label{eq:hconformal}
h_k''+2{\cal H} h_k'+k^2h_k\approx{k_p^2}\delta\varphi_p^2\,.
\end{equation}
We readily see that, if we focus on $\delta\varphi_p={\rm constant}$ and we evaluate $h_k$ at horizon crossing, that is when the mode $k$ is of the size of the comoving horizon, i.e., $k=aH$, then the amplitude of induced GWs is proportional to $h_k\sim \delta\varphi_p^2$ for $k\sim k_p$, as used in our previous estimate \eqref{eq:spectraldensityestimate}. Now, to derive the spectral features we can focus on two interesting limits: $(i)$ the infra-red (IR) tail for modes with $k\ll k_p$ and $(ii)$ the near peak behaviour for modes with $k\sim k_p$.

\subsubsection{The IR tail}
If we focus on modes with $k\ll k_p$ there will be no competition between the $k^2$ and $k_p^2$ terms nor any resonance. Thus, we can peacefully solve the tensor modes in the superhorizon ($k\tau\ll1$) and subhorizon ($k\tau\gg1$) regimes. First, in the superhorizon regime we find that
\begin{equation}\label{eq:hsuper}
h(k\tau\ll1)\approx\left\{
\begin{aligned}
&\left(k_p\tau\right)^2 & c_sk_p\tau\ll 1 \\
&{\rm constant}+\left(k_p\tau\right)^{-2b}& c_sk_p\tau\gg 1 
\end{aligned}
\right.\,.
\end{equation}
What Eq.~\eqref{eq:hsuper} tells us is that tensor modes initially grow as $\tau^2$ due to a constant source. Then, after the scalar source enters the sub(sound)-horizon regime, i.e., at $c_sk_p\tau_p=1$, the tensors developed a constant amplitude. Now, if $b>0$ ($w<1/3$) the scalar source decays much faster than the cosmological background, that is $\delta\varphi_k^2/{\cal H}^2\propto (k\tau)^{-b}\to 0$, and the tensor growth stops. However, for $b<0$ ($w>1/3$) the scalar source decays slower than the background expansion which yields a second superhorizon growth of the tensor modes \cite{Domenech:2020kqm}. To follow tensor mode evolution, the second line of Eq.~\eqref{eq:hsuper} has to be matched at horizon crossing for the tensor mode ($k\tau=1$) with the subhorizon solution. By doing so, we arrive at
\begin{equation}\label{eq:hsub}
h(k\tau\gg 1)\approx\left(k\tau\right)^{-1-b}{e^{ik\tau}}\left({\rm constant}+\left({k_p}/{k}\right)^{-2b}\right)\,.
\end{equation}
The final ingredient is that somehow the universe transitions to a radiation dominated universe to recover the standard cosmology. For simplicity, we take an instantaneous reheating at $\tau=\tau_{\rm rh}$ with the corresponding reheating scale $k_{\rm rh}={\cal H}_{\rm rh}$. From that moment on the spectral density of induced GWs stays constant and we can derive our predictions. Inserting Eq.~\eqref{eq:hsub} evaluated at $\tau=\tau_{\rm rh}$ into Eq.~\eqref{eq:spectumdensity} we find that the IR slope of the induced GW spectrum is in general given by
\begin{equation}\label{eq:IRtail}
\frac{d\ln\Omega_{\rm GW}^{\rm induced}}{d\ln k}(k_{\rm rh}\ll k\ll k_p)\approx 3-2|b|\,.
\end{equation}
We later find that for $\beta\in\mathbb{Z}$ there is a logarithmic correction \cite{Domenech:2020kqm}. Also, if the scalar spectrum of fluctuations is really a Dirac delta, the spectrum slope is given by $2-2|b|$ due to the unphysical divergence of $\delta\varphi_p$ at $k=k_p$. Note that for tensor modes which entered during the radiation domination era we have to match the first line of \eqref{eq:hsuper} to the subhorizon solution. This yields a spectral index given by
\begin{equation}\label{eq:IRtail2}
\frac{d\ln\Omega_{\rm GW}^{\rm induced}}{d\ln k}(k\ll k_{\rm rh})\approx 3\,,
\end{equation}
in agreement with the general results of Ref.~\cite{Cai:2019cdl}. These naive derivations will be later recovered from the analytical formulas in Sec.~\ref{sec:analytical}. It is interesting to see that for $b>3/2$ ($w<-1/15$) the induced GW spectrum peaks at $k\sim k_{\rm rh}$ which is clearly different than the peak in the primordial spectrum at $k\sim k_p$.

The intuitive physical picture behind \eqref{eq:IRtail} is the following. In a universe with constant $b$, modes of a massless field (e.g., the tensor modes $h_{ij}$ or the scalar field $\delta\varphi$) which enter the horizon early get more diluted than modes which enter later. For example, consider that right after horizon crossing, i.e., when $k={\cal H}_k$, we have that the energy density of the massless field starts to dilute as $\rho\propto (a_k/a)^{4}$. Since we have that $a_k\propto \tau_k^{1+b}\propto k^{-1-b}$ then $\rho\propto a_k^{4}\propto k^{-4b}$ and $\rho/H^2\propto a^{-2b/(1+b)}\propto k^{-2b}$. Thus, if the IR tail of induced GWs in radiation domination goes as $k^3$, we expect $\Omega_{\rm GWs}\propto k^{3-2b}$ due to dilution \cite{Cai:2019cdl}. However, as we have shown, this is incomplete for induced GWs from a peaked spectrum. For $b<0$, the density fraction of the scalar field $\rho_{\delta\varphi}$ grows as $a^{-2b/(1+b)}$. This means that induced tensor modes which are generated later have a larger source and a larger amplitude than those generated earlier. In other words, for $b<0$ small $k$ have a larger amplitude than large $k$ by a factor $a_k^{-4}\propto k^{4b}$. Then, for $b<0$ the density fraction of induced GWs goes as $\Omega_{\rm GWs}\propto k^{3-2b+4b}\propto k^{3+2b}$ \cite{Domenech:2020kqm}. This gives the general formula $\Omega_{\rm GWs}\propto  k^{3-2|b|}$ that we derived in \eqref{eq:IRtail}. Whether this absolute value of $b$ is unique to induced GWs is something yet to explore.

Before moving on to the near peak limit, let us comment on the implications of Eq.~\eqref{eq:IRtail}. While the far most IR tail of the induced GW spectrum indeed must go as $k^3$, we may have an intermediate regime where $\Omega_{\rm GWs}\propto  k^{3-2|b|}$. This has interesting consequences when comparing with GWs from other sources. Let us model a general GW spectrum by a broken power-law around a characteristic scale $k_c$, as done in Ref.~\cite{Kuroyanagi:2018csn}. Then, for example, the IR spectrum of induced GWs for $b=2$ ($w=-1/9$) resembles to that of GWs generated by a strong first order phase transition which have $\Omega_{\rm GWs}\propto  k^{2.8}$ for $k<k_c$ and $\Omega_{\rm GWs}\propto  k^{-1}$ for $k>k_c$. This gives another very good reason to be open about a primordial universe with a different equation of state than radiation.

\subsubsection{The near peak regime}
Looking closely at Eq.~\eqref{eq:hconformal}, we might suspect that there could be a resonance if the wavenumber $k$ of a tensor mode approaches that of twice the scalar mode $2k_p$. This resonance may be understood in two ways. First, we know that a harmonic oscillator with an external force has a resonance when the period of the force matches that of the harmonic oscillator. If that occurs, the amplitude of the oscillations diverges in time. In short, the external force is kicking the oscillator at the right times. The second way to see it is that two scalar modes with physical momentum $c_s k_p$ have a preferred window to produce a tensor mode with physical momentum equal to twice the scalar momenta, i.e., $k=2c_s k_p$, if allowed by momentum conservation. To see this mathematically, let us redefine the tensor modes in Eq.~\eqref{eq:hconformal} to remove the Hubble friction by
\begin{equation}\label{eq:hkv}
h_k=v_k/a\,.
\end{equation}
With this new variable $v_k$ Eq.~\eqref{eq:hconformal} now reads
\begin{equation}\label{eq:hkresonance}
v''_k+\left({k^2}-\frac{b(1+b)}{\tau^2}\right)v_k=\frac{k_p^2}{a}\delta\varphi_p^2{e^{2ic_sk_p\tau}}\,,
\end{equation}
where we only focused on the oscillatory behaviour of the scalar field for $c_sk_p\tau\gg1$. The particular solution to $v_k$ can be found by the Green's function method (see App.~\ref{sec:green} for the details).  Neglecting the term proportional to $1/\tau^2$ in Eq.~\eqref{eq:hkresonance}, the homogeneous solutions to $v_k$ are $e^{\pm ik\tau}$. With such homogeneous solutions, the particular solution by the Green's method reads
\begin{equation}\label{eq:hkresonance2}
v_k\approx\int_{\tau_i}^\tau d\tilde\tau\,\sin\left(k(\tau-\tilde\tau)\right)\tilde\tau^{-1-b}e^{2ic_sk_p\tilde\tau}\,.
\end{equation}
We clearly see from the integrand in Eq.~\eqref{eq:hkresonance2} that there is a resonance when the frequency of the sine equals that of the exponential. Then the oscillations interfere and leave the integral of a power-law. Since we are only interested in how the amplitude diverges, let us only focus on the divergent part of \eqref{eq:hkresonance2} which is given by
\begin{equation}\label{eq:hkresonance3}
v_k\propto\int d\tilde\tau\,\tilde\tau^{-1-b}e^{-i\tilde\tau(k-2c_sk_p)}\propto  \Gamma[-b](k-2c_sk_p)^{b}\,.
\end{equation}
We see that when $b\leq0$ the resonance is effective and the amplitude diverges. Note that for $b=0$ the divergence is logarithmic. When $b>0$ the resonance is not effective enough to cause a divergence as the amplitude of the integrand decays faster than $\tau^{-1}$. However, since we so far used a Taylor expansion around $k\sim 2c_sk_p$ we expect that for $1>b>0$ Eq.~\eqref{eq:hkresonance3} still yields the leading contribution after the constant term. Thus, for $1>b>0$ we also expect a peak at around $k\sim 2c_sk_p$, albeit not divergent. For $b>1$ the situation is less clear and a detailed analysis is necessary. Nevertheless for $b<1$, we can roughly write that the tensor modes in the near peak regime have an amplitude given by
\begin{equation}\label{eq:hkvsol}
h_k(k\sim 2c_sk_p,b<1)\propto {\rm constant}-(k-2c_sk_p)^{b}\,.
\end{equation}
This also implies that the induced GW spectrum has a peak near $k\sim 2c_sk_p$ with a growth rate proportional to
\begin{align}\label{eq:omegakv}
\Omega_{\rm GW}^{\rm induced}(k\sim 2c_sk_p)\propto \left\{
\begin{aligned}
&\left|k-{2c_sk_p}\right|^{-2|b|} & b<0\\
&{\rm constant}-\left|k-{2c_sk_p}\right|^{b} & 1>b>0
\end{aligned}
\right.\,.
\end{align}
These estimates will be later checked by the general analytical solutions in Sec.~\ref{sec:analytical}. Thus, the resonant peak in the induced GW spectrum has information on the sound horizon $c_sk_p$ by the position of the peak and on the expansion history $b$ by the growth rate around the peak. This type of resonant structure is characteristic of induced GWs. Thus, the observation of such resonant peak could give strong evidence for induced GWs. Then, the sound speed of fluctuations $c_s$ and the peak $k_p$ could be disentangled by the simultaneous observation of the PBH counterpart, which we proceed to discuss. Note that in \eqref{eq:omegakv}
the radiation domination case $b=0$ is ``special'' because it is exactly when the amplitude of the source term is proportional to the background density.

\subsection{Primordial black hole counterpart \label{sec:PBH}}

In this section we briefly introduce the PBH counterpart to the induced GWs. It is by no means a review on PBH formation and it does not intend to be so. A detailed review on PBHs can be found in Ref.~\cite{Sasaki:2018dmp}. Other interesting reviews are listed at the \hyperref[sec:structure]{beginning of this review}. PBHs from large primordial fluctuations will form in those Hubble patches where the density contrast $\delta\rho/\rho$ is above a certain threshold\footnote{Here we define the critical threshold in terms of $\delta\rho/\rho$. However, it may also be defined through the peak of the so-called compaction function \cite{Shibata:1999zs}. See Refs.~\cite{Nakama:2013ica,Harada:2015yda,Escriva:2021pmf} for recent studies.} $\delta_c$~\cite{Zeldovich:1967lct,Hawking:1971ei,Carr:1974nx,Meszaros:1974tb,Carr:1975qj,Khlopov:1985jw,Niemeyer:1999ak}, which depends on the EoS parameter $w$ at horizon re-entry~\cite{Musco:2004ak,Musco:2008hv,Musco:2012au,Harada:2013epa,Escriva:2019phb,Musco:2020jjb,Young:2020xmk,Escriva:2020tak}. Since fluctuations generated during inflation closely follow a Gaussian distribution centred around zero, only the tail of the distribution has a large enough amplitude to collapse to PBH. This means that PBH formation is an extremely rare event. The fraction of Hubble patches where PBH forms is exponentially suppressed and sensitive to the root mean squared of the distribution, which is related to the power spectrum. This is why we generally need ${\cal P}_{\cal R}\sim 10^{-2}$ to form a substantial fraction of PBHs. Precise estimations of the fraction of PBH is an active and ongoing research field, e.g., see Refs.~\cite{Kehagias:2019eil,Atal:2019erb,Yoo:2019pma,Riccardi:2021rlf,Young:2019yug,Musco:2020jjb,Young:2020xmk,Escriva:2020tak,Escriva:2021pmf} and references therein. For our purposes, it is enough to know that only large induced GWs have a large associated fraction of PBH and, inversely, the absence of large induced GWs rules out PBHs from large primordial fluctuations.

For simplicity, we continue with the assumptions of the previous section and take a primordial spectrum with a Dirac delta peak at $k_p$. This essentially leads to an almost monochromatic\footnote{From the critical collapse we know there will be some dispersion in the masses. For simplicity, we neglect it here.} PBH mass function. When the peak in the scalar spectrum enters the horizon, those fluctuations which are above $\delta_c$ collapse into a PBH. By the end of the collapse, only a fraction $\gamma$ of the total mass enclosed inside the horizon at horizon crossing ($H_\circledast a_\circledast=k_p$) will end up as a PBH. We use the symbol $\circledast$ to refer to evaluation at PBH formation time. In mathematical terms, we have that PBHs mainly form with a mass equal to the mass enclosed inside the Hubble sphere, that is
\begin{equation}\label{eq:MPBH}
M_{{\rm PBH},\circledast}=\gamma\frac{4\pi}{3}\frac{\rho_\circledast}{H_\circledast^3}=\frac{4\pi\gamma M_{\rm pl}^2}{H_\circledast}\,.
\end{equation}
To grasp the magnitude of the actual PBH mass, let us write it in terms of a solar mass $M_\odot$ by referring all quantities in \eqref{eq:MPBH} in terms of the size of the horizon at matter-radiation equality which is very well measured by Planck \cite{Aghanim:2018eyx}. The numerical factors and formulas used to derive this and the following estimates can be found App.~\ref{app:formulasuseful}. After some numerical manipulations we arrive at
\begin{align}
M_{{\rm PBH},\circledast}\approx 3.97\times 10^{-12}M_\odot\left(\frac{k_p}{k_{\rm rh}}\right)^{-2-b}&\left(\frac{k_{\rm rh}}{10^{12}{\rm Mpc}^{-1}}\right)^{-2}\nonumber\\&\times\left(\frac{\gamma}{0.2}\right)\left(\frac{g_*(T_{\rm rh})}{106.75}\right)^{1/2}\left(\frac{g_{*s}(T_{\rm rh})}{106.75}\right)^{-2/3}\,,
\end{align}
where we kept our ignorance on the primordial universe by leaving a free $b$
(or $w$) \eqref{eq:b}. We took $\gamma\sim 0.2$ in a radiation dominated universe \cite{Carr:1975qj} but it may differ for different expansion histories \cite{Musco:2004ak,Musco:2008hv,Musco:2012au,Harada:2013epa}. $g_{*}(T_{\rm rh})$ and $g_{*s}(T_{\rm rh})$ are the effective degrees of freedom in the energy density and entropy evaluated at reheating. It should be noted that we also assumed an instantaneous reheating. In that case, the horizon scale at reheating $k_{\rm rh}$ can be related to the reheating temperature by
\begin{align}\label{eq:krh}
k_{\rm rh}=1.2\times 10^{12}\,{\rm Mpc}^{-1}\,\left(\frac{T_{\rm rh}}{5\times 10^4\,{\rm GeV}}\right)\left(\frac{g_*(T_{\rm rh})}{106.75}\right)^{1/2}\left(\frac{g_{*s}(T_{\rm rh})}{106.75}\right)^{-1/3}\,.
\end{align}

As we have seen in Sec.~\ref{sec:induced1}, the peak in the primordial curvature power spectrum induces GWs. These GWs have a typical frequency $f_p$ associated to the wavenumber $k_p$. Since the typical mass and the typical frequency are related by the same typical scale $k_p$, we can write one in terms of the other. In this way, the typical frequency evaluated today reads
\begin{equation}\label{eq:fpM}
f_p=\frac{k_{p}}{2\pi a_0}\approx 1.54\times 10^{-3}\,{\rm Hz}\left(\frac{M_{{\rm PBH},\circledast}}{3.97\times 10^{-12}M_\odot}\right)^{\frac{-1}{2+b}}\left(\frac{f_{\rm rh}}{1.54\times 10^{-3}\,{\rm Hz}}\right)^{\frac{b}{2+b}}\,.
\end{equation}
For a radiation dominated universe ($b=0$) we have a one-to-one correspondence between the frequency and the mass. If we fix the reheating scale, the relation \eqref{eq:fpM} roughly tells us that PBH with a given mass have an induced GW counterpart with a peak frequency at around $f_p$. The exact relation between $f_p$ and $M_{{\rm PBH},\circledast}$ depends on the value of $b$. For instance, if $b<0$ ($w>1/3$) the expansion rate was faster and for a fixed $k_p$ corresponds to smaller PBH mass at formation. The opposite holds for $b>0$ ($w<1/3$). It should also be noted that for $b>3/2$ ($w<-1/15$) there is another peak in the induced GWs at $k\sim k_{\rm rh}$. This may break the correspondence between the PBH mass and the peak of the induced GWs if not all the GW spectrum is seen.


\section{General formalism\label{sec:general}}

In this section we present the general formulation of induced GWs. This includes the derivation of the equations of motion in Sec.~\ref{sec:derivation} and the general form of the solutions in Sec.~\ref{eq:generalsolutions}. Then in Sec.~\ref{sec:ng} we include the leading terms due to local-type non-Gaussianity. In Sec.~\ref{sec:derivation}, we derive the equations of motion using the action formalism and so we do not follow the conventional derivation, e.g., of Refs.~\cite{Matarrese:1992rp,Ananda:2006af}. The final result is obviously the same, but the derivation is rather quick and clear. Throughout this section and on, we shall neglect the effect of vector perturbations as they typically decay. Anyone interested in the approximations and applications may jump directly to Sec.~\ref{sec:analytical}.\\

\noindent\faBook\,\,\textit{Main references:} In Sec.~\ref{sec:derivation} we closely follow Ref.~\cite{Domenech:2017ems} although we work at the level of the Lagrangian. For alternative derivations starting from the Einstein equations see for example Refs.~\cite{Matarrese:1992rp,Ananda:2006af,Baumann:2007zm}. Then, Sec.~\ref{eq:generalsolutions} is extracted from \cite{Domenech:2019quo} which is built upon and uses the notation of Ref.~\cite{Kohri:2018awv} except for the numerical factor ${2}$ involving $h_{ij}$. Lastly, Sec.~\ref{sec:ng} on the primordial non-Gaussianity is mainly based on Ref.~\cite{Atal:2021jyo} updated and corrected by Ref.~\cite{Adshead:2021hnm}.

\subsection{Derivation from the action \label{sec:derivation}}
We need to arrive at the equations of motion for the transverse-traceless component of the spatial metric at second order in cosmological perturbation theory. Although it is a second order calculation, we shall derive it quickly from the action formalism; if we work in a particular gauge and we only focus on the interaction of tensor with scalars. It is particularly convenient to work in a gauge similar to the Newtonian gauge\footnote{Although we will abuse the notation ``Newtonian gauge'', the Newtonian gauge is actually defined as the shear-free slicing at the linear level. Our gauge choice reduces to the definition of the Newtonian gauge at first order but there might be some subtleties if one wants to relate them at second order.} with the exponential notation typical of Misner, Throne \& Wheeler \cite{Misner:1974qy}. With this choice and assuming a perturbed FLRW universe, the metric reads
\begin{equation}\label{eq:confdecom}
   ds^2=g_{\mu\nu}dx^\mu dx^\nu
   =-{\rm e}^{2\Psi}dt^2+a^2(t){\rm e}^{2\Phi}Y_{ij}dx^idx^j\,,
\end{equation}
where $g_{\mu\nu}$ is the metric of our space-time, $i=\{1,\,2,\,3\}$ are the spatial components, $a(t)$ is the scale factor and we have further used the conformal decomposition of the spatial metric such that
\begin{equation}\label{eq:derdet}
   \frac{\partial}{\partial t}\det Y = Y^{ij}\frac{\partial}{\partial t}Y_{ij}=0\,.
\end{equation}
Note that for a flat FLRW background in Cartesian coordinates we have $Y_{ij}=\delta_{ij}$.
The conformal decomposition of the spatial metric is very convenient and simplifies calculations. It is also used, e.g., in Numerical Relativity \cite{Gourgoulhon:2007ue}. In short, the conformal decomposition completely splits the trace (volume changing) from the trace-free (volume preserving) degrees of freedom. In the Newtonian gauge this decomposition directly splits the scalar and tensor modes of the spatial metric. This means that $Y_{ij}$ only contains the transverse-traceless degrees of freedom. As we shall see, it also leaves a clean splitting between $Y_{ij}$ and $\Phi$ at the level of the action.
   
To consider a sensible cosmological set up, we include a canonical scalar field $\varphi$ in the perturbed FLRW metric. We can later generalize it to a perfect fluid quite straightforwardly. The action, without any decomposition, is given by
\begin{equation}\label{eq:action1}
      S= \int d^4x \sqrt{-g} \left\{\frac{1}{2}R-\frac{1}{2}g^{\mu\nu}\partial_\mu\varphi\partial_\nu\varphi-V(\varphi)\right\}\,,
\end{equation}
where $g$ is the determinant of $g_{\mu\nu}$, $R$ is the 4D Ricci scalar, $\partial_\mu\equiv\partial/\partial x^\mu$ and $V(\varphi)$ is the potential of $\varphi$.
In the (3+1) conformal decomposition \eqref{eq:confdecom}, after some algebra and integration by parts, the action becomes
\begin{equation}\label{eq:actionsim}
      \begin{split}
         S=\int d^3x \,dt &\bigg\{ a\,{\rm e}^{\Psi+\Phi}\Bigg(\frac{1}{2}R^{(3)}[Y_{ij}]-2Y^{ij}D_iD_j\Phi-Y^{ij} D_i \Phi D_j \Phi-\frac{1}{2}Y^{ij} D_i \varphi D_j \varphi\\&
         +a^3{\rm e}^{3\Phi-\Psi}\left(\frac{1}{8}Y^{ij}Y^{kl}\dot Y_{ik}\dot Y_{jl}-{3}\left(H+\dot\Phi\right)^2+\frac{1}{2}\dot\varphi^2\right)-a^3{\rm e}^{3\Phi+\Psi}V(\varphi)\bigg\}\,,
      \end{split}
\end{equation}
where $R^{(3)}[Y_{ij}]$ and $D_i$ are respectively the 3D Ricci scalar and the covariant derivative associated to $Y_{ij}$. Since we work in Cartesian coordinates, we already used that $\det Y=1$. In going from \eqref{eq:action1} to \eqref{eq:actionsim} we took a big leap in the (3+1) and conformal decomposition of the 4D Ricci scalar. Some steps can be found in App.~\ref{app:ADM}. For more details, the interested reader is referred to E.~Poisson's book \cite{Poisson:2009pwt} for the (3+1) or ADM decomposition \cite{Arnowitt:1962hi}, and to the appendix of R.~Wald's book \cite{Wald:1984rg} for the conformal transformation rules.

Now, we could take the variation of \eqref{eq:actionsim} with respect to $Y_{ij}$, having in mind that the result should still be transverse and traceless, and obtain the transverse-traceless spatial component of Einstein Equations. However, this would not be very illuminating. Before taking the variation, let us use cosmological perturbation theory and expand the action. A smart way to decompose the spatial metric and to keep the requirement \eqref{eq:derdet} is to take the exponential matrix, i.e.,
\begin{align}\label{eq:exph}
Y_{ij}=(e^h)_{ij}=\delta_{ij}+h_{ij}+\frac{1}{2}\delta^{kl}h_{ik}h_{jl}+O(h^3)\,,
\end{align}
where $h_{ij}\ll1$ are the transverse-traceless (tensor) modes and satisfy $\delta^{ij}h_{ij}=\partial^i h_{ij}=0$. This is used, e.g., in Maldacena's work on non-gaussianities \cite{Maldacena:2002vr}. For the details of the expansion in general gauges see \cite{Domenech:2017ems}. From now on, spatial indices are contracted with the background spatial metric $\delta_{ij}$, for example $h^{ij}h_{ij}\equiv\delta^{ij}\delta^{kl}h_{ik}h_{jl}$. The inverse metric is then $Y^{ij}=\delta^{ik}\delta^{jl}(e^{-h})_{kl}$. With this expansion, the 3D Ricci scalar up to second order is given by
\begin{equation}
   R^{(3)}[e^h]=-\frac{1}{4}\partial_ih_{kl}\partial^ih^{kl}+O(h^3)\,.
\end{equation}
We stopped at second order since we are only interested in scalar-scalar-tensor interactions. We split the other variables as
\begin{align}\label{eq:otherexp}
\Psi=\Psi(t,\mathbf{x})\quad,\quad \Phi=\Phi(t,\mathbf{x})\quad,\quad \varphi=\bar\varphi(t)+\delta\varphi(t,\mathbf{x})\,,
\end{align}
where $\bar\varphi(t)$ is the background solution and $\Phi$, $\Psi$ and $\delta\varphi$ are the perturbations. Using the perturbative expansions \eqref{eq:exph} and \eqref{eq:otherexp} and only selecting the terms with two scalars and one tensor, we arrive at
\begin{equation}\label{eq:actionsim2}
      \begin{split}
         S=\int d^3x \,dt &\bigg\{\frac{a^3}{8}\dot h^{ij}\dot h_{ij} -\frac{a}{8}\partial_ih_{kl}\partial^ih^{kl}-2ah^{ij}\partial_i(\Phi+\Psi)\partial_j\Phi+ah^{ij} \partial_i \Phi \partial_j \Phi+\frac{a}{2}h^{ij} \partial_i \delta\varphi \partial_j \delta\varphi
         \bigg\}\,,
      \end{split}
\end{equation}
where the first two terms correspond to the second order Lagrangian of the tensor modes.
By taking the variation with respect to $h_{ij}$ we obtain the equations of motion of induced GWs, namely
\begin{equation}\label{eq:eominducedGW}
\ddot h_{ij}+3H\dot h_{ij}-a^{-2}\Delta h_{ij}=a^{-2}{\cal P}^{ab}\,_{ij}\left\{-8\partial_a(\Phi+\Psi)\partial_b\Phi+4\partial_a \Phi \partial_b \Phi+2\partial_a \delta\varphi \partial_b \delta\varphi\right\}\,,
\end{equation}
where $H=\dot a/a$ is the Hubble parameter and ${\cal P}^{ab}\,_{ij}$ is the transverse-traceless projector, so that the equation is consistent with the transverse-traceless degrees of freedom. For the moment, it is enough to know that it satisfies the requirements of a transverse-traceless object for both pairs of indices. Also note that in the current case of study, we have no source of anisotropic stress. Thus, for the sake of simplicity and consistency, we momentarily cheat and use the traceless part of the spatial component of the linear Einstein equations in App.~\ref{app:generalgauge}, which in the absence of anisotropic stress yields
\begin{equation}
\Psi+\Phi=0\,.
\end{equation}
Note that this is not strictly true in the early universe where neutrinos have a large mean free path and give a tiny contribution to the anisotropic stress \cite{Baumann:2007zm,Mangilli:2008bw}.

Before going to the general solutions of the induced GWs, let us translate the current calculation to the perfect fluid picture. A perfect fluid with energy density $\rho$ and pressure $P$ is described by the following energy-momentum tensor:
\begin{equation}\label{eq:Tmunu}
T_{\mu\nu}=\left(\rho+P\right)u_\mu u_\nu+P g_{\mu\nu}\,,
\end{equation}
where $u^\mu$ is the fluid 4-velocity. The perfect fluid is specified once the equation of state $w=P/\rho$ is given. In the perturbative expansion, one takes $\rho=\bar\rho+\delta\rho$ and $u_i=\partial_i v$ where $v$ is the velocity perturbation and we neglected vector modes. The energy momentum tensor \eqref{eq:Tmunu} has to be compared with the one for the scalar field, which reads
\begin{equation}\label{eq:Tmunuphi}
T^\varphi_{\mu\nu}=\partial_\mu\varphi\partial_\nu\varphi-g_{\mu\nu}\left(\frac{1}{2}\partial_\alpha\varphi\partial^\alpha\varphi+V(\varphi)\right)\,.
\end{equation}
The perfect fluid description of the scalar field follows from the identification
\begin{equation}
u_\mu=\frac{\partial_\mu \varphi}{\sqrt{-\partial_\alpha\varphi\partial^\alpha\varphi}}\,.
\end{equation}
If we simply focus on the spatial component $u_i$ we find that
\begin{equation}
\delta\varphi\leftrightarrow v\sqrt{\rho+P}\,.
\end{equation}
Thus, in terms of the perfect fluid description we have that
\begin{equation}\label{eq:eominducedGW2}
\ddot h_{ij}+3H\dot h_{ij}-a^{-2}\Delta h_{ij}=a^{-2}{\cal P}^{ab}\,_{ij}\left\{4\partial_a \Phi \partial_b \Phi+2(\rho+P)\partial_a v \partial_b v\right\}\,.
\end{equation}
This result coincides with the equations derived in Ref.~\cite{Ananda:2006af,Baumann:2007zm}. Note that Eq.~\eqref{eq:eominducedGW2} will slightly change in the case of a multi-perfect fluid system. We expect contributions from the relative velocities of the fluids. For a radiation-matter dominated universe, the source term to induced GWs can be found, e.g., in Ref.~\cite{Domenech:2020ssp}.

\subsection{General solutions \label{eq:generalsolutions}}

To solve for the induced GWs, we must solve beforehand the first order equations of motion. Continuing with the scalar field fluctuations $\delta\varphi$, we find that they are related by the momentum constraint to $\Phi$ by \cite{Koyama:2010xj}
\begin{equation}\label{eq:momentumconstraint}
\delta\varphi=-\sqrt{\frac{2}{\epsilon}}\left(\Phi+\frac{\Phi'}{\cal H}\right)\,.
\end{equation}
Thus, we must only solve for the only scalar degree of freedom $\Phi$, which we will do briefly. For the moment, we take a general approach and split $\Phi$ into an initial value $\Phi_{\mathbf{k}}$ and a transfer function $T_\Phi$ as
\begin{equation}\label{eq:transferphi}
\Phi({\mathbf{k}},\tau)=T_\Phi(k,\tau)\Phi_{\mathbf{k}}\,.
\end{equation}
If we are considering primordial fluctuations, the initial spectrum for $\Phi_{\mathbf{k}}$ is set on superhorizon scales ($k\tau\ll1$) by quantum fluctuations during inflation. However, in general, any source of fluctuations to the curvature perturbation $\Phi$ will lead to induced GWs. We will discuss this possibility later in Sec.~\ref{sec:dust}.  With Eqs.~\eqref{eq:momentumconstraint} and \eqref{eq:transferphi} we can formally solve the equations of motion for induced GWs \cite{Ananda:2006af,Baumann:2007zm}, which in Fourier space read
\begin{equation}\label{eq:eominducedGW3}
h''_{\mathbf{k},\lambda}+2{\cal H} h'_{\mathbf{k},\lambda}+k^2 h_{\mathbf{k},\lambda}={\cal S}_{\mathbf{k},\lambda}\,,
\end{equation}
with
\begin{equation}
{\cal S}_{\mathbf{k},\lambda}=4\int \frac{d^3q}{(2\pi)^3} e_\lambda^{ij}(k)q_iq_j \Phi_{\mathbf{q}}\Phi_{|\mathbf{k-q}|} f(\tau, q,|\mathbf{k-q}|)\,,
\end{equation}
and
\begin{align}\label{eq:f}
f(\tau, q,|\mathbf{k-q}|)=&T_\Phi(q\tau)T_\Phi(|\mathbf{k}-\mathbf{q}|\tau)\nonumber\\&+\frac{1+b}{2+b}\left(T_\Phi(q\tau)+\frac{T'_\Phi(q\tau)}{\cal H}\right)\left(T_\Phi(|\mathbf{k}-\mathbf{q}|\tau)+\frac{T'_\Phi(|\mathbf{k}-\mathbf{q}|\tau)}{\cal H}\right)\,,
\end{align}
where we wrote $\epsilon$ in terms of $b$ using Eqs.~\eqref{eq:epsilon} and \eqref{eq:b}.
To derive Eq.~\eqref{eq:eominducedGW3} from Eq.~\eqref{eq:eominducedGW2}, we used the Fourier transform for $h_{ij}$ in terms of the polarization tensors $e_{ij}(k)$ which are defined in App.~\ref{app:polarization}. Then, we used that the projection of $e_{ij}$ with the Fourier transform of the transverse-traceless projector $\tilde {\cal P}^{ab}\,_{ij}$ is trivial, i.e., $e^{ij}\tilde {\cal P}^{ab}\,_{ij}=e^{ab}$. Now, with the Green's method we can find a formal solution to \eqref{eq:eominducedGW3} given by
\begin{equation}\label{eq:greensolution}
h_{\mathbf{k},\lambda}(\tau)=\int_{\tau_i}^{\tau}d\tilde\tau G_h(\tau,\tilde \tau){\cal S}_{\mathbf{k},\lambda}(\tilde\tau)\,,
\end{equation}
where $G_h(\tau,\tilde \tau)$ is the Green's function of the homogeneous solutions to \eqref{eq:eominducedGW3} given in App.~\ref{sec:green}. The formal solution \eqref{eq:greensolution} is defined such that at the initial time $\tau_i$ we have $h_{\mathbf{k},\lambda}(\tau_i)=h'_{\mathbf{k},\lambda}(\tau_i)=0$. Any GW with primordial origin, i.e., generated during inflation, may be simply added to the solution \eqref{eq:greensolution}.

Since we are interested in the main observable of the SGWB, that is the GW power, let us compute the 2-point correlation function of the induced GWs which is given by
\begin{align}\label{eq:h2pt}
\langle h_\lambda(k,\tau)h_\lambda(k',\tau)\rangle=\int_0^\tau d\tau_1\int_0^\tau d\tau_2G(\tau,\tau_1)G(\tau,\tau_2)\langle S_\lambda(k,\tau_1)S_\lambda(k',\tau_2)\rangle\,.
\end{align}
Here we have neglected any primordial signal but we could add one without further issue. In fact, this was considered in Ref.~\cite{Garcia-Bellido:2017aan} where they may have non-trivial cross-correlated spectrum by tensor-scalar-scalar couplings during inflation. We will not pursue this possibility further in this review. Continuing with Eq.~\eqref{eq:h2pt}, we can compute the 2-point function of the source term by
\begin{align}\label{eq:sl}
\langle S_\lambda(k,\tau_1)S_\lambda(k',\tau_2)\rangle=&4^2\int \frac{d^3q}{(2\pi)^3}\int \frac{d^3q'}{(2\pi)^3}e_\lambda^{ij}(k)q_iq_je_\lambda^{ij}(k')q'_iq'_j\nonumber\\&\times f(\tau_1,q,|\mathbf{k}-\mathbf{q}|)f(\tau_2,q',|\mathbf{k}'-\mathbf{q}'|)
\langle\Phi_q\Phi_{|\mathbf{k}-\mathbf{q}|}\Phi_{q'}\Phi_{|\mathbf{k}'-\mathbf{q}'|}\rangle\,.
\end{align}
We see that the induced GW spectrum depends on the 4-point function of the scalar fluctuations. Such 4-point function may be decomposed in general into a disconnected, i.e., the product of 2-point functions, and a connected piece \cite{Cai:2018dig}. If we consider that the stochastic fluctuations are drawn from a Gaussian distribution, the connected 4-point function vanishes and we are led to\footnote{As in the tensor modes we have that the dimensionless power spectrum is given by
\begin{equation*}\label{eq:PP}
\langle \Phi_{\bm{k}}\Phi_{\bm{k}'}\rangle= \frac{2\pi^2}{k^3}{\cal P}_{\Phi}(k)\times (2\pi)^3\delta^{3}(\bm{k}+\bm{k}')\,.
\end{equation*}
}
\begin{align}\label{eq:gaussian}
\langle\Phi_q\Phi_{|\mathbf{k}-\mathbf{q}|}\Phi_{q'}\Phi_{|\mathbf{k}'-\mathbf{q}'|}\rangle&=\frac{2\pi^2}{q^3}{\cal P}_\Phi(q)\frac{2\pi^2}{|\mathbf{k}-\mathbf{q}|^3}{\cal P}_\Phi(|\mathbf{k}-\mathbf{q}|)\nonumber\\&\times(2\pi)^6\delta^{(3)}(\mathbf{q}+\mathbf{q}')\delta^{(3)}(\mathbf{k}+\mathbf{k}'-\mathbf{q}-\mathbf{q}')+(\mathbf{q}\leftrightarrow\mathbf{k}-\mathbf{q})\,.
\end{align}
We discuss the case of mildly non-Gaussian fluctuations in Sec.~\ref{sec:ng}.
With further manipulations of Eqs.~\eqref{eq:h2pt} and \eqref{eq:hh}, integrating one internal momentum using a Dirac delta and writing the remaining integral in spherical (momentum) coordinates, we arrive at
\begin{equation}\label{eq:Phgaussian}
\overline{{\cal P}_h}=8\int_0^\infty dv\int_{|1-v|}^{1+v}du\left(\frac{4v^2-(1-u^2+v^2)^2}{4uv}\right)^2\overline{I^2(\tau,k,v,u)}{{\cal P}_{\Phi}(ku)}{{\cal P}_{\Phi}(kv)}\,,
\end{equation}
where we followed the notation of Kohri and Terada \cite{Kohri:2018awv}. In Eq.~\eqref{eq:Phgaussian} we already summed over polarizations (see App.~\ref{app:polarization} for the details) and we have introduced for convenience
\begin{equation}\label{eq:uv}
v\equiv\frac{q}{k}\quad,\quad u\equiv\frac{|\mathbf{k}-\mathbf{q}|}{k}\,.
\end{equation}
We also took the oscillation average, i.e., we integrated over half period and divided by $\pi$, to take into account that observations of the SGWB actually measure an average over many wavelengths. Furthermore, we have inserted all the time dependence into a ``kernel'' or ``transfer function'' defined by
\begin{equation}\label{eq:kernelApp}
I(\tau,k,u,v)\equiv \int_{\tau_i}^{\tau}d\tilde\tau \,G(\tau,\tilde\tau)f(\tilde\tau,k,u,v)\,.
\end{equation}
The (averaged) power spectrum \eqref{eq:Phgaussian} is the main quantity needed for the calculations of induced GWs. Knowing \eqref{eq:Phgaussian} we can compute the spectral density of GWs by Eq.~\eqref{eq:spectraldensity}. Note that the induced tensor power spectrum \eqref{eq:Phgaussian} is valid for any curvature perturbation regardless of its origin as long as it is Gaussian. In the next section, we discuss in more detail the case when the fluctuations $\Phi$ are generated during inflation.

\subsection{Inclusion of primordial non-Gaussianity \label{sec:ng}}

Primordial fluctuations generated out of quantum fluctuations during inflation are very close to being Gaussian, although not exactly. Small departures of a Gaussian distribution are expected due to gravitational interactions \cite{Maldacena:2002vr,Koyama:2010xj}. These perturbative departures from the Gaussian distribution are often referred to as Non-Gaussianity (NG).\footnote{Although the name is not informative at all, as strictly speaking any distribution which is not a Gaussian is non-Gaussian, in cosmology the term "non-Gaussianity" is often used to indicate small departures from a Gaussian distribution, e.g., by a small but non-vanishing 3-point correlation function. For a detailed explanation, I recommend to read E.~A.~Lim's notes on primordial NG: \url{https://nms.kcl.ac.uk/eugene.lim/AdvCos/lecture2.pdf}.} Depending on the formation mechanism, e.g., due to large interactions among fields, we may obtain some significant level of NG. The possibility of observing NG of primordial fluctuations is very exciting, as it might provide information on the particle content during inflation. For this reason, the study of NG is sometimes referred to as cosmological collider physics \cite{Arkani-Hamed:2015bza}.

For practical convenience, the predictions of primordial fluctuations generated during inflation are often given in terms of the curvature perturbation ${\cal R}$. This is because ${\cal R}$ is a non-linearly conserved quantity on superhorizon scales \cite{Lyth:2004gb}. This makes it easy to set well-defined initial conditions on superhorizon scales for whatever scalar variable one might want to consider, independently of what happened previously (e.g., how inflation ended). For instance, since we are working in the Newtonian gauge, we have to relate ${\cal R}$ with $\Phi$. It can be shown that on superhorizon scales the linear relation is given by \cite{Mukhanov:2005sc}
\begin{equation}\label{eq:Rtophi}
{\cal R}=\frac{5+3w}{3(1+w)}\Phi=\frac{2b+3}{b+2}\Phi\,.
\end{equation}
In the case at hand, we are interested on NGs generated during inflation, i.e., primordial NG. These primordial NGs are often parametrized by the amplitude and shape of 3-point function or the bispectrum \cite{Komatsu:2010hc,Bartolo:2010qu}. For simplicity, we usually consider only the local type NG which can be expressed as a local perturbative expansion around the Gaussian curvature perturbation ${\cal R}^g$ by
\begin{equation}\label{eq:FNLRExpansion}
{\cal R}(x)={\cal R}^g(x)+\frac{3}{5}f_{NL}\left({\cal R}^g(x)\right)^2\,,
\end{equation}
where the factor $3/5$ is by convention.\footnote{This factor is actually to compensate the factor in \eqref{eq:Rtophi} for $w=0$ since the original definition \cite{Komatsu:2003iq} was in terms of $\Phi$ not ${\cal R}$.} Note, however, that this is just a particular shape of the NG. Currently though, there is no study on the impact of other shapes of NG on the induced GW spectrum. Now, if we go to Fourier space the squared term in \eqref{eq:FNLRExpansion} becomes a convolution. Then, we can write that for the $\Phi_q$ modes the perturbative expansion is given by
\begin{equation}\label{eq:FNLPHIExpansion}
\Phi_q=\Phi_q^g+F_{NL}\int \frac{d^3l}{(2\pi)^3}\Phi_l^g\Phi_{|\mathbf{q}-\mathbf{l}|}^g\,,
\end{equation}
where to avoid unnecessary numerical factors we have introduced
\begin{equation}
F_{NL}\equiv\frac{3}{5}\frac{2b+3}{b+2}f_{NL}\,.
\end{equation}
We can then expand perturbatively the 4-point function in \eqref{eq:sl} as
\begin{align}\label{eq:perturbative4}
&\langle\Phi_q\Phi_{|\mathbf{k}-\mathbf{q}|}\Phi_{q'}\Phi_{|\mathbf{k}'-\mathbf{q}'|}\rangle=\langle\Phi^g_q\Phi^g_{|\mathbf{k}-\mathbf{q}|}\Phi^g_{q'}\Phi^g_{|\mathbf{k}'-\mathbf{q}'|}\rangle\nonumber\\&+ F_{NL}^2\int\frac{d^3l}{(2\pi)^3}\int\frac{d^3l'}{(2\pi)^3}\langle\Phi^g_{q}\Phi^g_l\Phi^g_{|\mathbf{k}-\mathbf{q}-\mathbf{l}|}\Phi^g_{q'}\Phi^g_{l'}\Phi^g_{|\mathbf{k}'-\mathbf{q}'-\mathbf{l}'|}\rangle\nonumber\\&+(|\mathbf{k}-\mathbf{q}|\leftrightarrow q)+(|\mathbf{k'}-\mathbf{q'}|\leftrightarrow q')+(q\leftrightarrow q';|\mathbf{k}-\mathbf{q}|\leftrightarrow|\mathbf{k'}-\mathbf{q'}|)\nonumber\\&+ F_{NL}^2\int\frac{d^3l}{(2\pi)^3}\int\frac{d^3l'}{(2\pi)^3}\langle\Phi^g_{l}\Phi^g_{|\mathbf{q}-\mathbf{l}|}\Phi^g_{l'}\Phi^g_{|\mathbf{k}-\mathbf{q}-\mathbf{l}'|}\Phi^g_{q'}\Phi^g_{|\mathbf{k}'-\mathbf{q}'|}\rangle\nonumber\\&+(q\leftrightarrow q';|\mathbf{k}-\mathbf{q}|\leftrightarrow|\mathbf{k'}-\mathbf{q'}|)+O(F_{NL}^4)\,.
\end{align}
The first term on the right hand side of \eqref{eq:perturbative4} is already computed in Eq.~\eqref{eq:gaussian}. If we focus on the second line in the NG contribution of \eqref{eq:perturbative4}, we see that there are 6 possible non-vanishing Wick contractions in the first 6 point function of $\Phi^g$ which, together with the 4 remaining possibilities of the third line, makes 24 terms for the first NG contribution. In the fourth line of \eqref{eq:perturbative4} there are 8 possible non-vanishing contractions, which make 16 possibilities counting the fifth line. The total is then 40 non-vanishing contractions.\footnote{Since there are more possible contractions in the non-gaussian case with respect to the gaussian one, the numerical factors of the NG contribution will be larger.} The NG contribution can be further classified into three different terms following the notation of Ref.~\cite{Adshead:2021hnm}: the ``H'', the ``C'', the ``Z''. Using Eq.~\eqref{eq:perturbative4} in \eqref{eq:h2pt} and \eqref{eq:hh}, these NG contributions to the induced tensor spectrum can be respectively written as
\begin{align}\label{eq:PH}
\overline{{\cal P}^{\rm H}_h}=2^5 \,F_{NL}^2k^3\sum_\lambda\int \frac{d^3q}{2\pi}  \left(e_\lambda^{ij}(k)q_iq_j\right)^2\,\,&\overline{I^2(\tau,q,|\mathbf{k}-\mathbf{q}|)}\nonumber\\&\times\frac{{\cal P}_\Phi(q)}{q^3}
\int \frac{d^3l}{2\pi}\frac{{\cal P}_\Phi(l)}{l^3}\frac{{\cal P}_\Phi(|\mathbf{k}-\mathbf{q}-\mathbf{l}|)}{|\mathbf{k}-\mathbf{q}-\mathbf{l}|^3}\,,
\end{align}
\begin{align}\label{eq:PC}
\overline{{\cal P}^{\rm C}_h}=2^6F_{NL}^2k^3\sum_\lambda\int \frac{d^3q}{2\pi}\int \frac{d^3l}{2\pi} e_\lambda^{ij}(k)q_iq_j&e_\lambda^{ij}(k)l_il_j \,\overline{I(\tau,q,|\mathbf{k}-\mathbf{q}|)I(\tau,l,|\mathbf{k}-\mathbf{l}|)}\nonumber\\&
\times\frac{{\cal P}_\Phi(|\mathbf{k}-\mathbf{l}|)}{|\mathbf{k}-\mathbf{l}|^3}\frac{{\cal P}_\Phi(l)}{l^3}\frac{{\cal P}_\Phi(|\mathbf{q}-\mathbf{l}|)}{|\mathbf{q}-\mathbf{l}|^3}\,,
\end{align}
and
\begin{align}\label{eq:PZ}
\overline{{\cal P}^{\rm Z}_h}=2^6\,F_{NL}^2k^3\sum_\lambda\int \frac{d^3q}{2\pi}\int \frac{d^3l}{2\pi} e_\lambda^{ij}(k)q_iq_j&e_\lambda^{ij}(k)l_il_j  \,\overline{I(\tau,q,|\mathbf{k}-\mathbf{q}|)I(\tau,l,|\mathbf{k}-\mathbf{l}|)}\nonumber\\&
\times\frac{{\cal P}_\Phi(q)}{q^3}\frac{{\cal P}_\Phi(l)}{l^3}\frac{{\cal P}_\Phi(|\mathbf{k}-\mathbf{q}-\mathbf{l}|)}{|\mathbf{k}-\mathbf{q}-\mathbf{l}|^3}\,.
\end{align}
These derived formulas agree with those of Ref.~\cite{Adshead:2021hnm}. We will assume that we are in the perturbative regime, so that the NG contributions \eqref{eq:PH}-\eqref{eq:PC} should be thought of as a correction to the Gaussian contribution \eqref{eq:Phgaussian}. The first contribution \eqref{eq:PH} is also referred to as ``hybrid'' \cite{Unal:2018yaa}. Also note that the second line in \eqref{eq:PH} can be reabsorbed into a redefinition of the primordial power spectrum to include non-Gaussianities. For instance, we could write
\begin{align}\label{eq:redefNG}
{\cal P}_{\cal R}^{NL}(q)={\cal P}_{\cal R}(q)+F^2_{NL}q^3\int \frac{d^3l}{2\pi} \frac{{\cal P}_{\cal R}(l)}{l^3} \frac{{\cal P}_{\cal R}(|\mathbf{q}-\mathbf{l}|)}{|\mathbf{q}-\mathbf{l}|^3}\,,
\end{align}
which follows from the computation of $\langle\Phi(k)\Phi(k')\rangle$ using Eq.~\eqref{eq:FNLPHIExpansion}. This is the term considered by Cai, Pi and Sasaki \cite{Cai:2018dig} and acts as a local redefinition of the primordial spectrum. With the form of Eq.~\eqref{eq:redefNG} it is much easier to compute the induced GW spectrum up to $O(F_{NL}^4)$ since it just accounts for a replacement of ${\cal P}_{\cal R}$ in \eqref{eq:Phgaussian} for ${\cal P}_{\cal R}^{NL}$.The second contribution \eqref{eq:PC} was first considered by Unal \cite{Unal:2018yaa} and the third \eqref{eq:PZ} by Ref.~\cite{Atal:2021jyo}. Unal also considered the terms of $O(F_{NL}^4)$ \cite{Unal:2018yaa}. Most recently, Adshead, Lozanov and Weiner \cite{Adshead:2021hnm} presented a detailed and complete analysis of all the contributions up to $O(F_{NL}^4)$. The interested reader is referred to Ref.~\cite{Adshead:2021hnm} for the detailed formulas. We discuss the main features of the NG corrections in Sec.~\ref{sec:typical}. It should be noted that the effect of primordial NG might compete with the terms due to non-linearities, e.g., from studied third order contributions to the source term of induced GWs. The reader is referred to Refs.~\cite{Yuan:2019udt,Yuan:2021qgz} for the details on the third order expansion. Lastly, it should also be noted that for simplicity we assumed a scale invariant $f_{NL}$ in \eqref{eq:FNLRExpansion}. But, in general situations one expects that $f_{NL}$ is scale dependent. For a recent study on the impact of the scale dependence see Ref.~\cite{Ragavendra:2021qdu}.


\section{Analytical transfer functions \label{sec:analytical}}

In general situations the kernel \eqref{eq:kernelApp} and the power spectrum \eqref{eq:Phgaussian} have to be computed numerically. The problem is that the time integral in \eqref{eq:kernelApp} is extremely demanding for $k\tau\gg 1$, as the integrand is the product of three oscillating functions with frequencies that depend on the tensor and scalar mode wavenumbers. In fact, $k\tau\gg1$ is the range of scales we are interested in. If we extend the integral until today, say at $\tau=\tau_0\propto H_{0}^{-1}$, and we look at scales much smaller than the horizon, we trivially have $k\tau_0\gg1$. But there is more we can learn. For instance, scales accessible to future GW detectors range from $k\sim 10^7$ to $10^{18}\,{\rm Mpc}^{-1}$. These very small scales entered the horizon much before BBN, which roughly correspond to the time when modes with $k\sim 10^{3}\,{\rm Mpc}^{-1}$ entered the horizon.\footnote{For reference note that the radiation-matter equality corresponds to $k\sim 10^{-2}\,{\rm Mpc}^{-1}$. This means that at the transition to the cold dark matter dominated epoch such induced GWs are already freely propagating GWs and can be treated as radiation.} Luckily, this implies that we can evaluate the kernel \eqref{eq:kernelApp} some time prior to BBN when induced GWs are really inside the horizon and in a radiation dominated universe. Then, from that moment on, we may simply treat these induced GWs as a radiation fluid with $w=1/3$. Note that we will also consider the case that the modes of interest enter the horizon in an epoch where $w\neq1/3$. In that case, we must follow induced GWs until the universe transitions to the standard radiation dominated era. We give more details later.

Surprisingly, there are relevant regimes where the integrals in \eqref{eq:kernelApp} can be done analytically. This is the case of a constant equation of state parameter $w$ and constant propagation speed of fluctuations $c_s$. For example, this includes a perfect fluid whether it is made of a scalar field\footnote{For instance, one has arbitrary constant $w$ and $c_s=1$ for a canonical scalar field in an exponential potential \cite{Lucchin:1984yf}} or an adiabatic perfect fluid, which are commonly used in cosmology. Even when there are transitions between one perfect fluid domination to another, there will be periods in which $w$ and $c_s$ are almost constant. Thus, the constant $w$ and $c_s$ limit is a good approximation in a cosmological set up if the modes of interest enter the horizon outside the transition periods. With these assumptions, we present the solutions to first order cosmological perturbations in the Newtonian gauge in Sec.~\ref{sec:firstorder} and derive a very simple form of the source term \eqref{eq:f} for the induced GWs. In the general situation where $w\neq1/3$, the integral \eqref{eq:kernelApp} can be analytically done for modes which either enter the horizon before or after the transition. In this way, in Secs.~\ref{sec:subhorizonkernel} and \ref{sec:superhorizonkernel} we respectively derive the kernel for modes which are subhorizon and superhorizon before the transition. When then follow our solutions to the radiation dominated epoch.\\

\noindent\faBook\,\,\textit{Main references:} The whole section is a generalisation of Refs.~\cite{Domenech:2019quo,Domenech:2020kqm} to account for not only general constant $w$ but general constant $c_s$. The notation used here improves and unifies those of \cite{Domenech:2019quo,Domenech:2020kqm}. The main base of Refs.~\cite{Domenech:2019quo,Domenech:2020kqm} was started in \cite{Baumann:2007zm,Kohri:2018awv} in the radiation dominated universe.

\subsection{First order solutions \label{sec:firstorder}}

At the moment, the formal solution \eqref{eq:greensolution} for the induced tensor modes does not tell us much. To extract some meaningful information, we have to solve the equations of motion for first order perturbations, which are given in detail in App.~\ref{app:generalgauge}. After some simplifications, the equation for the Newtonian potential for general $w$ and $c_s$ and in the absence of isocurvature\footnote{Isocurvature fluctuations are fluctuations that leave the total energy density of matter homogeneous. Thus, they are only possible in multi-fluid systems where the density fluctuations of one fluid can be compensated by the density fluctuations of another. \label{footnote:isocurvature}} fluctuations reads\footnote{Actually to derive Eq.~\eqref{eq:phieom} we used the equations of motion for the curvature perturbation ${\cal R}$ for general $w$ and $c_s$ as can be found, e.g., in \cite{Garriga:1999vw,Koyama:2010xj} and then we changed variables according to Eq.~\eqref{eq:Rtophi}.}
\begin{equation}\label{eq:phieom}
\Phi''+\left(2\epsilon-\eta\right){\cal H}\Phi'-\left(\eta+2s\left(1+\epsilon-\eta-2s+\frac{\dot s}{Hs}\right)\right){\cal H}^2\Phi+c_s^2k^2\Phi=0\,,
\end{equation}
where we have also introduced $\epsilon$ as in Eq.~\eqref{eq:epsilon} as well as
\begin{equation}\label{eq:eta}
\eta\equiv \frac{\dot\epsilon}{H\epsilon}\quad{\rm and}\quad s\equiv\frac{\dot c_s}{Hc_s}\,.
\end{equation}
The notation $\epsilon$, $\eta$ and $s$ is typical of inflationary models and simplifies the form of the equations. One may of course write Eq.~\eqref{eq:phieom} in a more conventional notation, e.g., that in Mukhanov's book \cite{Mukhanov:2005sc} which for constant $c_s$ reads
\begin{equation}
\Phi''+3{\cal H}\left(1+c_w^2\right)\Phi'+\left(2{\cal H}'+(1+3c_w^2){\cal H}^2\right)\Phi+c_s^2k^2\Phi=0\,,
\end{equation}
where $c_w^2=\dot P/\dot \rho$ and $c_s^2=\delta P/\delta\rho$. In the case of an adiabatic perfect fluid one has $c_s^2=c_w^2=w$. In the case of a canonical scalar field we have $c_w^2=w$ and $c_s^2=1$. As we already mentioned, for analytical viability we are mostly interested in the case of constant $w$ and $c_s$. This requirement in turn implies $\eta=s=0$, so that the Newtonian potential in Eq.~\eqref{eq:phieom} behaves as a massless field. The solution to \eqref{eq:phieom} for general $c_s$ and $w$ is given in terms of Bessel functions of the first and second kind, explicitly
\begin{equation}\label{eq:phigeneralsolution}
\Phi(k\tau)=(c_sk\tau)^{-b-3/2}\left(C_1 J_{b+3/2}(c_sk\tau)+C_2 Y_{b+3/2}(c_sk\tau)\right)\,,
\end{equation}
where $b$ in terms of $w$ is defined in Eq.~\eqref{eq:b}. Now, we must give some kind of initial conditions to $\Phi$. If the primordial spectrum was generated by quantum fluctuations during inflation, the initial conditions are well-defined and constant on super(sound)horizon scales, that is when $c_sk\tau\ll1$. By picking up the asymptotically constant term in \eqref{eq:phigeneralsolution}, we find that $C_2=0$ and
\begin{equation}\label{eq:phisolsol}
\Phi(k\tau)=\Phi_{\mathbf{k}}\,2^{b+3/2}\Gamma[b+5/2](c_sk\tau)^{-b-3/2}J_{b+3/2}(c_sk\tau)\,,
\end{equation}
where $\Phi_{\mathbf{k}}$ is the primordial (stochastic) value set by inflation. Note that the general solution \eqref{eq:phigeneralsolution} is not valid for $c_s=0$ where the gradient term in \eqref{eq:phieom} is absent. Nevertheless, the solution for constant $w$ is just $\Phi=\Phi_{\mathbf{k}}$ on all scales. The case of $c_s=0$ is quite particular as the fact that $\Phi$ does not decay implies a constant source to induced GWs. This makes the detailed dynamics of the transition to radiation domination very important for the final GW spectrum \cite{Inomata:2019zqy,Inomata:2019ivs}, when the source term experiences drastic changes. We dedicate Sec.~\ref{sec:dust} to this particular case. 

With an exact analytical formula for the first order perturbations, we can attempt to compute the resulting induced GWs. For that, we first need to know  the two independent homogeneous solutions to the tensor mode equations \eqref{eq:eominducedGW3}, say $h_1$ and $h_2$. For constant $w$ they are given as well in terms of Bessel functions with one order less than in $\Phi$, concretely
\begin{equation}\label{eq:hgeneralsolution}
h_1(k\tau)=(k\tau)^{-b-1/2}J_{b+1/2}(k\tau)\quad{\rm and}\quad h_2(k\tau)=(k\tau)^{-b-1/2} Y_{b+1/2}(k\tau)\,.
\end{equation}
In Eq.~\eqref{eq:hgeneralsolution}, $h_1$ and $h_2$ respectively are the growing and decaying modes.
Note that there is no $c_s$ since in general relativity tensor modes propagate at the speed of light. This yields a Green's function (see App.~\ref{sec:green}) given by
\begin{equation}\label{eq:hgreen}
G(\tau,\tilde\tau)=\frac{\pi}{2k}\frac{(k\tilde\tau)^{b+3/2}}{(k\tau)^{b+1/2}}\left(J_{b+1/2}(k\tilde \tau)Y_{b+1/2}(k\tau)-J_{b+1/2}(k\tau)Y_{b+1/2}(k\tilde\tau)\right)\,.
\end{equation}
This finishes the list of necessary ingredients for our induced GWs calculations. In passing, since there always appears the combination $k\tau$ or $k\tilde\tau$ we introduce the useful notation
\begin{equation}\label{eq:x}
x\equiv k\tau\,,
\end{equation}
which we will use from now on. Also, a tilde on $x$ implies a tilde on $\tau$, i.e., $\tilde x=k\tilde\tau$.

Thus far it seems that we may have to deal with several integrals containing at least three Bessel functions, which is challenging to do analytically in general. However, the source term to the induced GWs ends up acquiring a simple enough form after several simplifications. For instance, using the Bessel functions properties (see App.~\ref{app:bessel}), we find that Eq.~\eqref{eq:f} reduces to
\begin{align}\label{eq:fsimple}
f(x,u,v)=&\frac{2^{2b+3}\Gamma^2[b+5/2]}{(2b+3)(b+2)}(c_sx)^{-2b-1}(uv)^{-b-1/2}\nonumber\\&\times\left(J_{b+1/2}(c_svx)J_{b+1/2}(c_sux)+\frac{b+2}{b+1}J_{b+5/2}(c_svx)J_{b+5/2}(c_sux)\right)\,,
\end{align}
where we used $x$ from \eqref{eq:x} and $u$ and $v$ from \eqref{eq:uv}. To arrive at \eqref{eq:fsimple} one needs to write the Bessel function $J_{b+3/2}$ that appears in $\Phi$ in a combination of $J_{b+1/2}$ and $J_{b+5/2}$ which can be found in App.~\ref{app:bessel}.
So far, it is not entirely clear whether there is any physical reason behind the simplification \eqref{eq:fsimple} which turns out to render the source term as a sum of the squared of Bessel functions of the same order. This simple form of the source term is crucial to obtain analytical formulas for general $b$ and $c_s$ that we derive below.

Replacing \eqref{eq:fsimple} and \eqref{eq:x} into Eq.~\eqref{eq:kernelApp} we find that the kernel can be written as
\begin{align}\label{eq:Isimple}
I(x,u,v)=&\pi 4^{b}\Gamma^2[b+3/2]\frac{2b+3}{b+2}(c_s^2uvx)^{-b-1/2}\left(J_{b+1/2}(x){\cal I}_{Y}-Y_{b+1/2}(x){\cal I}_{J}\right)\,,
\end{align}
where we defined for compactness
\begin{align}\label{eq:Isimpledef}
{\cal I}_{J/Y}\equiv\int_0^xd\tilde x& \,\tilde x^{1/2-b}\begin{Bmatrix}J_{b+1/2}(\tilde x)\\ Y_{b+1/2}(\tilde x)
\end{Bmatrix}\nonumber\\&\times\left(J_{b+1/2}(c_sv\tilde x)J_{b+1/2}(c_su\tilde x)+\frac{b+2}{b+1}J_{b+5/2}(c_sv\tilde x)J_{b+5/2}(c_su\tilde x)\right)\,.
\end{align}
Unfortunately, we are not aware of a general analytical formula for the time integrals \eqref{eq:Isimpledef} unless $b\in \mathbb{Z}$, in which case the Bessel functions reduce to spherical Bessel functions and can be written in terms of transcendental functions. This is for example the case of radiation domination $b=0$ ($w=1/3$), where the general kernel can be found in Ref.~\cite{Kohri:2018awv}. Nevertheless, we can integrate \eqref{eq:Isimpledef} in two relevant limits, $x\gg1$ (subhorizon) and $x\ll1$ (superhorizon).

Let us explain the assumptions needed for the approximations before jumping into the actual calculations. We are considering that right after inflation there is a period when the universe has a free $b$ (or $w$) and free $c_s$. However, to recover the standard hot big bang cosmology, such epoch must end, and the universe must enter a radiation domination stage. Thus, whatever solution we find in this section for the induced GWs must be followed until we reach the radiation dominated stage. For simplicity, we will assume an instantaneous transition. Then, we will refer to the time of transition as ``reheating'' time $\tau_{\rm rh}$ and the comoving size of the horizon at that time as ``reheating'' scale $k_{\rm rh}$. The value of $k_{\rm rh}$ in terms of the ``reheating'' temperature $T_{\rm rh}$ can be found in Eq.~\eqref{eq:krh}. The assumption of an instantaneous transition might be important for those tensor modes with wavenumber $k\sim k_{\rm rh}$. Namely, those modes which entered the horizon around the transition. If the transition is gradual, then our approximations will not be very good for modes which entered during the last stages of the transition. In any case, as we shall see, the instantaneous approximation gives a very good idea of the shape of the induced GW spectrum even for $k\sim k_{\rm rh}$. On top of that, we will assume that the scale corresponding to the peak of the primordial spectrum enters the horizon well before reheating so that during and after reheating there is no significant source of induced GWs. In the case of a top-hat primordial spectrum, this means that the scale corresponding to the IR cut-off enters the horizon much before reheating. Under these two assumptions, i.e., sudden transition and ``localized'' primordial spectrum, we can proceed to derive our analytical formulas. Note that this does not apply to the case when $c_s=0$ which we discuss in Sec.~\ref{sec:dust}.

\subsection{General subhorizon kernel \label{sec:subhorizonkernel}}

Let us start with the simplest situation. This corresponds to the case when all modes of interest entered the horizon much before reheating, that is $k\gg k_{\rm rh}$. Then, we can safely send the upper integration limit in \eqref{eq:Isimpledef} to infinity, i.e., $x=k\tau\gg1$. This is because when we eventually match our solution at reheating to a free GW propagating in a radiation dominated universe, the upper limit would be $x_{\rm rh}=k\tau_{\rm rh}\sim k/k_{\rm rh}\gg 1$. The integrals \eqref{eq:Isimpledef} for $x\to \infty$ are derived by Gervois and Navelet \cite{threebesselI} and given in App.~\ref{app:integralbessel}. Here we just give the form of \eqref{eq:Isimple} after integration, which reads
\begin{align}\label{eq:Isimple2}
I(x\gg1,u,v)=&x^{-b-1} 4^{b}\Gamma^2[b+3/2]\frac{2b+3}{b+2}\frac{|1-y^2|^{b/2}}{c_s^2 uv}\nonumber\\&
\times\Bigg\{\cos\left(x-\tfrac{b\pi}{2}\right)\left(\mathsf{P}^{-b}_{b}(y)+\frac{b+2}{b+1}\mathsf{P}^{-b}_{b+2}(y)\right)\Theta[c_s(u+v)-1]\nonumber\\&
+\frac{2}{\pi}\sin\left(x-\tfrac{b\pi}{2}\right)\left(\mathsf{Q}^{-b}_{b}(y)+\frac{b+2}{b+1}\mathsf{Q}^{-b}_{b+2}(y)\right)\Theta[c_s(u+v)-1]\nonumber\\&
-\frac{2}{\pi}\sin\left(x-\tfrac{b\pi}{2}\right)\left({\cal Q}^{-b}_{b}(-y)+2\frac{b+2}{b+1}\mathsf{\cal Q}^{-b}_{b+2}(-y)\right)\Theta[1-c_s(u+v)]\Bigg\}\,,
\end{align}
where $\Theta[x]$ is the Heaviside function, $\mathsf{P}^{-b}_{b}(y)$ and $\mathsf{Q}^{-b}_{b}(y)$ are the Ferrer's function of the first and second kind, valid for $|y|<1$, and ${\cal Q}^{-b}_{b}(y)$ is the associated Legendre function of the second kind, valid for $|y|>1$. Their explicit expressions in terms of Hypergeometric functions can be found in App.~\ref{app:legendre}. We also defined for simplicity
\begin{equation}\label{eq:y}
y=1-\frac{1-c_s^2(u-v)^2}{2c_s^2 uv}=-1-\frac{1-c_s^2(u+v)^2}{2c_s^2 uv}\,,
\end{equation}
and we Taylor expanded the Bessel functions for large argument. Note that $y$ is related to the area of the triangle made by the three momenta, $c_s|\mathbf{k}-\mathbf{q}|$, $c_s|\mathbf{q}|$ and $k$.
The presence of the Heaviside function in \eqref{eq:Isimple2} signals the possible resonance we anticipated in Sec.~\ref{sec:induced1} when the wavenumber of a tensor modes equals the sum of two typical scalar modes $2c_sk_p\sim k$. In more practical terms, when $c_s(u+v)\sim 1$ the three Bessel functions in the definite integral \eqref{eq:Isimpledef} for $x\to\infty$ conspire (by rendering their product as coherent oscillations) and yield a divergent integral. Also note that Eq.~\eqref{eq:Isimple2} has the same time dependence as a subhorizon tensor mode propagating in a constant $b$ and $c_s$ universe, e.g., see Eq.~\eqref{eq:hgeneralsolution}. In other words, it decays as $a^{-2}$ and oscillates with a frequency proportional to $x$. This was expected as the source term for induced GWs decays once inside the horizon and, therefore, induced GWs behave as a freely propagating GW. This does not apply in the case of a dust dominated universe though.

In the current form Eq.~\eqref{eq:Isimple2} is general enough so that it can also be used in the C \eqref{eq:PC} and Z \eqref{eq:PZ} NG terms. However, in this review we are more interested with the Gaussian contribution \eqref{eq:Phgaussian}. Thus, we are more concerned about the averaged square kernel. After some simple algebra, that is squaring \eqref{eq:Isimple2} and calculating the oscillation average, we arrive at
\begin{align}\label{eq:kernelaverage}
\overline{I^2(x,u,v)}=&x^{-2(b+1)} 4^{2b}\Gamma^4[b+3/2]\left(\frac{2b+3}{b+2}\right)^2\frac{|1-y^2|^{b}}{2c_s^4 u^2v^2}\nonumber\\&
\times\Bigg\{\left(\mathsf{P}^{-b}_{b}(y)+\frac{b+2}{b+1}\mathsf{P}^{-b}_{b+2}(y)\right)^2\Theta[c_s(u+v)-1]\nonumber\\&
+\frac{4}{\pi^2}\left(\mathsf{Q}^{-b}_{b}(y)+\frac{b+2}{b+1}\mathsf{Q}^{-b}_{b+2}(y)\right)^2\Theta[c_s(u+v)-1]\nonumber\\&
+\frac{4}{\pi^2}\left({\cal Q}^{-b}_{b}(-y)+2\frac{b+2}{b+1}\mathsf{\cal Q}^{-b}_{b+2}(-y)\right)^2\Theta[1-c_s(u+v)]\Bigg\}\,.
\end{align}
With the averaged kernel \eqref{eq:kernelaverage} we are ready to compute the induced GWs in quite general situations. We are only left with the integral over momenta which can be easily computed numerically. Analytical formulas for the final GW spectrum are discussed in Sec.~\ref{sec:typical}. Let us note that although \eqref{eq:kernelaverage} might look technically complicated to implement, it is by far more useful than attempting a numerical integration of three Bessel functions. Furthermore, the superhorizon approximation is very good for modes with $k\gg k_{\rm rh}$ regardless of the shape of the primordial spectrum. This is because induced GWs with $k\gg k_{\rm rh}$ are essentially already free GWs. Thus, soon after horizon re-entry and thereon they will behave as relativistic particles in an expanding universe. This also means that we can evaluate \eqref{eq:kernelaverage} and \eqref{eq:spectraldensity} at the instantaneous reheating time, i.e., $\tau=\tau_{\rm rh}$, to estimate the observable spectral density of induced GWs. We will do this shortly. If the transition is gradual, the kernel \eqref{eq:kernelaverage} is a good approximation for modes which enter during a period with almost constant $b$.

In the next subsections, we will have a closer inspection to the kernel \eqref{eq:kernelaverage}. After matching to radiation domination, we will look at the behaviour of the kernel near the resonant point and in the IR tail.

\subsubsection{Matching to radiation domination}

Let us discuss the matching to radiation domination in more technical terms. After matching the background quantities at $\tau=\tau_{\rm rh}$, that is the scale factor $a$ and the Hubble parameter ${\cal H}$ we find that $a$ and ${\cal H}$ in the radiation dominated period follows the typical solution (found in \eqref{eq:b} with $w=1/3$) with a shifted conformal time, which reads
\begin{align}\label{eq:bartau}
\bar\tau\equiv\tau-\frac{b}{1+b}\tau_{\rm rh}\,,
\end{align}
where $b$ is related to the equation of state of the previous epoch by \eqref{eq:b}. Now, for the matching of tensor modes it is important to realize that the continuity of $h_{ij}$ and its first derivative imply the continuity of the kernel \eqref{eq:kernelApp} and its first derivative. Thus, from now on we simply deal with the matching of the kernel \eqref{eq:kernelApp}. Since we are dealing with subhorizon scales we have that the kernel before the transition is
\begin{align}\label{eq:kernelsubh}
I(x\gg1,u,v)\approx & x^{-b-1}\left(A_{1,b}\sin\left(x-\frac{b\pi}{2}\right)+A_{2,b}\cos\left(x-\frac{b\pi}{2}\right)\right)\,,
\end{align}
where the coefficients, not important at the moment, can be extracted from Eq.~\eqref{eq:Isimple2}. The kernel after the transition is also for subhorizon modes, which must have the form of a freely propagating GW in a radiation dominated universe, namely
\begin{align}\label{eq:kernelsubhRD}
I_{RD}(x\gg1,u,v)\approx & A_{1,RD}\frac{\sin \bar x}{\bar x}+A_{2,RD}\frac{\cos \bar x}{\bar x}\,.
\end{align}
It is easy to convince oneself that after matching \eqref{eq:kernelsubh} and \eqref{eq:kernelsubhRD} at $\tau=\tau_{\rm rh}$ one finds that
\begin{align}
A_{1,RD}^2+A_{2,RD}^2=\left(\frac{k\tau_{\rm rh}}{1+b}\right)^2\left(A_{1,b}^2+A_{2,b}^2\right)\,.
\end{align}
This implies that although we should be calculating 
\begin{equation}\label{eq:spectraldensitysubRD}
\Omega_{\rm GW}(k\gg k_{\rm rh}, \tau\gg \tau_{\rm rh})=\frac{k^2}{12a^2H^2}\overline{{\cal P}^{RD}_{h}}\,,
\end{equation}
with
\begin{equation}\label{eq:PhgaussiansubRD}
\overline{{\cal P}^{RD}_h}=8\int_0^\infty dv\int_{|1-v|}^{1+v}du\left(\frac{4v^2-(1-u^2+v^2)^2}{4uv}\right)^2\overline{I_{RD}^2(x\gg1,v,u)}{{\cal P}_{\Phi}(ku)}{{\cal P}_{\Phi}(kv)}\,,
\end{equation}
it is equivalent to evaluate
\begin{equation}\label{eq:spectraldensitysubRD2}
\Omega_{\rm GW}(k\gg k_{\rm rh}, \tau\gg \tau_{\rm rh})=\frac{k^2}{12a^2H^2}\overline{{\cal P}_{h}}\Big|_{\tau=\tau_{\rm rh}}\,,
\end{equation}
with
\begin{equation}\label{eq:PhgaussiansubRD2}
\overline{{\cal P}_h}=8\int_0^\infty dv\int_{|1-v|}^{1+v}du\left(\frac{4v^2-(1-u^2+v^2)^2}{4uv}\right)^2\overline{I^2(x\gg1,v,u)}{{\cal P}_{\Phi}(ku)}{{\cal P}_{\Phi}(kv)}\,,
\end{equation}
where the kernel is that given in \eqref{eq:kernelaverage}.

It is important to note that our subhorizon approximation is valid for modes with $k\tau_{\rm rh}\gg 1$. Later, we may try to join the subhorizon approximation with the superhorizon approximation valid for $k\tau_{\rm rh}\ll 1$ by extrapolating up to $k\tau_{\rm rh}\sim 1$. Let us call such scale $k_m$. However, there is an important subtlety in joining the two approximations. If we look at the scale that last crossed the horizon at reheating, namely
\begin{align}
k_{\rm rh}={\cal H}_{\rm rh}=\frac{1+b}{\tau_{\rm rh}}\,,
\end{align}
we see that if we compare with $k_m$ we have that
\begin{align}
\frac{k_m}{k_{\rm rh}}=\frac{1}{1+b}\,.
\end{align}
For $b>0$ we have that $k_m<k_{\rm rh}$. This means that our subhorizon approximation is even valid for $k\sim k_{\rm rh}$ and, therefore, we expect a good matching with the superhorizon approximation at $k\sim k_{\rm rh}$ (since for $k<k_{\rm rh}$ the modes are superhorizon before the transition). However, for $b<0$ we find $k_m>k_{\rm rh}$. This time the subhorizon approximation breaks down for $k>k_{\rm rh}$ and so we expect the matching with the superhorizon approximation to be at $k\sim k_m$. This is what we eventually find in Sec.~\ref{sec:analytical}. In addition to the above, we have that the subhorizon approximation has corrections of $O(x^{1+b})$ \cite{Domenech:2020kqm}. So for $b<0$ the corrections quickly become important for $x\sim 1$. A detailed studied for $b<0$, perhaps with numerical methods, is needed to clarify the exact shape of the GW spectrum. Nevertheless, our results should be a good order of magnitude estimate.

Let us remind the reader that when implementing \eqref{eq:kernelaverage} inside the tensor spectrum \eqref{eq:Phgaussian}, we must bear in mind that if we are considering primordial curvature fluctuations from inflation, we have to rescale the spectrum of $\Phi$ accordingly, namely
\begin{equation}\label{eq:pphitopR}
{\cal P}_{\Phi}=\left(\frac{b+2}{2b+3}\right)^2{\cal P}_{\cal R}\,.
\end{equation}

\subsubsection{Resonances}

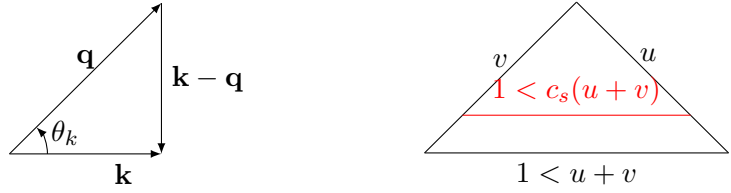
\begin{figure}
\centering
\begin{tikzpicture}
  \coordinate (O) at (0, 0);

  \draw[-latex] (O) -- (2, 0) coordinate (P1) node[pos = .75, below]
  {\(\mathbf{k}\)};
  \draw[-latex] (O) -- (2,2) node[pos = .5, above] {\(\mathbf{q}\)};
  \draw[-latex] (2,2) -- (2,0) node[pos = .5, right] {\(\mathbf{k-q}\)};
  \draw[-latex] (0.5,0) arc[radius = 0.5, start angle = 0, end angle = 45] node[pos = .75, right] {\(\theta_k\)};
\end{tikzpicture}
\hspace{2cm}
\begin{tikzpicture}
  \draw (0,0) -- (4, 0) coordinate (P1) node[pos = .5, below]
  {\(1<u+v\)};
  \draw (0,0) -- (2,2) node[pos = .5, above] {\(v\)};
  \draw (2,2) -- (4,0) node[pos = .35, right] {\(u\)};
  \draw[red] (0.5,0.5) -- (3.5, 0.5) coordinate (P1) node[pos = .5, above]
  {\(1<c_s(u+v)\)};
\end{tikzpicture}
\caption{On the left, relation between the three vectors involved in the integrals of induced GWs for the Gaussian part. On the right, illustration of the triangle inequality. We see that when $1<c_s(u+v)$ and $c_s<1$ we are below the triangle inequality and the resonance is allowed by momentum conservation.  \label{fig:triangle}}
\end{figure}

Here we focus on the general behaviour near the resonant point for the averaged kernel \eqref{eq:kernelaverage} that we discussed in  Sec.~\ref{sec:induced1}. We start by noting that we have three vectors, the one from the tensor mode $\mathbf{k}$ and the two scalar modes, $\mathbf{q}$ and $\mathbf{k}-\mathbf{q}$, which of course satisfy momentum conservation. Recall that from Eq.~\eqref{eq:uv} $q$ corresponds to $v$ and $|\mathbf{k}-\mathbf{q}|$ to $u$. This means that our three variables satisfy a triangle inequality, that is
\begin{align}
|k-q|<|\mathbf{k}-\mathbf{q}|<k+q\quad{\rm or}\quad ||\mathbf{k}-\mathbf{q}|-q|<k<|\mathbf{k}-\mathbf{q}|+q\,,
\end{align}
which translates into
\begin{align}
|1-v|<u<1+v \quad {\rm or}\quad |u-v|<1<u+v\,.
\end{align}
The area $A$ of the triangle made by $k$, $q$ and $|\mathbf{k}-\mathbf{q}|$ is related to the angle between $\mathbf{k}$ and $\mathbf{q}$ by
\begin{align}
\sin \theta_k=\frac{2A}{kq}\,.
\end{align}
This means that the projection with the polarization tensor (see App.~\ref{app:polarization}) is proportional to the area, namely
\begin{align}
e^{ij}_\lambda q_iq_j\propto\sin^2\theta_k\propto A^2\,.
\end{align}
This implies that the saturation of the triangle inequality at $u+v=1$ corresponds to zero area and, therefore, does not contribute to the integral \eqref{eq:Phgaussian}. Now, from the arguments in Sec.~\ref{sec:induced1} we expect a resonance when 
\begin{align}
c_s(u+v)=1\,.
\end{align}
Thus, if we have $c_s^2=1$ the resonant point coincides with the saturation of the triangle inequality and, thus, has vanishing contribution. In this way we see that, by momentum conservation, there is no resonance for $c_s^2=1$ in the induced GW spectrum. Nevertheless, for $c^2_s<1$ we do have a resonance inside the integration region (see Fig.~\ref{fig:triangle}).

We can confirm these expectations by expanding the kernel \eqref{eq:kernelaverage} around $c_s(u+v)\sim 1$ which corresponds to $y\sim -1$, where the associated Legendre functions might have a divergence. The asymptotic behaviours around $y\sim -1$ are given in App.~\ref{app:legendre}. Focusing first on the case $b<0$, we find that the kernel diverges as
\begin{align}\label{eq:kernelresbneg}
\overline{I_{\rm res}^2(x,u,v,b<0)}\approx&x^{-2(b+1)} 4^{2b}\Gamma^4[b+5/2]\frac{8}{c_s^4 u^2v^2}\frac{\csc^2(b\pi)}{\Gamma^2[3+b]}|1-|y||^{2b}\,.
\end{align}
This recovers Eq.~\eqref{eq:omegakv} which showed that a very peaked primordial spectrum, so that $u\sim v\sim k_p/k$, leads to
\begin{align}
\Omega^{\rm res}_{\rm GWs}(b<0)\propto |1-|y||^{2b}\propto |k-2c_sk_p|^{2b}\,.
\end{align}
The case $b=0$ is explicitly given in Eq.~\eqref{eq:w13}.
In the case when $b>0$, we arrive at
\begin{align}\label{eq:kernelresbpos}
\overline{I_{\rm res}^2(x,u,v,b>0)}\approx &x^{-2(b+1)} 4^{3b}\Gamma^4[b+5/2]\frac{32}{\pi^2c_s^4 u^2v^2}
\left(\frac{(1+b+b^2)}{(b+1)(b+2)}\frac{\Gamma[b]}{\Gamma[2b+3]}\right)^2\,,
\end{align}
which is independent of $y$. Thus, as expected there is no resonance for $b>0$. Nevertheless, as shown in \cite{Domenech:2019quo} for $1>b>0$ there is still a peak at $k=2c_sk_p$ in the induced GWs spectrum for a peaked primordial spectrum. We can roughly understand the peak by noting that since ${\cal Q}^{-b}_{b}(y)\sim (1-y)^{-b}$, the next leading order must be ${\cal Q}^{-b}_{b}(y)\sim (1-y)^{-b}+{\rm constant}$ since $b<1$. Then, inserting this into the kernel \eqref{eq:kernelaverage} we find that
\begin{align}
\Omega^{\rm res}_{\rm GWs}(1<b<0)\propto {\rm constant}-|1-|y||^{b}\propto {\rm constant}-|k-2c_sk_p|^{b}\,.
\end{align}
For $b>1$ and $c_s\neq1$ there is no detailed study of the kernel \eqref{eq:kernelaverage}. Nevertheless, by looking at the explicit expression for $b=2$ \eqref{eq:wm19} we see that there is nothing special at $c_s(u+v)=1$. Before ending this section, note that the kernel is symmetric around the resonant point $c_s(u+v)=1$. It displays the same behaviour when approaching from both directions, $c_s(u+v)\to1^{\pm}$.

\subsubsection{Infrared regime}
Here we show the behaviour of the kernel when $u\sim v\gg 1$ which is the relevant regime for the IR tail of the GW spectrum as $v\sim k_p/k$ and $k\ll k_p$. This limit means that, if there is any peak in the spectrum of scalar fluctuations, we are looking at scales much larger than the scale of the peak. Thus, such limit of the kernel \eqref{eq:kernelaverage} is relevant for wavenumbers $k_{\rm rh}\ll k\ll k_p$, where we included $k_{\rm rh}$ since it is the limit of the subhorizon approximation. The cases $b<0$ and $b>0$ exhibit different behaviours as anticipated in Sec.~\ref{sec:induced1}. In the limit $y\to 1$, the average kernel \eqref{eq:kernelaverage} is respectively given by
\begin{align}\label{eq:kernelIRbneg}
\overline{I_{\rm IR}^2(x\gg1,u,v,b<0)}\approx&x^{-2(b+1)} 4^{2b}\Gamma^4[b+5/2]\frac{8}{c_s^4 u^2v^2}\frac{\csc^2(b\pi)}{\Gamma^2[3+b]}|1-y|^{2b}\,,
\end{align}
for $b<0$ and
\begin{align}\label{eq:kernelIRbpos}
\overline{I_{\rm IR}^2(x\gg1,u,v,b>0)}\approx &x^{-2(b+1)} 4^{3b}\Gamma^4[b+5/2]\frac{32}{\pi^2c_s^4 u^2v^2}
\left(\frac{(1+b+b^2)}{(b+1)(b+2)}\frac{\Gamma[b]}{\Gamma[2b+3]}\right)^2\,,
\end{align}
for $b>0$. They are actually very similar to \eqref{eq:kernelresbneg} and \eqref{eq:kernelresbpos}, except for the fact that we are now looking at $u\sim v\gg 1$, so that $1-y\sim v^{-2}$. Evaluating \eqref{eq:kernelIRbneg} and \eqref{eq:kernelIRbpos} at the time of reheating and taking into account the factor $k^2$ in \eqref{eq:Phgaussian}, we see that whatever the IR scaling in the radiation dominated universe (which for localized sources is $k^3$ \cite{Cai:2019cdl}) we have to correct the exponent by a factor $-2|b|$, i.e., to $k^{3-2|b|}$ for a localized sourced. This is the same result as in Sec.~\ref{sec:induced1}. The absolute value is due to an extended superhorizon growth of tensor modes for $b<0$, as explained in Sec.~\ref{sec:induced1}.

\subsection{Superhorizon kernel approximation \label{sec:superhorizonkernel}}

In the previous subsections we have studied in detail the subhorizon kernel. However, there are modes which are still superhorizon at the time of reheating. Thus, for such modes we cannot take the upper limit in \eqref{eq:Isimpledef} to $x\to\infty$ as $x\ll1$ at reheating. In addition to the approximation $x\ll1$, we will use the assumption that the peak in the scalar spectrum ${\cal P}_{\Phi}$ is on scales $k_p\gg k_{\rm rh}$. This means that we are only concerned with the region of integration of \eqref{eq:Isimpledef} around the peak in ${\cal P}_{\Phi}$, where we have that $v\sim k_{p}/k\gg1$ and therefore $u\sim v\gg1$. Using this fact we can expand the first Bessel function for small argument in \eqref{eq:Isimpledef} and integrate in the limit of $u\sim v$. By doing so we arrive at Eq.~\eqref{eq:hsuper} but for the kernel, namely
\begin{align}\label{eq:kernelsuperh}
I(x\ll1,u,v)\approx & B_{1,b}+B_{2,b} x^{-2b}\,,
\end{align}
where the exact coefficients are given by
\begin{align}
B_{1,b}=-\frac{(3+2b)^2(1+b+b^2)}{4b(1+b)^2(2+b)}(c_sv)^{-2}\quad ,\quad
B_{2,b}=\frac{4^{1+b}\Gamma^2[b+5/2]}{b(1+b)(2+b)\pi}(c_sv)^{-2(1+b)}\,.
\end{align}
The details on the integrations can be found in App.~\ref{app:integralbessel}. Thus, we have found the kernel for modes which are superhorizon before reheating. However, these induced tensor modes are not yet GWs so we must follow them after they enter the horizon in the radiation domination era.

\subsubsection{Matching to radiation domination}

The kernel \eqref{eq:kernelsuperh} is valid for superhorizon modes during an arbitrary $b={\rm constant}$ period. After this period ends with a sudden reheating, we shall assume that the source term is shut off and tensor modes evolve as freely propagating massless tensor mode. As explained in Sec.~\ref{sec:superhorizonkernel} the continuity of $h_{ij}$ implies continuity of the kernel. This means that the kernel goes from \eqref{eq:kernelsuperh} to the kernel of a superhorizon tensor mode in radiation domination, which reads 
\begin{align}\label{eq:kernelsuperhRD}
I_{RD}(\bar x\ll1,u,v)\approx & B_{1,RD}+B_{2,RD} (k\bar\tau)^{-1}\quad {\rm with}\quad \bar\tau\equiv\tau-\frac{b}{1+b}\tau_{\rm rh}\,,
\end{align}
where we introduced $\bar\tau$ from the continuity of the background FLRW metric, i.e., we imposed continuity of $a$ and $\cal H$ at $\tau=\tau_{\rm rh}$.
Matching \eqref{eq:kernelsuperhRD} and its first derivatives with \eqref{eq:kernelsuperh}, we find that
\begin{align}
B_{1,RD}=B_{1,b}+\frac{1-b}{1+b}B_{2,b}(k\tau_{\rm rh})^{-2b}\quad,\quad B_{2,RD}=\frac{2b}{(1+b)^2}B_{2,b} (k\tau_{\rm rh})^{1-2b}\,.
\end{align}
Thus, we have effectively continued the superhorizon kernel from the $b={\rm constant}$ era to the radiation era. We can now follow the kernel down to subhorizon scales, which must behave as a tensor mode in a radiation dominated universe as there is no source, that is Eq.~\eqref{eq:hgeneralsolution} with $b=0$. Then, taking the averaged square kernel yields
\begin{align}\label{eq:kernelsuperhave1}
\overline{I^2_{RD}(k\ll k_{\rm rh},\tau\gg \tau_{\rm rh})}\approx & \frac{1}{2\bar x^2}\left(B_{1,RD}^2+B_{2,RD}^2\right)\,.
\end{align}
After picking up the most relevant contributions, that is the leading terms for $v\gg1$, we arrive at
\begin{align}\label{eq:kernelsuperhave2}
&\overline{I^2_{RD}(k\ll k_{\rm rh},\tau\gg \tau_{\rm rh})}\approx \frac{1}{2\bar x^2}\left(\frac{(3+2b)^2}{4b(1+b)^2(2+b)}\right)^2 \nonumber\\&\times\left({(1+b+b^2)}(c_sv)^{-2}-4^{1+b}\Gamma^2[b+3/2]\frac{(1-b)}{\pi}(c_sv)^{-2(1+b)}\left((1+b)\frac{k}{k_{\rm rh}}\right)^{-2b}\right)^2\,.
\end{align}
This is the kernel squared that should be used in the induced GW spectrum \eqref{eq:Phgaussian} for scales $k\ll k_{\rm rh}$. In this way, the kernel \eqref{eq:kernelsuperhave2} yields the right IR scaling for localized sources in a radiation dominated universe, i.e., $\Omega^{\rm IR}_{\rm GWs}\propto k^3$ \cite{Cai:2019cdl}. The matching between the superhozion \eqref{eq:kernelsuperhave2} and subhorizon \eqref{eq:kernelaverage} approximations at $k\sim k_{\rm rh}$ has shown to be very good in the instantaneous reheating case \cite{Domenech:2020kqm}. Note that for a gradual transition, the superhorizon approximation should still be good enough except for those modes which enter during the transition. We can then calculate the spectral density of induced GWs by
\begin{equation}\label{eq:spectraldensitysuperRD}
\Omega_{\rm GW}(k\gg k_{\rm rh}, \tau\gg \tau_{\rm rh})=\frac{k^2}{12a^2H^2}\overline{{\cal P}^{RD}_{h}}\,,
\end{equation}
with
\begin{equation}\label{eq:PhgaussiansuperRD}
\overline{{\cal P}^{RD}_h}=8\int_0^\infty dv\int_{|1-v|}^{1+v}du\left(\frac{4v^2-(1-u^2+v^2)^2}{4uv}\right)^2\overline{I_{RD}^2(x\gg1,v,u)}{{\cal P}_{\Phi}(ku)}{{\cal P}_{\Phi}(kv)}\,,
\end{equation}
and using Eq.~\eqref{eq:kernelsuperhave2}. Nevertheless, it is worth noting that for a Dirac delta or a fairly sharp peak and $b>0$ a good approximation to the full induced GW spectrum is to stop the subhorizon spectral density \eqref{eq:spectraldensitysubRD2} at $k\sim 3/4 k_{\rm rh}$ and match with a power-law $\Omega_{\rm GW}=\Omega_{\rm GW,rh} (4k/3k_{\rm rh})^2$ \cite{Domenech:2020kqm}. The amplitude of $\Omega_{\rm GW,rh}$ can be found by matching the two.

It should be noted that Eq.~\eqref{eq:kernelsuperhave2} does not recover the logarithmic correction typical of induced GWs in a radiation dominated universe \cite{Yuan:2019wwo}. The main reason is that we assumed the source term to stop at reheating and, therefore, that superhorizon tensor modes at that time behave as free tensor modes. This is a good approximation taken for simplicity. However, the precise matching would require to first match the Newtonian potential $\Phi$ from a $b={\rm constant}$ universe to radiation domination and then continue the kernel \eqref{eq:kernelApp} after reheating with the matched $\Phi$. This would give a more accurate calculation and would most likely recover the logarithmic correction in the IR tail.


\section{Typical induced GW spectra \label{sec:typical}}

In Sec.~\ref{sec:analytical} we have derived analytical approximations for the kernel (or transfer functions) of induced GWs. However, we still have the integral over the scalar momenta. While the kernel was not so sensitive to the shape of the primordial spectrum, i.e., the approximation is valid as long as all the relevant primordial spectrum enters the horizon before reheating, the integral over momenta crucially depends on the shape of the primordial spectrum. In this section, we will present the final form of the induced GW spectrum and discuss their main features. We will not be so concerned in the series of approximations to achieve analytical formulas of the GW spectrum, which can be found in the papers. Before doing so, we will carefully derive  in Sec.~\ref{sec:densitytoday} what is the spectral density of induced GWs measured today. Then we will focus on the simplest types of motivated primordial spectra. These are a very sharp peak \ref{sec:sharp}, a log-normal peak \ref{sec:broad}, a broken power-law \ref{sec:powerlaw} and a scale invariant spectrum \ref{sec:flat}. We will also comment on possible oscillatory features \ref{sec:oscillatory} on top of the basic shapes and the impact of primordial NG on the final GW spectrum \ref{sec:nongaussian}.\\

\noindent\faBook\,\,\textit{Main references:} Although there are many studies on the induced GW spectrum for various primordial spectra, we focus on the cases where analytical formulas have been derived. In this way, Sec.~\ref{sec:densitytoday} can be derived using the formulas in App.~\ref{app:formulasuseful} and can also be found in \cite{Domenech:2019quo,Domenech:2020kqm}. Sec.~\ref{sec:sharp} is mainly based on \cite{Domenech:2020kqm}. Sec.~\ref{sec:broad} follows \cite{Domenech:2020kqm,Pi:2020otn}. Then, Sec.~\ref{sec:powerlaw} is based on \cite{Atal:2021jyo}. Sec.~\ref{sec:oscillatory} briefly describes the results of Refs.~\cite{Fumagalli:2020nvq,Fumagalli:2021cel}. Lastly, Sec.~\ref{sec:nongaussian} reports the main findings from Refs.~\cite{Cai:2018dig,Unal:2018yaa,Adshead:2021hnm}.

\subsection{The GW spectral density today \label{sec:densitytoday}}
To find the power of GWs that we would observe we have to evaluate \eqref{eq:spectraldensity} today, namely
\begin{equation}\label{eq:spectraldensitytoday}
\Omega_{\rm GW,0}=\frac{1}{3 M_{\rm pl}^2 H_0^2}\frac{d \rho_{\rm GW,0}}{d\ln k}\,.
\end{equation}
However, our computations in Sec.~\ref{sec:analytical} were done up to and including the transition to a radiation dominated universe. Since deep inside the horizon we have that $\rho_{\rm GW}\propto \rho_r\propto a^{-4}$ then after reheating we have that $\Omega_{\rm GW}=\Omega_{\rm GW,\rm rh}={\rm constant}$. To relate $\Omega_{\rm GW,\rm rh}$ with $\Omega_{\rm GW,0}$ we can make use of the fact that GWs behave as radiation and write
\begin{align}\label{eq:spectraldensitytoday2}
\Omega_{\rm GW,0}h^2&=\Omega_{r,0}h^2\frac{1}{\rho_{r,0}}\frac{d \rho_{\rm GW,0}}{d\ln k}\nonumber\\&=1.62\times 10^{-5}\left(\frac{\Omega_{r,0}h^2}{4.18\times 10^{-5}}\right)\left(\frac{g_*(T_{\rm rh})}{106.75}\right)\left(\frac{g_{*s}(T_{\rm rh})}{106.75}\right)^{-4/3}\Omega_{\rm GW,rh}\,.
\end{align}
where $\Omega_{r,0}h^2\approx4.18\times 10^{-5}$ is the density fraction of radiation today given by Planck \cite{Aghanim:2018eyx}. We also traced $\rho_{r,0}$ and $\rho_{GW,0}$ back to the reheating time taking into account that the effective degrees of freedom change with temperature. In the case that the induced GWs are generated during a radiation domination era we should replace the subscript ``rh'' for the evaluation at a time when the induced tensor modes start to behave as a GW. This is denoted with a subscript ``c'' by Inomata et al.~\cite{Inomata:2016rbd}.

\subsection{Dirac delta peak \label{sec:sharp}}

\begin{figure}
\centering
\includegraphics[width=0.55\columnwidth]{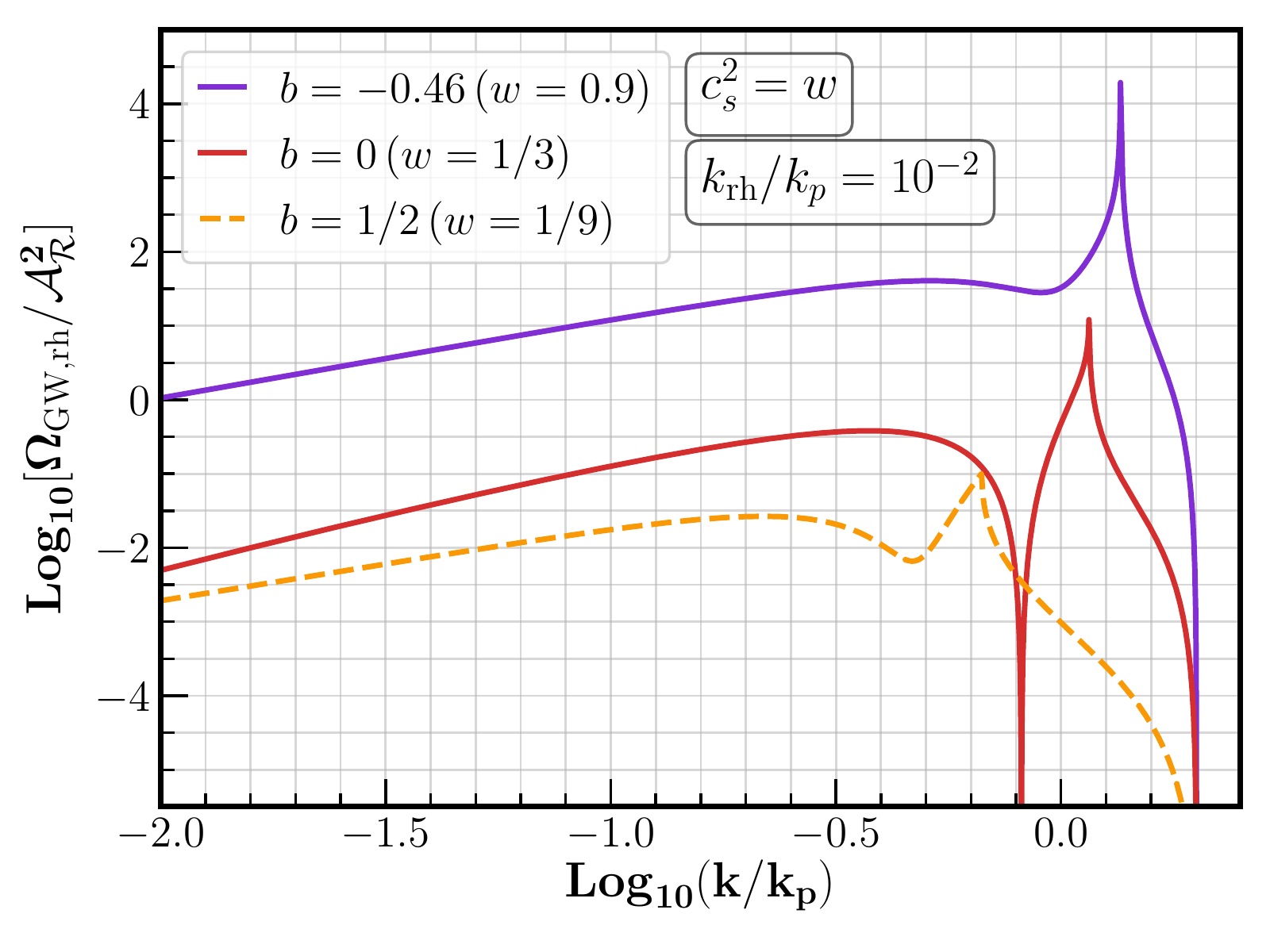}
\caption{Induced GW spectral density for a Dirac delta primordial spectrum, in terms of wavenumber $k/k_p$. We restricted the plot to $k>k_{\rm rh}$. We show the induced GW generated in a radiation dominated universe (red line), in a stiff fluid domination with $w\sim 0.9$ (purple line) and in a soft fluid domination with $w=1/9$ (orange line). All cases have $c_s^2=w$ and, thus, the spectrum presents the resonant peak at $k\sim 2c_s^2 k_p$. We see that for softer $w$, the position of the peak moves to the left. We also see that stiff $w$ enhances the overall amplitude of the induced GW spectrum and has a sharper resonant peak. \label{fig:plotscs}}
\end{figure}

\begin{figure}
\centering
\includegraphics[width=0.55\columnwidth]{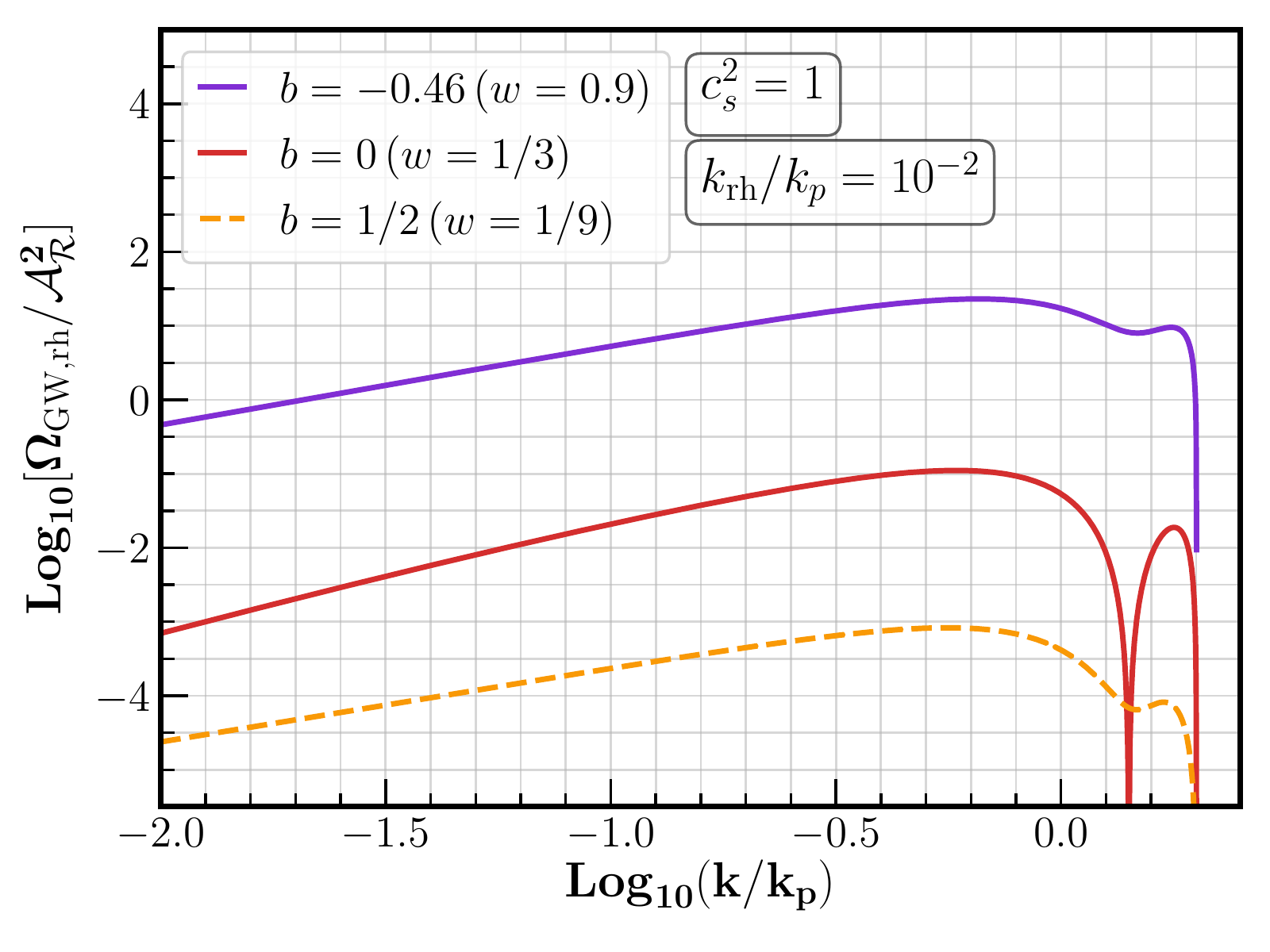}
\caption{Induced GW spectral density for a Dirac delta primordial spectra, in terms of wavenumber $k/k_p$. This time we chose $c_s^2=1$ and, therefore, the resonant peaks of Fig.~\ref{fig:plotscs} are absent. We show in red $w=1/3$, in purple $w\sim0.9$ and in orange $w=1/9$. We see how the stiffer the $w$, the larger the overall amplitude of the induced GW spectrum.\label{fig:plotsnocs}}
\end{figure}

The simplest case is that of a very sharp peak in the primordial spectrum. While this type of spectrum is not possible to generate during single field inflation \cite{Byrnes:2018txb,Carrilho:2019oqg,Ozsoy:2019lyy}, it may be achieved in multi-field models of inflation \cite{Kawasaki:1997ju,Frampton:2010sw,Kawasaki:2012wr,Inomata:2017okj,Pi:2017gih,Cai:2018tuh,Cai:2019jah,Chen:2019zza,Chen:2020uhe}. If the peak is extremely sharp, it is often modelled by a Dirac delta as
\begin{align}
{\cal P}_{\cal R}(k)={\cal A}_{\cal R}\delta\left(\ln(k/k_p)\right)\,.
\end{align}
Although this might be an unrealistic situation, it captures the essence of sharp peaks and gives insight on the kernel. We discuss a more realistic situation in section \ref{sec:broad}. Now, for the power spectrum inside the integral \eqref{eq:Phgaussian} we have
\begin{align}
{\cal P}_{\cal R}(vk)={\cal A}_{\cal R}\delta\left(\ln(vk/k_p)\right)={\cal A}_{\cal R}\frac{k_p}{k}\delta\left(v-\frac{k_p}{k}\right)\,,
\end{align}
and similarly for $u$. Recall, however, that the integral over $u$ is bounded by momentum conservation to $|1-v|<u<1+v$. So, when we integrate over $v$ and $u$ using the Dirac delta, the range is now $|k-k_p|<k_p<k+k_p$. This translates to a range in $k$ given by $0<k<2k_p$. After taking this finite range of $k$ into consideration, we may simply evaluate at $v=u=k_p/k$ the integrand \eqref{eq:PhgaussiansubRD} (and \eqref{eq:PhgaussiansuperRD}) and insert it into \eqref{eq:spectraldensitysubRD} (and \eqref{eq:spectraldensitysuperRD}) to arrive at
\begin{align}\label{eq:omegadelta}
\Omega_{\rm GW,rh}(k)=\frac{2}{3}\left(\frac{k_p}{k_{\rm rh}}\right)^{2}\left(1-\frac{k^2}{4k_p^2}\right)^2\overline{I_{RD}^2\left({k}/{k_{\rm rh}},{k}/{k_p}\right)}\Theta(2k_p-k)\,.
\end{align}
Note that we are being general by allowing a $b={\rm constant}$ phase before the standard radiation domination. Since we are assuming an instantaneous transition at $\tau=\tau_{\rm rh}$ from the $b={\rm constant}$ to radiation domination, we have to include the scale $k_{\rm rh}$ corresponding to the last scale that entered the horizon at the transition. In more detail, $k_{\rm rh}$ is related to $\tau_{\rm rh}$ by
\begin{align}
k_{\rm rh}={\cal H}_{\rm rh}=\frac{1+b}{\tau_{\rm rh}}\,.
\end{align}
Thus, we have that $x_{\rm rh}=k\tau_{\rm rh}=(1+b)k/k_{\rm rh}$. If we are considering the induced GWs in a radiation dominated universe we may simply replace $k_{\rm rh}\to k_p$ and recover the standard formula \cite{Kohri:2018awv}. In Figs.~\ref{fig:plotscs} and \ref{fig:plotsnocs} we see the GW spectrum for different values of $b$ and respectively for $c_s^2=w$ and $c_s^2=1$, so that the resonance is evident.

Despite the fact that \eqref{eq:omegadelta} is already a completely analytical formula, the functional complication does not provide much insight yet. Nevertheless, we can check one interesting limit which is the IR tail of the spectrum. Expanding \eqref{eq:omegadelta} for $k\ll k_p$ we arrive at a general analytical approximation for the low frequency tail of the GW spectrum.  First, for $b<0$ we have that
\begin{align}\label{eq:OMbn}
\Omega_{\rm GW,rh}(b<0,k\ll k_p)=&\frac{{\cal A}^2_{\cal R}}{12}\left(\frac{2^{1+b}(2+b)\Gamma^2[3/2+b]}{\pi c_s^{2(1+b)}\left(1+b\right)^{1+b}}\right)^2\left(\frac{k_{\rm rh}}{k_p}\right)^{2+4b}\nonumber\\&\times\left\{
\begin{aligned}
&\frac{2^{3+2b}}{\pi\left(1+b\right)^{2b}}\left(\frac{k}{k_{\rm rh}}\right)^{2} &(k\lesssim k_{\rm rh}/(1+b))\\
&\left(\frac{\pi}{\sin(b\pi)\Gamma[2+b]}\right)^2\left(\frac{k}{k_{\rm rh}}\right)^{2+2b}  &(k\gtrsim k_{\rm rh}/(1+b))
\end{aligned}
\right.\,,
\end{align}
where we set the broken power-law knee at $k\sim k_{\rm rh}/(1+b)$ since it is when the subhorizon approximation breaks down for $b<0$. A more detailed discussion can be found in Sec.~\ref{sec:subhorizonkernel}.
Second, for $b>0$ we find
\begin{align}\label{eq:OMbp}
\Omega_{\rm GW,rh}(b>0,k\ll k_p)=&\frac{{\cal A}^2_{\cal R}}{{24\pi}}\left(\frac{(2+b)(1+b+b^2)}{c_s^2b\left(1+b\right)^{2}}\right)^2\left(\frac{k_{\rm rh}}{k_p}\right)^{2}\nonumber\\&\times
\left\{
\begin{aligned}
&\left(\frac{k}{k_{\rm rh}}\right)^{2}\quad &(k\lesssim k_{\rm rh})\\
&\frac{1}{2}\left(\frac{{2^{1+b}}\Gamma[b+3/2]}{\left(1+b\right)^{1+b}}\right)^2\,\left(\frac{k}{k_{\rm rh}}\right)^{2-2b}\quad &(k\gtrsim k_{\rm rh})
\end{aligned}
\right.\,,
\end{align}
where this time the matching is roughly at $k\sim k_{\rm rh}$.
The above formulas agree with those in Ref.~\cite{Domenech:2020kqm}. The difference between $b>0$ and $b<0$ is the continued superhorizon growth in the case of $b<0$ which is explained in Sec.~\ref{sec:estimates}. The most important aspect of Eqs.~\eqref{eq:OMbn} and \eqref{eq:OMbp} is that it clearly shows how the IR spectral tilt of the induced GW spectrum is directly related to the equation of state parameter at the time of induced GW generation. Thus, the observation of the IR tail gives insight on the expansion history of the primordial universe. Interestingly, for $b>1$ ($w<0$) the spectrum has a peak at $k\sim k_{\rm rh}$ which is larger than that at $k\sim k_p$. This means that induced GWs for $b<1$ peaks at a different scale than the peak of the primordial spectrum. This has been used, e.g., in \cite{Domenech:2020ers} to fit the NANOGrav results \cite{Arzoumanian:2020vkk} and at the same time predict PBH with a mass smaller than the solar mass. If instead one assumes induced GWs in radiation domination to fit the NANOGrav frequency range, the associated PBH counterpart must be of the order of solar mass. We discuss more about these possibilities in Sec.~\ref{sec:observations}. Another important point is that the amplitude of the spectrum at $k\sim k_p$ strongly depends on the ratio $k_{\rm rh}/k_p$ and often mildly on $b$. We thus see that the smaller the ratio $k_{\rm rh}/k_p$, the more suppressed is the spectrum at $k\sim k_{\rm rh}$. This is because the main induced GW generation occurs when the scalar mode $k_p$ enters the horizon. Then, since the modes with $k\sim k_{\rm rh}$ have been superhorizon until the transition they can only grow by a factor $(k_{\rm rh}/k_p)^2$. For $b<0$ the argument is slightly different because of the second superhorizon growth and so the suppression is minor compared to $b>0$ with a power index that depends on $b$.

\subsection{Log-normal peak \label{sec:broad}}

We now turn to a more realistic situation where the peak in the primordial spectrum has a finite width. Inspired by several models of multi-field inflation the peak of the primordial spectrum may be parametrized by a log-normal \cite{Pi:2020otn}, namely
\begin{align}\label{eq:lognormalpeak}
\mathcal{P}_{\cal R}(k)=\frac{\mathcal{A_{\cal R}}}{\sqrt{2\pi}\Delta}\exp\left[-\frac{\ln^2(k/k_p)}{2\Delta^2}\right],
\end{align}
where $\Delta$ is the dimensionless width of the peak. Note that the power spectrum \eqref{eq:lognormalpeak} is normalized, i.e., $\int_{-\infty}^\infty d\ln k\, \mathcal{P}_{\cal R}(k)=\mathcal{A_{\cal R}}$. The log-normal parametrization includes sharp peaks where $\Delta\ll 1$ \cite{Kawasaki:1997ju,Frampton:2010sw,Kawasaki:2012wr,Inomata:2017okj,Pi:2017gih,Cai:2018tuh,Cai:2019jah,Chen:2019zza,Ashoorioon:2019xqc,Chen:2020uhe} and broad peaks where $\Delta\gtrsim1$ \cite{Garcia-Bellido:1996mdl,Yokoyama:1998pt,Kohri:2012yw,Clesse:2015wea,Cheng:2016qzb,Espinosa:2017sgp,Inomata:2017okj,Kannike:2017bxn,Garcia-Bellido:2017mdw,Ando:2017veq,Cheng:2018yyr,Ando:2018nge,Espinosa:2018eve,Inomata:2018cht,Braglia:2020eai,Palma:2020ejf,Fumagalli:2020adf}. For $\Delta\to 0$ we recover the Dirac delta case.

\begin{figure}
\centering
\includegraphics[width=0.55\columnwidth]{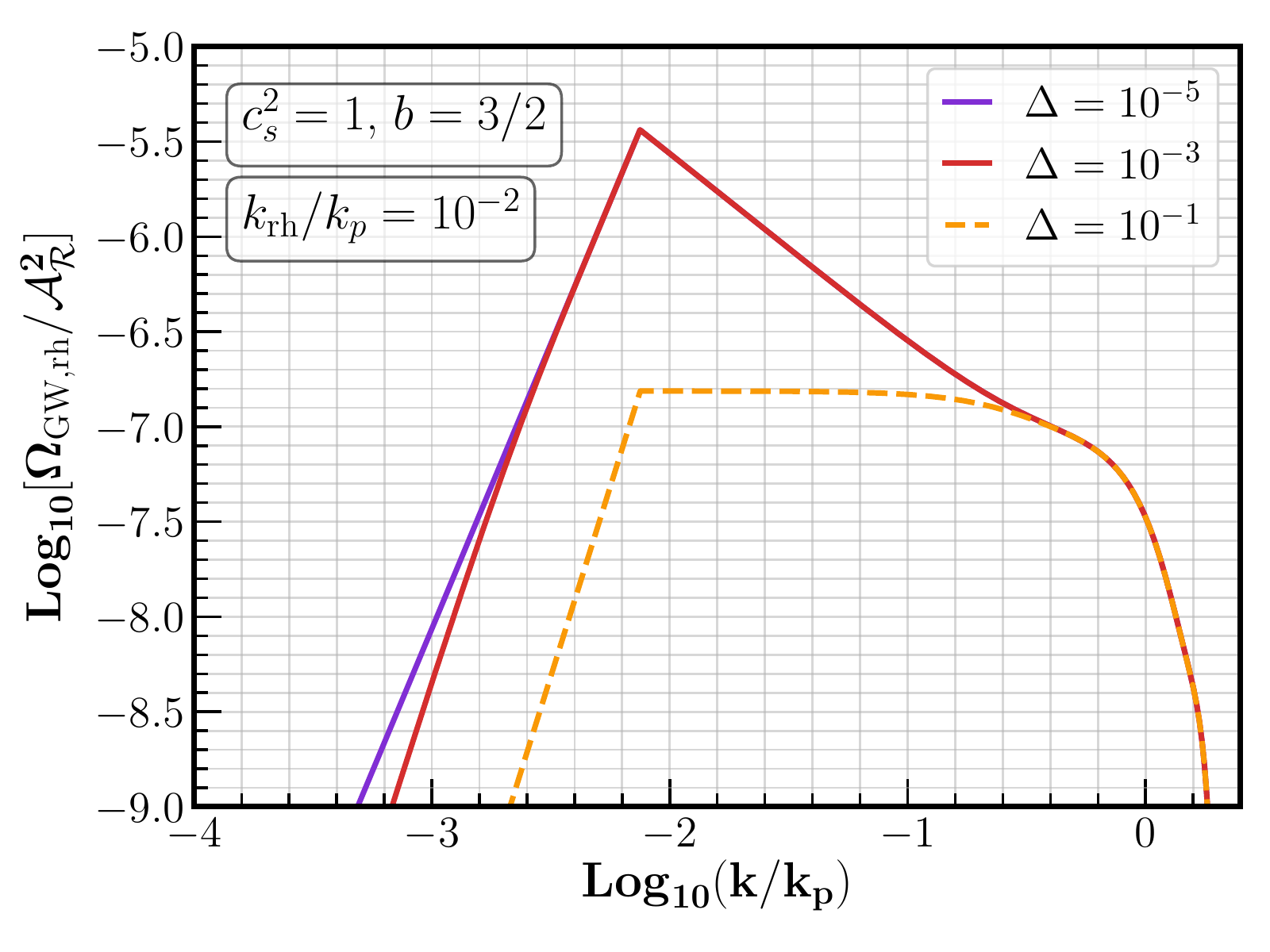}
\caption{Induced GW spectral density for the log-normal primordial spectrum. We used $c_s^2=1$, so there is no resonant peak. We also chose $b=3/2$ ($w=-1/15$) to show that for $b<1$ the largest peak of the induced GW spectrum for a Dirac delta is at $k\sim k_{\rm rh}$. In purple we show $\Delta=10^{-5}$ which for the range of $k$ shown in the plot is effectively the induced GW spectrum of a Dirac delta. For $k<k_{\rm rh}$ the GW spectrum goes as $k^2$. In red we show $\Delta=10^{-3}$ so that the effects of the finite width occur within the plot range. Around $k/k_p\sim 5\times 10^{-2}$ the GW spectrum goes from $k^2$ to $k^3$. Lastly, we show in orange $\Delta=10^{-1}$. The finite width becomes important now for $k_{\rm rh}<k<k_p$ and the GW spectrum transition from $k^{2-2b}=k^{-1}$ to $k^{3-2b}=k^{0}$.\label{fig:plotsLN}}
\end{figure}

In this review we are mainly interested in the case of a narrow peak ($\Delta\ll1$). The reasons are its simplicity and that we can use our results for general $b$ of Sec.~\ref{sec:sharp}. The interested reader for analytical approximation in the broad peak case ($\Delta\gtrsim1$) in radiation domination is referred to the work of Pi and Sasaki \cite{Pi:2020otn}. Here we may directly borrow their result for $\Delta\ll 1$ which reads
\begin{align}\label{eq:omegafinitewidth}
\Omega_{\text{GW},\Delta}(k)=\text{erf}\left(\frac{1}{\Delta}\sinh^{-1}\frac{k}{2k_p}\right)
\Omega_{\text{GW},\delta}(k),
\end{align}
where ${\rm erf}(x)$ is the error function and $\Omega_{\text{GW},\delta}(k)$ is the GW spectrum induced by a $\delta$-function peak given in Sec.~\ref{sec:sharp}. Then, Eq.~\eqref{eq:omegafinitewidth} provides the correction to the Dirac delta spectrum of Sec.~\ref{sec:sharp} due to the small but finite width of the spectrum which leads to the $k^3$ IR scaling in radiation domination \cite{Cai:2019cdl}. The most important effect then appears when the ratio of scales $k/k_p$ is smaller than the width $\Delta$. Indeed, when $k\ll k_p$ we have that
\begin{align}\label{eq:omegafinitewidth2}
\Omega_{\text{GW},\Delta}(k)=\text{erf}\left(\frac{k}{2k_p\Delta}\right)
\Omega_{\text{GW},\delta}(k),
\end{align}
which gives a correction of $\text{erf}\left({k}/({2k_p\Delta})\right)\sim {k}/({2k_p\Delta})$ when $\Delta\gg{k}/({2k_p})$. Thus, due to the presence of $\Delta$ our IR tail of Sec.~\ref{sec:sharp} changes respectively for ${2 k_p \Delta}< k_{\rm rh}$ and ${2 k_p \Delta}>k_{\rm rh}$ to
\begin{align}\label{eq:OMGWRbroad}
\Omega_{\rm GWs}(k\ll k_p,{2 k_p \Delta}< k_{\rm rh})\propto{\cal A}^2_{\cal R}\left\{
\begin{aligned}
&\left(\frac{k}{k_p}\right)^3&k\ll {2 k_p \Delta} \ll k_{\rm rh}\\
&\left(\frac{k}{k_p}\right)^{2}& {2 k_p \Delta}\ll k \ll k_{\rm rh}\\
&\left(\frac{k}{k_p}\right)^{2-2|b|}& k\ll {2 k_p\Delta}\ll k_p
\end{aligned}
\right.\,,
\end{align}
or
\begin{align}\label{eq:OMGWRbroad2}
\Omega_{\rm GWs}(k\ll k_p,{2 k_p \Delta}>k_{\rm rh})\propto{\cal A}^2_{\cal R}\left\{
\begin{aligned}
&\left(\frac{k}{k_p}\right)^3&k\ll k_{\rm rh}\\
&\left(\frac{k}{k_p}\right)^{3-2|b|}& k_{\rm rh}\ll k\ll {2 k_p\Delta}\\
&\left(\frac{k}{k_p}\right)^{2-2|b|}& {2k_p\Delta}\ll k\ll k_p
\end{aligned}
\right.\,.
\end{align}
Note that since $k_{\rm rh}/k_p\ll 1$ the case ${2 k_p \Delta}>k_{\rm rh}$ still allows for a very sharp peak with $k_{\rm rh}/(2k_p)<\Delta\ll 1$. See how the inclusion of the finite width increases the richness of the induced GWs spectrum. We illustrate this in Fig~\ref{fig:plotsLN}. Also note that if $b=0$ then the induced GW spectrum only has two different slopes and $k_{\rm rh}$ becomes meaningless. 

Before ending this subsection, we briefly discuss the case when $\Delta>1$. It is shown in Ref.~\cite{Pi:2020otn} that in the induced GW spectrum in radiation domination for $\Delta>1$ and near $k\sim k_p$ is well approximated by
\begin{align}
\Omega^{\rm peak}_{\rm GWs}(\Delta>1)\approx 0.125 \frac{{\cal A}_{\cal R}^2}{\Delta^2}e^{-\frac{\ln^2(k/k_p)}{\Delta^2}}\,.
\end{align}
This means that the GW spectrum induced by a log-normal peak in the primordial spectrum also follows a log-normal peak but with a shallower width given by $\Delta/\sqrt{2}$. The factor $2$ is a reflection of the secondary nature of induced GWs \cite{Pi:2020otn}.

\subsection{Broken power-law \label{sec:powerlaw}}

In Sec.~\ref{sec:broad} we have studied the log-normal sharp peak which is inspired in multi-field models of inflation. However, if the curvature perturbation is enhanced during single field inflation it often takes a broken power-law shape \cite{Byrnes:2018txb,Atal:2018neu,Carrilho:2019oqg,Ozsoy:2019lyy}. The induced GWs counterpart for a broken power-law primordial spectrum have been studied in Refs.~\cite{Clesse:2018ogk,Xu:2019bdp,Liu:2020oqe,Riccardi:2021rlf,Atal:2021jyo}. Here we present the results of Ref.~\cite{Atal:2021jyo}, which contain analytical approximations for the amplitude and spectral tilt of both the IR and UV tails of the induced GW spectrum. A detailed discussion on the approximations used can be found in \cite{Atal:2021jyo}. At the end of this subsection, we also briefly discuss the case of a scale invariant primordial spectrum.

Let us parametrize the primordial curvature power spectrum as a strict broken power-law given by
\begin{align}\label{eq:PRbpl}
{\cal P}_{\cal R}={\cal A}_{\cal R}\left\{
\begin{aligned}
&\left(\frac{k}{k_p}\right)^{n_{\rm IR}}\quad &k\leq k_p\\
&\left(\frac{k}{k_p}\right)^{-n_{\rm UV}}\quad &k\geq k_p
\end{aligned}
\right.\,,
\end{align}
where $n_{\rm IR},\,n_{\rm UV}>0$ respectively are the IR and UV spectral tilts of the spectrum which peaks at $k=k_p$. In single field models one often finds that $n_{\rm IR}\sim 4$ \cite{Byrnes:2018txb}. If the enhancement of the curvature fluctuations is due to a bump in the inflationary potential, then $n_{\rm UV}$ can be related to the second derivative of the potential at the bump \cite{Atal:2018neu,Liu:2020oqe}. In such models $n_{\rm UV}$ is also related to the local non-gaussianity parameter $f_{NL}$ \cite{Atal:2018neu}. In what follows, we will only focus on induced GWs in the radiation dominated universe and later discuss its generalisations to other values of the equation of state parameter. Currently, only analytical approximations for radiation domination are available. Now, using \eqref{eq:PhgaussiansubRD2} and \eqref{eq:PRbpl} we can compute the associated induced GW spectrum. Doing so for the IR tail, i.e., for modes with $k\ll k_p$, we find that
\begin{align}\label{eq:OGWIR}
\Omega_{\rm GW,rh}(k\ll k_{p})\approx 12{\cal A}_{\cal R}^2\left(\frac{1}{2n_{\rm IR}-3}+\frac{1}{2n_{\rm UV}+3}\right)\left(\frac{k}{k_p}\right)^{3}\ln^2\left(\frac{k}{k_p}\right)\,.
\end{align}
Eq.~\eqref{eq:OGWIR} is valid for $n_{\rm IR}>3/2$ and $n_{\rm UV}>0$ and gives a good approximation of the IR tail. The condition $n_{\rm IR}>3/2$ comes from requiring convergence of the integral at large internal momenta ($u\sim v\gg 1$).

\begin{figure}
\centering
\includegraphics[width=0.55\columnwidth]{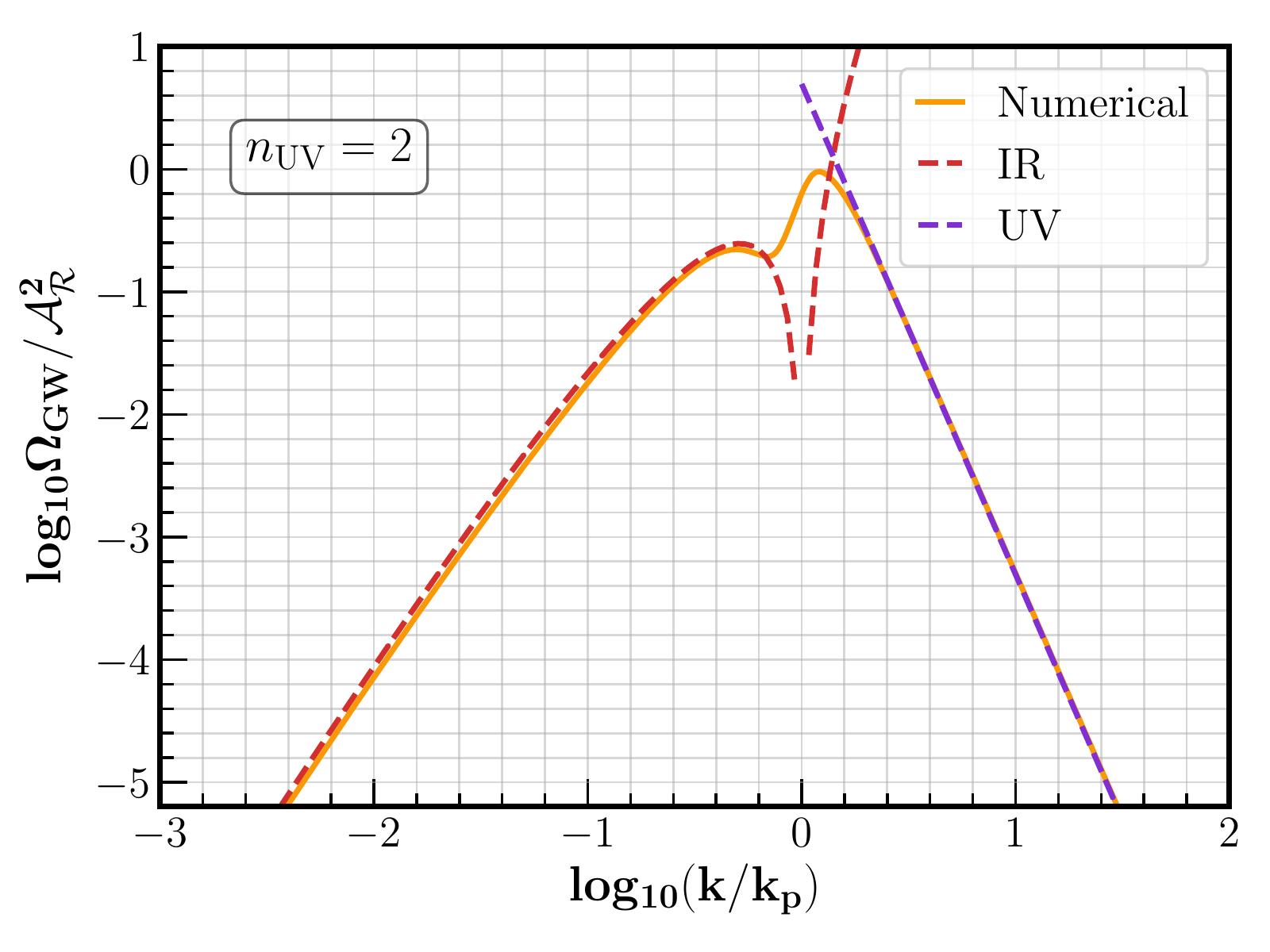}
\caption{Induced GW spectral density for a broken power-law primordial spectrum with $n_{\rm IR}=4$ and $n_{\rm UV}=2$. In orange we show the result of numerical integration. In red we show the IR approximation \eqref{eq:OGWIR}. In purple we show the UV approximation \eqref{eq:OGWUV}. See how the analytical estimates fit well the numerical integration for $k\ll k_p$ and $k\gg k_p$. They even give a good order of magnitude estimate of the amplitude of the GW spectrum around $k\sim k_p$. \label{fig:plotPL}}
\end{figure}

The UV tail of the induced GW spectrum presents two different behaviours. If $0<n_{\rm UV}<4$ the integral converges even in the strict limit $v\to 0$. This means that we can pull out of the integral \eqref{eq:PhgaussiansubRD2} all the $k$ dependence in ${\cal P}_{\cal R}$ and have that $\Omega_{\rm GW,rh}\sim {\cal P}_{\cal R}^2$. The remaining piece is just a numerical factor. In this way, we find that the induced GW spectrum in the UV reads 
\begin{align}\label{eq:OGWUV}
\Omega_{\rm GW,rh}(k\gg k_{p},n_{\rm UV}<4)\approx \frac{1}{12}{\cal A}_{\cal R}^2F(n_{\rm UV})\left(\frac{k}{k_p}\right)^{-2n_{\rm UV}}\,,
\end{align}
where $F(n_{\rm UV})$, for $0<n_{\rm UV}<4$, is given by
\begin{align}\label{eq:fuv}
F(n_{\rm UV})=8\int_0^\infty dv\int_{|1-v|}^{1+v}du\left(\frac{4v^2-(1-u^2+v^2)^2}{4uv}\right)^2 (uv)^{n_{\rm UV}} \overline{I^2(n_{\rm UV},k,v,u)}\,,
\end{align}
and converges everywhere. The range of $n_{\rm UV}$ for which this approximation is valid depends on the value of $b$. We discuss this after presenting the results for radiation domination ($b=0$). Since \eqref{eq:fuv} has to be computed numerically, we present here a numerical fit for $n_{\rm UV}<4$, which reads
\begin{align}\label{eq:f_UV}
F(n_{\rm UV})\approx 41+\frac{16n_{\rm UV}^2}{\sqrt{16-n_{\rm UV}^2}}\,.
\end{align}
The form of \eqref{eq:f_UV} has been found noting that the integral diverges in the exact limit $n_{\rm UV}\to 4$. After several trial and error attempts, a reasonable fit was found to be \eqref{eq:f_UV}. A detailed study of \eqref{eq:fuv} might yield better analytical insights than \eqref{eq:f_UV}. For the moment Eq.~\eqref{eq:f_UV} gives a good order of magnitude estimate. Note that when $n_{\rm UV}=4$ the induced GW spectrum present a logarithmic divergence \cite{Liu:2020oqe,Atal:2021jyo}. Lastly, for $n_{\rm UV}>4$ the primordial spectrum has a sharp decay and, in analogy with the sharp peak, we do not expect that $\Omega_{\rm GW,rh}\sim {\cal P}_{\cal R}^2$. Instead, there should be a fast fall off at around $k\sim 2k_p$ by momentum conservation. Indeed when $n_{\rm UV}>4$ we find that
\begin{align}\label{eq:OGWUV2}
\Omega_{\rm GW,rh}(k\gg k_{p},n_{\rm UV}>4)\approx \frac{16}{3}{\cal A}_{\cal R}^2\left(\frac{1}{n_{\rm UV}-4}+\frac{1}{n_{\rm IR}+4}\right)\left(\frac{k}{k_p}\right)^{-4-n_{\rm UV}}\,.
\end{align}
We show the analytical estimates against the numerical results in Fig.~\ref{fig:plotPL}. See how even if the approximations are extrapolated to $k\sim k_p$ they still yield a sensible order of magnitude estimate of the induced GW spectrum. 

With the above analytical estimates for the amplitude and the spectral tilt in the IR and UV limits we conclude that the induced GW spectrum has the following broken power-law shape
\begin{align}\label{eq:OMGWR}
\Omega_{\rm GWs}(k)\propto{\cal A}^2_{\cal R}\left\{
\begin{aligned}
&\left(\frac{k}{k_p}\right)^3&k\ll k_p\\
&\left(\frac{k}{k_p}\right)^{-\Delta}& k\gg k_p
\end{aligned}
\right.\,,
\quad
{\rm with}
\quad
\Delta=\left\{
\begin{aligned}
&2n_{\rm UV} & 0<n_{\rm UV}<4\\
&4+n_{\rm UV}& n_{\rm UV}>4
\end{aligned} 
\right.\,.
\end{align}
Eq.~\eqref{eq:OMGWR} tells us that the UV spectral tilt of the induced GW has information the UV tail of the primordial power spectrum. This is to be contrasted with the IR spectral tilt, which does not tell us much about the inflationary model unless we measure very well the amplitude. Lastly, if we focus on the UV tail of the induced GW spectrum \eqref{eq:OMGWR}, we see that for example it may be degenerate with the GW spectrum from the sound waves in a first order phase transition \cite{Kuroyanagi:2018csn}.

\subsubsection{Alternative expansion histories}
Here we briefly present the spectral tilts of the induced GW spectrum if instead of radiation the universe was dominated by another perfect fluid. In this case, we find that \eqref{eq:OMGWR} changes to \cite{Atal:2021jyo}
\begin{align}\label{eq:OMGWGB}
\Omega_{\rm GW}(k)\propto{\cal A}^2_{\cal R}\left\{
\begin{aligned}
&\left(\frac{k}{k_p}\right)^{3}&k\ll k_{\rm rh}\\
&\left(\frac{k}{k_p}\right)^{3-2|b|}&k_{\rm rh}\ll k\ll k_p\\
&\left(\frac{k}{k_p}\right)^{-\Delta-2b}& k\gg k_p
\end{aligned}
\right.\,,
\end{align}
where $b$ is defined in Eq.~\eqref{eq:b} and this time
\begin{align}
\Delta=\left\{
\begin{aligned}
&2n_{\rm UV} & 0<n_{\rm UV}+b<4\\
&4+n_{\rm UV}& n_{\rm UV}+b>4
\end{aligned} 
\right.\,.
\end{align}
Previously, in Secs.~\ref{sec:induced1} and \ref{sec:analytical}, we derived the IR broken power-law behaviour of the induced GW spectrum for general equation of state parameter. Now, in this section we have a power-law UV tail in the induced GW coming from the UV tail of the primordial spectrum. The additional factor $-2b$ with respect to \eqref{eq:OMGWR} in the UV tail can be understood as follows. All modes with $k\gg k_p$ entered the horizon deep inside the $b={\rm constant}$ era and before the scalar peak at $k\sim k_p$. After that the induced tensor modes effectively behave as free GWs irrespective of whether $b<0$ or $b>0$. Thus, the free GWs experience the typical relative redshift with respect to the background which is the factor $k^{-2b}$. Note that if $b<0$ the UV tail might have a blue tilt.

\subsubsection{Scale invariant spectrum \label{sec:flat}}

A limiting case of the broken power-law studied \eqref{eq:PRbpl}, setting $n_{\rm UV}=n_{\rm IR}=0$, is a scale invariant spectrum typical of slow-roll inflation. In this case, it is easy to convince ourselves that the induced GW spectrum for general expansion histories is given by 
\begin{equation}\label{eq:Phflatfinal}
\Omega_{\rm GW,rh}={\cal A}^2_{\cal R}{\cal F}(b,c_s)\left(\frac{k}{k_{\rm rh}}\right)^{-2b}\,.
\end{equation}
where 
\begin{align}\label{eq:FFuv}
{\cal F}(b,c_s)=\frac{2}{3}\int_0^\infty dv\int_{|1-v|}^{1+v}du\left(\frac{4v^2-(1-u^2+v^2)^2}{4uv}\right)^2 \overline{I^2(b,c_s,k,v,u)}\,,
\end{align}
and $\overline{I^2}$ is given by \eqref{eq:kernelaverage}. Another explicit formula can be found at the summary section \ref{sec:summary}. The integral \eqref{eq:FFuv} converges everywhere for any value of $b$ and $c_s$. However, note that \eqref{eq:FFuv} with \eqref{eq:kernelaverage} is only valid for $k\gg k_{\rm rh}$. Nevertheless, since deep inside the radiation domination we expect a scale invariant induced GW spectrum it is reasonable to assume that for scales $k\sim k_{\rm rh}$ we can match the $k^{-2b}$ behaviour to the constant spectral density. A more detailed, perhaps numerical, analysis is needed to determine the spectrum for $k\sim k_{\rm rh}$ as induced GWs are being sourced before and after the transition.

\subsection{Oscillatory features \label{sec:oscillatory}}

In the subsections above we have discussed the GWs induced by the simplest shapes of the primordial spectrum, that is a log-normal peak and a broken power-law. However, in some situations, it is expected that the primordial power spectrum presents a series of oscillations usually referred to as ``primordial features''. For recent reviews on such features see, e.g., \cite{Chen:2010xka,Chluba:2015bqa,Slosar:2019gvt}. These oscillations can be classified according to the $k$ dependence in their frequency. If the frequency is linear in $k$, the oscillations resulted from sharp dynamics during inflation. For example, a step in the potential in single field inflation \cite{Starobinsky:1992ts,Adams:2001vc,Bean:2008na,Adshead:2011jq,Bartolo:2013exa,Palma:2014hra,Ballesteros:2018wlw,Kefala:2020xsx} or a sharp turn in field space \cite{Achucarro:2010da,Shiu:2011qw,Gao:2012uq,Palma:2020ejf,Fumagalli:2020adf,Fumagalli:2020nvq,Braglia:2020taf}. This case is also known as sharp feature. If the frequency is logarithmic in $k$, the oscillations are called resonant features. They are due to resonances during inflation on subhorizon scales. For instance, resonances occurs if the potential in single field inflation has wiggles or a massive scalar field is oscillating at the bottom of the potential \cite{Pahud:2008ae,Chen:2008wn,Silverstein:2008sg,Flauger:2009ab,Chen:2011zf,Chen:2014joa,Chen:2014cwa,Gao:2015aba,Fumagalli:2021cel}. In more general cases, such as general interactions \cite{Huang:2016quc,Domenech:2018bnf} or alternative scenarios to inflation \cite{Chen:2018cgg}, the frequency of the oscillations in the primordial spectrum could be a power-law of $k$.

\begin{figure}
\centering
\includegraphics[width=0.49\columnwidth]{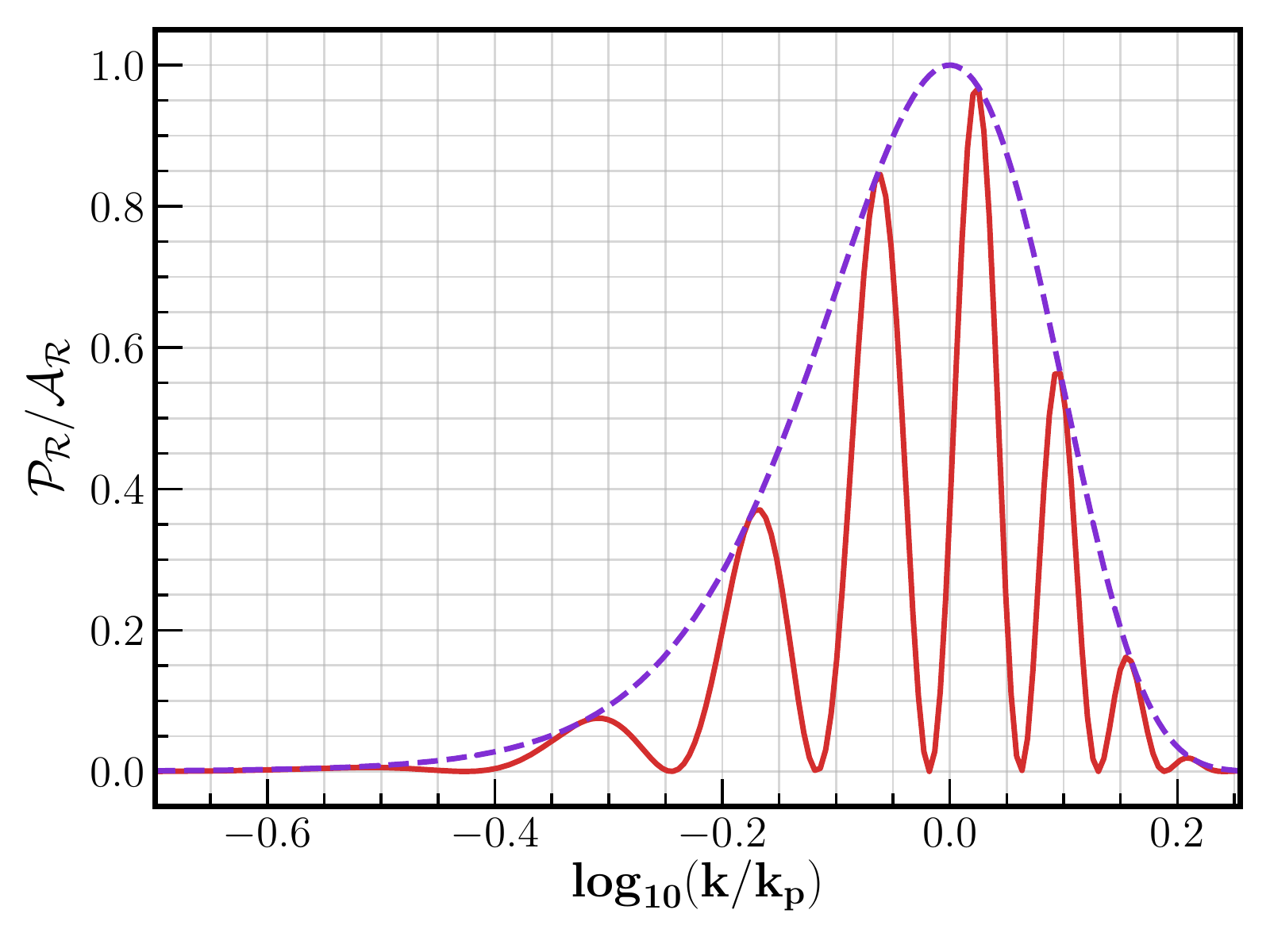}
\includegraphics[width=0.49\columnwidth]{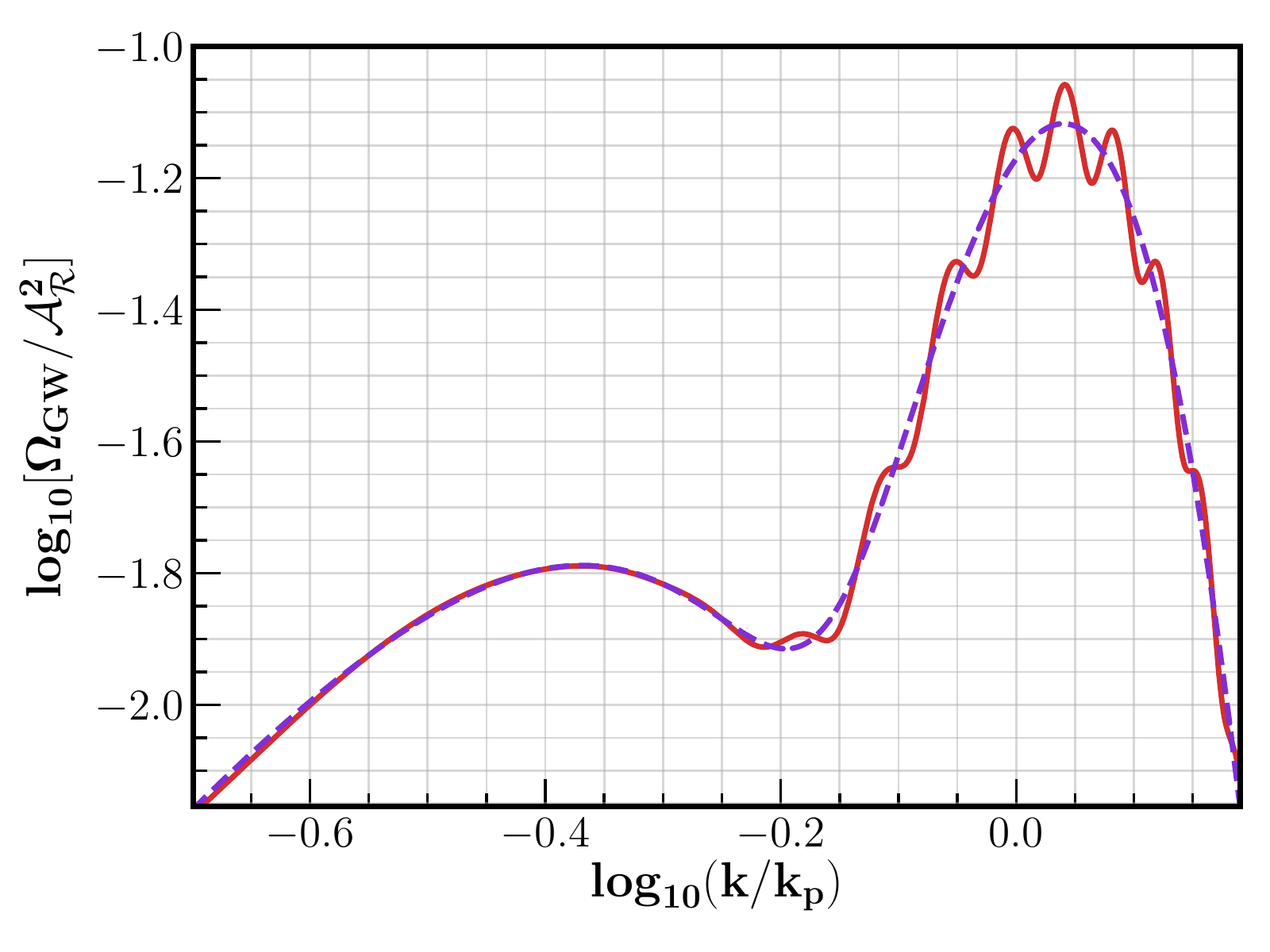}
\caption{Primordial spectrum and induced GWs from a sharp turn in the inflationary trajectory from the model of Ref.~\cite{Fumagalli:2020nvq}, courtesy of L.~Witkowski, J.~Fumagalli and S.~Renaux-Petel. Left: Normalized primordial spectrum in terms of wavenumber $k/k_p$. The solid line is the power spectrum resulting from the sharp feature. The dashed line is the envelope of the oscillations. They are respectively given by Eqs.~(2.25) and (2.26) of Ref.~\cite{Fumagalli:2020nvq} with $\delta = 0.5$ and $\eta_\perp = 20$. Right: Induced GW spectral density normalized by ${\cal A}_{\cal R}$. The solid line is the induced GW spectrum of the oscillatory power spectrum while the dashed line is the induced GWs from the envelope properly normalised. The oscillations have an amplitude of about $20\%$ larger than the envelope around the induced GW spectrum peak.\label{fig:lukas}}
\end{figure}

Remarkably, despite the induced GWs being a second order effect, oscillations are not completely washed out. The main difficulty is to derive analytical approximations to the induced GW spectrum. In this respect, Fumagalli, Renaux-Petel and Witkowski \cite{Fumagalli:2020nvq,Fumagalli:2021cel} and also Braglia, Chen and Hazra \cite{Braglia:2020taf} did a detailed numerical analysis of the induced GW spectrum and derived motivated semi-analytical templates. Perhaps the most interesting case which is when the oscillations in the primordial spectrum are $O(1)$. For example, this is the case when a sharp feature is responsible for the peak in the primordial spectrum \cite{Fumagalli:2020adf,Fumagalli:2020nvq}. Now, if the oscillations are large and separated linearly in $k$, we can gain intuition by looking at the sum of Dirac delta peaks \cite{Cai:2019amo,Fumagalli:2020nvq}. In this case if we call $\omega_{\rm lin}$ to the frequency of the Dirac delta peaks in the primordial spectrum, then the frequency in the induced GW spectrum is $\omega^{\rm GWs}_{\rm lin}=\omega_{\rm lin}/c_s$ \cite{Cai:2019amo,Fumagalli:2020nvq} (in radiation domination $c_s=1/\sqrt{3}$). Interestingly, in the more realistic situation studied \cite{Fumagalli:2020nvq} the induced GW spectrum follows an envelope as if there were no oscillations and, on top of that, oscillations appear near the resonant peak at $k\sim 2c_s k_p$ and reach at most an amplitude of about $20\%$ in radiation domination \cite{Fumagalli:2020nvq}. We illustrate this possibility in Fig.~\ref{fig:lukas}. Another example are $O(1)$ oscillations due to a resonant feature in the primordial spectrum. In this case, if we call $\omega_\textrm{log}$ the frequency of the resonant feature, the resulting induced GW spectrum exhibits two oscillatory components with frequencies $\omega_\textrm{log}$ and $2 \omega_\textrm{log}$. The relative amplitude of such components depend on both the frequency and the envelope of the primordial spectrum \cite{Fumagalli:2021cel}. For the analytical templates and a more detailed discussion see Refs.~\cite{Fumagalli:2020nvq,Fumagalli:2021cel}.

\subsection{Impact of non-Gaussianities \label{sec:nongaussian}}

Here we briefly discuss the effects that primordial non-gaussianities may have on the induced GW spectrum. We will not dwell into the details, which are well explained in Refs.~\cite{Cai:2018dig,Unal:2018yaa,Adshead:2021hnm}. The leading formulas for local-type non-gaussianity can be found in Sec.~\ref{sec:ng}. We describe the main effects of local-type non-gaussianity below.

First, let us focus on the very sharp peaks of sec.~\ref{sec:sharp} and \ref{sec:broad}. In this case, the most relevant effect is that the induced GW spectrum does not stop at $k\sim 2k_p$ but continues until $k\sim 3k_p$. This is roughly because when including primordial non-gaussianity, the source term to induced GWs has three scalars. Thus, we have a cut-off at $k\sim 3k_p$ by momentum conservation.  Note that this effect is characteristic of the sharp peak in the primordial spectrum. For large non-gaussianity, that is $F_{NL}\gg1$ in Eq.~\eqref{eq:FNLPHIExpansion}, the spectrum does not have the clear resonant peak of \ref{sec:sharp} but presents a few peaks with similar amplitude instead \cite{Cai:2018dig,Adshead:2021hnm}. Second, in the case where the peak is broad or a broken power-law, the non-gaussian contribution typically has the same IR and UV spectral tilts \cite{Atal:2021jyo,Adshead:2021hnm}. For large enough $F_{NL}\gtrsim 1$, the induced GW spectrum also shows an additional peak/bump at around $k\sim 3k_p$. If the non-gaussianity is large enough, i.e., ${\cal A}_{\cal R}F^2_{NL}>O(1)$, the non-gaussian contribution starts to dominate over the gaussian one \cite{Cai:2018dig,Adshead:2021hnm}. However, it is not so clear whether one can achieve ${\cal A}_{\cal R}F^2_{NL}>1$ in inflationary models as in this case the non-gaussian expression \eqref{eq:FNLPHIExpansion} cannot be regarded as a perturbative expansion \cite{Atal:2021jyo}. A more detailed and general analysis is necessary. It should be noted that if we include third order perturbation theory terms, the induced GW spectrum for a Dirac delta peak has three resonant peaks instead of one \cite{Yuan:2021qgz,Yuan:2019udt}.

The primordial non-gaussian contribution to the induced GW spectrum, as we have just presented it, might not yield much information on inflation. For example, one may design a gaussian primordial spectrum such that the resulting induced GW spectrum is exactly as that of a primordial log-normal peak with large non-gaussianities. Although this may be a very complicated task, it is in principle possible. In this sense, there is no telling primordial non-gaussianity apart just from the induced GW spectrum \cite{Yuan:2021qgz}. Nevertheless, the story becomes more interesting when considering the PBH counterpart. Since PBH formation is highly dependent on the tail of the distribution of primordial fluctuations, any tiny deviation from a normal distribution could have an exponential impact on the fraction of PBHs. For the details see the recent discussions in Refs.~\cite{Young:2013oia,Yoo:2019pma,Atal:2018neu,Atal:2019cdz,Kehagias:2019eil,Young:2019yug,DeLuca:2019qsy,Kawasaki:2019mbl,Riccardi:2021rlf}. Then, ideally one may use the two observations, the PBH mass function and the induced GW counterpart, to perhaps disentangle the primordial non-gaussianity from the primordial spectrum. 


\section{The dust dominated universe \label{sec:dust}}

In previous sections we have derived intuitive and analytical formulas for the spectrum of induced GWs under the assumption that the universe was not dominated by a pressure-less perfect fluid with vanishing propagation speed of fluctuations, i.e., $w=c_s^2=0$, so-called dust. The particularity of a dust dominated universe is that the Newtonian potential is constant on all scales. As can be seen from Eq.~\eqref{eq:phieom} with $c_s^2=0$, $\Phi={\rm constant}$ is a solution. This also implies that density fluctuations of the dust fluid grow. In fact, they grow as $\delta\rho/\rho\propto a$. This is the typical behaviour of a CDM dominated universe. This also means that there is a scale below which density fluctuations grow so much that they become non-linear, i.e., larger than $O(1)$. This occurs for modes with \cite{Assadullahi:2009nf}
\begin{align}\label{eq:kNL}
k>k_{NL}\sim \sqrt{\frac{3}{2}}{\cal P}^{-1/4}_{\Phi}(\tau_{\rm rh}){\cal H}(\tau_{\rm rh})\,.
\end{align}
Note that it does not mean that if there are fluctuations on $k>k_{NL}$ there will be no production of induced GWs. It means that we cannot completely trust our calculations since non-linear dynamics might be important, e.g., due to GWs from mergers. Nevertheless, we quite often have $\Phi<1$ and thus perturbation theory for $\Phi$ is sensible. 

In Sec.~\ref{sec:generaldust} we show and discuss the subtleties of the dust dominated universe and derive analytical estimates for the amplitude of induced GWs in the case of an instantaneous transition. We apply our results to PBH dominated epoch in Sec.~\ref{sec:PBHdom} and use overproduction of induced GWs to place stringent constraints on the scenario. \\

\noindent\faBook\,\,\textit{Main references:} Sec.~\ref{sec:generaldust} is based on Inomata et al.~\cite{Inomata:2019ivs,Inomata:2019zqy} with the notation of Refs.~\cite{Domenech:2020ssp,Domenech:2021wkk}. These works are built upon previous papers by Assadullahi and Wands \cite{Assadullahi:2009nf} and Kohri and Terada \cite{Kohri:2018awv}. Then, Sec.~\ref{sec:PBHdom} follows the pioneering work by Papanikolau, Vennin and Langlois \cite{Papanikolaou:2020qtd} further extended by \cite{Domenech:2020ssp} to account for the transition to radiation domination. Lastly, Ref.~\cite{Inomata:2020lmk} contains useful details on the transition of the PBH dominated universe to radiation domination.

\subsection{General dust domination \label{sec:generaldust}}

Let us now derive the kernels and some estimates for the induced GWs. First we note that the fact that $\Phi={\rm constant}$ in an early Matter Dominated (eMD) era seemingly makes naive calculations of induced GWs easier. Just take a transfer function equal to unity in \eqref{eq:f}, which yields
\begin{align}
f(\tau,q,|\mathbf{k}-\mathbf{q}|)=\frac{5}{3}\,,
\end{align}
where we took $b=1$ ($w=0$). Now we have to simply integrate \eqref{eq:kernelApp} with a constant source. In the limit where $x\gg1$, we have that the averaged kernel squared is just a constant \cite{Assadullahi:2009nf,Kohri:2018awv}. In our notation we have
\begin{align}\label{eq:IeMD}
\overline{I^2_{eMD}(x\gg1,q,|\mathbf{k}-\mathbf{q}|)}\approx \left(\frac{20}{3}\right)^2\,.
\end{align}
We can then use \eqref{eq:Phgaussian} to give a naive estimation of the induced GW during the early matter era. However, this is far from being a complete picture. As first realised by Inomata, Kohri, Nakama and Terada, induced GWs from a dust dominated are more subtle and depend on the nature of the transition to radiation domination \cite{Inomata:2019ivs,Inomata:2019zqy}. If the transition is gradual, the induced GW spectrum is actually suppressed \cite{Inomata:2019zqy}. The suppression can be understood from the fact that the initial enhancement in the dust dominated universe found in Ref.~\cite{Assadullahi:2009nf} is due to the constant $\Phi$. If $\Phi$ decays, the enhancement is not as efficient. Instead, if the transition from dust to radiation is instantaneous, the induced GW spectrum get very much amplified \cite{Inomata:2019ivs}. We will shortly explain the source of the amplification. Let us first note that the reason why we did not find such behaviour in Sec.~\ref{sec:analytical} comes down to the particular behaviour of $\Phi$ in dust domination. In Sec.~\ref{sec:analytical}, due to the fact that we assumed $c_s^2\neq0$, the general solution to $\Phi$ in a $w={\rm constant}$ universe is a decaying and oscillating function. After the transition to radiation domination, $\Phi$ is also decaying and oscillating, albeit with different frequency. Thus, the matching between the two periods is well described within the WKB approximation. In this case, only those scales close to $k\sim k_{\rm rh}$ might receive at most a $O(1)$ correction. 

What is special about the instantaneous reheating in the dust dominated universe is that we are suddenly going from $\Phi'=0$ to $\Phi'\neq 0$. Since $\Phi$ did not have time to decay and the frequency of the oscillations is proportional to the ratio of the largest wavenumbers $k$ with the reheating scale $k_{\rm rh}$, the amplitude of $\Phi'$ right after the sudden reheating will be huge.  A more physical picture was presented in Ref.~\cite{Inomata:2019ivs} and it goes as follows. Density fluctuations grow as $\delta\rho/\rho\propto a$ and attain a ``large'' amplitude by the end of the dust dominated era. These large fluctuations suddenly evaporate and are converted into fluctuations in a radiation fluid, which create strong pressure waves. These waves generate spacetime oscillations and are the source to induced GWs.

Let us explicitly show that $\Phi'$ becomes very large right after reheating. To do that, we first consider that $\Phi$ has an arbitrary amplitude during the eMD of $\Phi=\Phi_{eMD}(k)$. Then, the eMD ends abruptly and we enter a late Radiation Domination (lRD). Matching the general solution \eqref{eq:phigeneralsolution} and its first derivative for $b=1$ and $b=0$ we arrive at
\begin{align}
\Phi_{lRD}=(c_sk\bar\tau)^{-3/2}\left(C_1 J_{3/2}(c_sk\bar\tau)+C_2 Y_{3/2}(c_sk\bar\tau)\right)\,,
\end{align}
where $c_s=1/\sqrt{3}$ since we are in a radiation dominated universe, $\bar\tau$ is defined in \eqref{eq:bartau} and comes from requiring continuity of $a$ and $\cal H$ and we defined
\begin{align}
C_1&=-\Phi_{\rm eMD}\frac{\pi}{2}\left(\frac{c_sk\bar\tau_{\rm rh}}{2}\right)^{5/2}Y_{5/2}\left(\frac{c_sk\bar\tau_{\rm rh}}{2}\right)\,,\\
C_2&=\Phi_{\rm eMD}\frac{\pi}{2}\left(\frac{c_sk\bar\tau_{\rm rh}}{2}\right)^{5/2}J_{5/2}\left(\frac{c_sk\bar\tau_{\rm rh}}{2}\right)\,.
\end{align}
If we now compute the variation of $\Phi_{\rm lRD}$ per Hubble time we find that the leading contribution for $x> x_{\rm rh}\gg 1$ is given by
\begin{align}\label{eq:phiprimedust}
\frac{d\Phi_{\rm lRD}}{d\ln \bar x}\approx -\Phi_{\rm eMD}\frac{c_s\bar x_{\rm rh}^2}{\bar x}\sin \left(c_s(\bar x-\bar x_{\rm rh})\right)\,.
\end{align}
Thus, we see that the amplitude of the time derivative is roughly
\begin{align}
\frac{d\Phi_{lRD}}{d\ln \bar x}\propto\left(\frac{k}{k_{\rm rh}}\right)^2\,.
\end{align}
This means that if there are fluctuations on scales $k\gg k_{\rm rh}$ the amplitude of $\Phi_{\rm lRD}'$ is very large and so it will produce induced GWs with a large amplitude.

Let us move on to the GWs induced after the sudden transition. For simplicity, we only focus on the largest contribution to the source \eqref{eq:f}, which comes from the time derivatives and  reads
\begin{align}\label{eq:fdust}
f(\tau, q,|\mathbf{k-q}|)\approx&\frac{1}{2}\frac{dT_\Phi(q\tau)}{d\ln\tau}\frac{dT_\Phi(|\mathbf{k}-\mathbf{q}|\tau)}{d\ln\tau}\,.
\end{align}
Inserting Eqs.~\eqref{eq:fdust} and \eqref{eq:phiprimedust} into \eqref{eq:kernelApp} we arrive at the kernel during the lRD, namely
\begin{align}\label{eq:Idust}
I_{lRD}(\bar x,u,v)\approx \Phi_{\rm eMD}^2\frac{c_s^2{\bar x_{\rm rh}^4}}{2\bar x}uv\int^{\bar x}_{\bar x_{\rm rh}}d\tilde{\bar x}\, \frac{\sin(\bar x-\tilde{\bar x})}{\tilde{\bar x}}\sin \left(c_sv(\tilde{\bar x}-\bar x_{\rm rh})\right)\sin \left(c_su(\tilde{\bar x}-\bar x_{\rm rh})\right)\,.
\end{align}
The integral in \eqref{eq:Idust} may be done analytically in terms of sine and cosine integrals but it is not so illuminating. The interested reader is referred to Ref.~\cite{Inomata:2019ivs}. Using our experience of Sec.~\ref{sec:analytical}, we know that the integral often diverges when $c_s(u+v)\sim 1$. Thus, within our order of magnitude estimates, we shall focus solely on the neighbourhood of the resonance. In this case, the divergent piece is the cosine integral ${\rm Ci}[x]$ and the kernel is approximately given by
\begin{align}
I_{lRD,\rm res}(\bar x,u,v)\approx \Phi_{\rm eMD}^2\frac{c_s^2{\bar x_{\rm rh}^4}}{8\bar x}uv\,{\rm Ci}\left[|1-c_s(u+v)|\bar x_{\rm rh}\right]\,\sin\bar x\,.
\end{align}
The calculation of the averaged kernel squared is straightforward and yields
\begin{align}\label{eq:I2dust}
\overline{I^2_{lRD,\rm res}(\bar x,u,v)}\approx \Phi_{\rm eMD}^4\frac{c_s^4{\bar x_{\rm rh}^8}}{128\bar x^2}u^2v^2\,{\rm Ci}^2\left[|1-c_s(u+v)|\bar x_{\rm rh}\right]\,.
\end{align}
For other approximations regarding the IR tail, such as the large momentum contribution $u\sim v\gg1$, we refer the reader to Refs.~\cite{Inomata:2019ivs,Inomata:2020lmk}.

With the analytic approximation for the kernel \eqref{eq:I2dust} we shall turn our attention to the dominant contribution to the tensor spectrum \eqref{eq:Phgaussian}. For clarity, let us explicitly insert \eqref{eq:I2dust} into \eqref{eq:Phgaussian} to obtain the resonant part of the tensor spectrum, which leads us 
\begin{align}\label{eq:Phgaussiandust}
\overline{{\cal P}^{\rm res}_h}\approx\frac{c_s^4{\bar x_{\rm rh}^8}}{16\bar x^2}\int_0^\infty dv\int_{|1-v|}^{1+v}du&\left(\frac{4v^2-(1-u^2+v^2)^2}{4uv}\right)^2{{\cal P}_{\Phi}(ku)}{{\cal P}_{\Phi}(kv)}\nonumber\\&\times u^2v^2\,{\rm Ci}^2\left[|1-c_s(u+v)|\bar x_{\rm rh}\right]\,.
\end{align}
Before carrying out the integral, let us analyse some of the assumptions. First, we are assuming that \eqref{eq:Phgaussiandust} yields the dominant contribution. This will be the case if there is integration support for large $k$. This means that ${\cal P}_{\Phi}(k)$ should peak at large $k$. We also argued that there might be a limit to the magnitude of $k$ due to non-linearities given by \eqref{eq:kNL}. We should use such limit if there are no stronger motivations to consider $k>k_{NL}$. However, as we shall see, if PBH dominate the universe, there must be fluctuations above $k_{NL}$ \cite{Papanikolaou:2020qtd}. For these reasons, let us parametrize the scalar power spectrum as power-law with a UV cut-off $k_{UV}$ given by \cite{Domenech:2021wkk}
\begin{align}\label{eq:pphidust}
{\cal P}_{\Phi}={\cal A}_\Phi\left(\frac{k}{k_{\rm UV}}\right)^{-n}\Theta(k_{UV}-k)\,,
\end{align}
where ${\cal A}_\Phi$ is the amplitude and $n$ the spectral tilt. Looking at \eqref{eq:Phgaussiandust} we see that the integrand peaks at large $k$ if $n<2$. This includes an almost scale invariant spectrum ($n\sim 0$) and the PBH density fluctuations ($n\sim 1$) \cite{Papanikolaou:2020qtd,Domenech:2020ssp}.

With these assumptions we shall proceed to perform the integral in \eqref{eq:Phgaussiandust}. We shall use the fact that ${\rm Ci}[z]$ is divergent for $z\sim0$ to evaluate the integrand at the resonant point except for ${\rm Ci}[z]$. For simplicity, let us switch integration variables to
\begin{align}
z\equiv(1-c_s(u+v))\bar x_{\rm rh}\quad{\rm and}\quad s\equiv u-v\,,
\end{align}
which have a Jacobian given by $|J|=1/(2c_s\bar x_{\rm rh})$. Replacing $(u,v)$ for $(z,s)$ in \eqref{eq:Phgaussiandust}, evaluating integrand at $z=0$ except for ${\rm Ci}[z]$ and maintaining the $s$ dependence we arrive at
\begin{align}\label{eq:Phgaussiandust2}
\overline{{\cal P}^{\rm res}_h}\approx\frac{\pi{\bar x_{\rm rh}^7}}{512c_s\bar x^2}\left(1-c_s^2\right)^2\int_{-s_0(k)}^{s_0(k)} ds {(1-s^2)^2}{{\cal P}_{\Phi}\left(k\frac{1+c_s\,s}{2c_s}\right)}{{\cal P}_{\Phi}\left(k\frac{1-c_s\,s}{2c_s}\right)}\,,
\end{align}
where we already integrated the cosine integral around the divergence\footnote{We used that $\int_{-\infty}^\infty dz\,{\rm Ci}^2[z]=\pi$.
} and we have defined \cite{Inomata:2019ivs}
\begin{align}\label{eq:s0}
s_0(k)=\left\{
\begin{aligned}
&1\qquad &{k_{UV}}/{k}\geq\tfrac{1+c_s^{-1}}{2}\\
&2\tfrac{k_{UV}}{k}-c_s^{-1}\qquad & \tfrac{1+c_s^{-1}}{2}\geq\tfrac{k_{UV}}{k}\geq\tfrac{c_s^{-1}}{2}\\
&0\qquad &\tfrac{c_s^{-1}}{2}\geq\tfrac{k_{UV}}{k}
\end{aligned}
\right.\,.
\end{align}
These bounds on $s$ come from momentum conservation, i.e., $|1-v|<u<1+v$, evaluated at the resonant point $z=0$. To have an order of magnitude estimate of the amplitude of the peak of the induced GW spectrum, we can evaluate \eqref{eq:Phgaussiandust2} at $s=0$. By doing so, the amplitude of the induced GW spectrum at $k\sim k_{\rm UV}$ due to the sudden reheating is roughly given by
\begin{align}\label{eq:Phgaussiandust3}
\Omega^{\rm peak}_{\rm GWs}\approx\frac{\pi}{3\times 2^{12}c_s}\left(1-c_s^2\right)^2\left({4c^2_s}\right)^{n}\left(\frac{k_{\rm UV}}{k_{\rm rh}}\right)^7{\cal A}^2_{\Phi}\,,
\end{align}
where we used that $\bar x_{\rm rh}=k\tau_{\rm rh}/2=k/k_{\rm rh}$ and we used that the numerical evaluation of the integral in $s$ roughly yields an extra factor $1/2$ \cite{Domenech:2020ssp}. Let us remind the reader that $c_s=1/\sqrt{3}$ since we are in radiation domination. From \eqref{eq:Phgaussiandust3} we see how if $k_{\rm UV}\gg k_{\rm rh}$ the amplitude of the induced GWs is tremendously large by a power of $7$. Thus, not to backreact, we either need ${\cal A}_{\Phi}$ to be extremely small or we restrict the value of $k_{\rm UV}$ not to be too large. Now, knowing the amplitude of the peak \eqref{eq:Phgaussiandust3} we can approximate the induced GW spectrum by
\begin{align}\label{eq:Phgaussiandust4}
\Omega_{\rm GWs,rh}(k\sim k_{\rm UV})\approx\Omega^{\rm peak}_{\rm GWs}\left(\frac{k}{k_{\rm UV}}\right)^{7-2n}\Theta(k_{\rm UV}-k)\,,
\end{align}
where we used that although the cut-off in the induced GWs is at $k=2k_{\rm UV}$ the decay after $k\sim k_{\rm UV}$ is so fast that we can safely neglect it.

The resonant peak is not the only contribution after the sudden reheating induced GWs. There is also the IR tail which corresponds to the $u\sim v\gg 1$ region of integration. Nevertheless, this contribution is always suppressed by a factor $k_{\rm rh}/k_{\rm UV}$. The interested reader can find the corresponding formulas in Refs.~\cite{Inomata:2019ivs,Inomata:2020lmk}. In addition to that, there is the contribution of the GWs induced during eMD, i.e., using the kernel \eqref{eq:IeMD} into \eqref{eq:Phgaussian}. In the case of the sudden transition, we can take the amplitude of the induced GWs as initial conditions for the subsequent radiation domination era. Then the total spectrum is well approximated by the sum of the two contributions \cite{Inomata:2019ivs}. This contribution will be small compared to the resonant part \eqref{eq:Phgaussiandust4} but since it may present a large plateau \cite{Assadullahi:2009nf,Inomata:2019ivs,Papanikolaou:2020qtd} it may dominate on the very low frequency band. The case of the gradual transition requires a more careful treatment which can be found in Ref.~\cite{Inomata:2019zqy}. We shall proceed with a very interesting application of \eqref{eq:Phgaussiandust4}.

\subsection{PBH dominated era \label{sec:PBHdom}}

An early matter dominated period could have been due to PBHs. Once formed, PBHs basically behave like a dust fluid, i.e., a fluid with no pressure and no propagation speed of fluctuations, like non-relativistic matter. Furthermore, their energy density practically redshifts as that of non-relativistic matter. You can see this from the following. Since the number density of PBH, $n_{\rm PBH}$ , is conserved (unless they substantially merge or evaporate) we have that $n_{\rm PBH}\propto a^{-3}$. The total energy density contained in these PBHs is $\rho_{\rm PBH}=M_{\rm PBH} n_{\rm PBH}$. Then in periods where PBH evaporation is negligible we have that $\rho_{\rm PBH}\propto a^{-3}$. For concreteness, let us assume that PBH formed in a radiation dominated era, with an initial fraction of PBH given by
\begin{align}
\beta\equiv\frac{\rho_{{\rm PBH},i}}{\rho_{r,i}}\approx \frac{\rho_{{\rm PBH},i}}{3H_i^2M^2_{\rm pl}}\,,
\end{align}
where in the last step we used the first Friedmann equation \eqref{eq:friedman}. If $\beta$ is large enough, then PBH will eventually dominate the universe since we have that $\rho_r\propto a^{-4}$. Now, for simplicity, let us assume that these PBHs formed by the collapse of large primordial fluctuations with a very peaked primordial spectrum at a scale $k_p$. This has two implications. First, the PBH mass function is almost monochromatic. Then, we can use Eq.~\eqref{eq:MPBH} to estimate the peak mass in terms of the peak scale $k_p$ by using that $H_i\equiv k_p/a_{i}$. We refer the reader to Sec.~\ref{sec:PBH} for more details. Second, the initial fraction of PBH $\beta$ is determined by the amplitude of the primordial spectrum. In this section, we take for convenience $M_{\rm PBH}$ and $\beta$ as the free parameters of the model. One can then relate the scale and the amplitude of the primordial spectrum to $M_{\rm PBH}$ and $\beta$, if necessary. Interestingly, in this situation the PBH evaporation occurs almost instantaneously \cite{Inomata:2020lmk} and, therefore, we can use the estimates we derived in Sec.~\ref{sec:dustestimate} with some corrections that we describe below. It should be noted that the results we will derive here only apply to a sharply peaked PBH mass function. If we considered an extended PBH mass function, induced GWs would be very much suppressed \cite{Inomata:2020lmk}.

A remarkable fact is that, as first realised by Papanikolaou, Vennin and Langlois \cite{Papanikolaou:2020qtd}, due to the inhomogeneous distribution, PBH themselves create density fluctuations which might later source induced GWs. To see this, note that PBH formation by the collapse of large fluctuations is a rare event \cite{Sasaki:2018dmp} and so each PBH form uncorrelated of the others. Also, clustering is often negligible. This means that, as a good approximation PBHs will be randomly distributed uniformly in space. In other words, PBHs are distributed according to Poisson statistics. Such Poisson statistics are dictated by the mean inter-PBH comoving separation. If PBH follow a monochromatic mass function, the mean inter-PBH comoving separation at formation is given by
\begin{align}\label{eq:kuv}
d_i=\left(\frac{3}{4\pi n_{{\rm PBH},i}}\right)^{1/3}\equiv k_{\rm UV}^{-1}\,.
\end{align}
Note that in Eq.~\eqref{eq:kuv} we identify the comoving inter-PBH separation as the UV cut-off in Eq.~\eqref{eq:pphidust}. This is because the fluid description of the PBH gas is valid only for $k<k_{\rm UV}$, i.e., at distances much larger than $d_i$. In this coarse grained regime the initial dimensionless power spectrum of PBH density fluctuations reads \cite{Papanikolaou:2020qtd}
\begin{align}\label{eq:Ppbh}
{\cal P}_{\delta,{\rm PBH}}=\frac{2}{3\pi}\left(\frac{k}{k_{\rm UV}}\right)^3\Theta(k_{\rm UV}-k)\,.
\end{align}

We have showed with Eq.~\eqref{eq:Ppbh} that PBHs give rise to density fluctuations. However, in the current model, PBHs form in an early Radiation Dominated stage (eRD) and, therefore, such density fluctuations correspond to isocurvature fluctuations (for the definition see footnote \ref{footnote:isocurvature}). Thus, they do not source induced GWs yet. Induced GWs are mainly generated by curvature fluctuations. If one is interested in the latter aspect, the general source term in a two fluid system can be found in Ref.~\cite{Domenech:2020ssp}. The isocurvature nature of the PBH density fluctuations can be understood from the following \cite{Domenech:2020ssp}. PBH formation in the fluid picture may be regarded as a transition of a fraction of the homogeneous radiation fluid into dust. Yet, the total energy density of radiation remains homogeneous, while PBH are distributed randomly. Thus, the inhomogeneity due to PBHs must be isocurvature as the total energy density is homogeneous. Interestingly, such isocurvature is converted into curvature fluctuations when PBH dominate the universe. This type of transition from radiation to dust was studied analytically in detail by Kodama and Sasaki \cite{Kodama:1986fg,Kodama:1986ud}. Thus, we can directly borrow their results which provide the transfer function for a vanishing initial curvature perturbation $\Phi$ and non-zero isocurvature deep inside the eMD era as
\begin{align}\label{eq:conversionfinal}
T^{{\rm iso}\to{\rm curv}}_{{\rm eRD}\to{\rm eMD}}(k;a\gg a_{\rm eq})=\left\{
\begin{aligned}
&\frac{1}{5}\qquad  & k\ll k_{\rm eq}\\
&\frac{3}{4}\left(\frac{k_{\rm eq}}{k}\right)^2\qquad & k\gg k_{\rm eq}
\end{aligned}
\right.\,.
\end{align}
The subscript ``eq'' refers to the early radiation-PBH equality. For an intuitive re-derivation, see Ref.~\cite{Papanikolaou:2020qtd}. Eq.~\eqref{eq:conversionfinal} tells us that the initial isocurvature has been transferred to a constant curvature perturbation. Modes which entered the horizon before the early radiation-PBH equality have decayed substantially and, hence, the suppression factor $({k_{\rm eq}}/{k})^2$. This means that the curvature power spectrum due to early PBH fluctuations in the eMD on the smallest scales, i.e., ($k_{\rm UV}>k\gg k_{\rm eq}$) is given by
\begin{align}\label{eq:PpbhPhi}
{\cal P}_{\Phi,{\rm PBH}}=\frac{3}{8\pi}\left(\frac{k_{\rm eq}}{k_{\rm UV}}\right)^4\left(\frac{k_{\rm UV}}{k}\right)\Theta(k_{\rm UV}-k)\Theta(k-k_{\rm eq})\,.
\end{align}
It is the curvature perturbation spectrum given by \eqref{eq:PpbhPhi} which sources induced GWs during the eMD. However, we are most interested in the induced GWs generated right at the start of the lRD. As we shall see, since PBH evaporation has a finite duration even for the monochromatic case, we must take into account an important suppression. Then we shall use our estimate Eq.~\eqref{eq:Phgaussiandust3}. Before that, we need to understand and quantify the PBH domination a bit further.

Let us derive the conditions to have a PBH dominated era in the early universe allowed by current observations. We start with our PBHs given an initial mass $M_{{\rm PBH},i}$ \eqref{eq:MPBH} and an initial fraction $\beta$. After formation, these PBHs will evaporate emitting Hawking radiation. This means that $\beta$ cannot be too small, otherwise PBH evaporate before they dominate. To quantify this requirement it is enough to look at the evaporation time which for a non-spinning black hole reads \cite{Hawking:1974sw,Hooper:2019gtx,Inomata:2020lmk}
\begin{align}\label{eq:teva}
t_{\rm eva}\approx\frac{160}{3.8\pi g_H(T_{\rm PBH})}\frac{M_{{\rm PBH},i}^3}{M_{\rm pl}^4}\,,
\end{align}
where $g_H(T_{\rm PBH})$ are the spin-weighted degrees of freedom and
\begin{align}\label{eq:TPBH}
T_{\rm PBH}\equiv M_{\rm pl}^2/M_{{\rm PBH},i}\approx1.06\times 10^9{\rm GeV}\left(\frac{M_{\rm PBH,f}}{10^4{\rm g}}\right)^{-1}\,.
\end{align}
If we assume only the standard model of particle physics we have that for PBHs which evaporated before BBN then $g_H(T_{\rm PBH})\approx 108$. If we include a possible PBH spin the evaporation time decreases reaching a factor $1/2$ for the extremal BH case \cite{Dong:2015yjs,Arbey:2019jmj,Kuhnel:2019zbc,Arbey:2021ysg,Masina:2021zpu}. Thus, we can take into account the PBH spin by replacing $t_{\rm eva}\to t_{\rm eva}/2$. This effect is considered in the induced GWs by Ref.~\cite{Domenech:2021wkk}. We will not consider PBH spin in what follows. From Eq.~\eqref{eq:teva} we see that the ``reheating'' temperature, i.e., the temperature of the radiation fluid which dominates the universe after PBH evaporate, is given solely in terms of $M_{{\rm PBH},i}$. Using the current cosmological parameters, which are given in App.~\ref{app:formulasuseful}, we find that the evaporation temperature can be written as
\begin{align}\label{eq:Teva}
T_{\rm eva}\approx 2.76\times 10^4{\rm GeV}\left(\frac{M_{{\rm PBH},i}}{10^4{\rm g}}\right)^{-3/2}
\left(\frac{g_H(T_{\rm PBH})}{108}\right)^{1/2}\left(\frac{g_*(T_{\rm eva})}{106.75}\right)^{-1/4}\,,
\end{align}
where $g_*$ are the effective degrees of freedom in the energy density of radiation \cite{Husdal:2016haj,Saikawa:2018rcs}. Here we find our first constraint on the model: PBH cannot evaporate too late if they are to reheat the universe. For a successful BBN the reheating temperature must be $T_{\rm eva}>4 \,{\rm MeV}$ \cite{Kawasaki:1999na,Kawasaki:2000en,Hannestad:2004px,Hasegawa:2019jsa}. This puts an upper bound on the PBH mass as
\begin{align}\label{eq:boundonmassBBN}
M_{{\rm PBH},i}<5\times 10^8{\rm g}\,.
\end{align} 
Using Eq.~\eqref{eq:Teva}, the comoving scale corresponding to the horizon size at the time of evaporation is given by
\begin{align}\label{eq:keva}
k_{\rm eva}\approx 4.7\times 10^{11}{\rm Mpc}^{-1}\left(\frac{g_H(T_{\rm PBH})}{108}\right)^{1/2}\left(\frac{g_*(T_{\rm eva})}{106.75}\right)^{1/4}\left(\frac{g_{*,s}(T_{\rm eva})}{106.75}\right)^{-1/3}\left(\frac{M_{{\rm PBH},i}}{10^4{\rm g}}\right)^{-3/2}\,,
\end{align}
where $g_{*,s}$ are the effective degrees of freedom in the entropy. With the above formulas we have that the end of the PBH domination era is fixed by the PBH mass. The second constraint on the model comes from requiring that PBHs dominate before they evaporate. This is a constrain on the initial fraction of PBHs in terms of the PBH mass, which is found to be \cite{Inomata:2020lmk,Domenech:2020ssp} 
\begin{align}\label{eq:betamin}
\beta>6.35\times10^{-10}\left(\frac{g_H(T_{\rm PBH})}{108}\right)^{1/2}
\left(\frac{\gamma}{0.2}\right)^{-1/2}\left(\frac{M_{{\rm PBH},f}}{10^4{\rm g}}\right)^{-1}\,.
\end{align}

Before we compute the final amplitude of the induced GW spectrum, it is very important to estimate the effects of the finite duration of the evaporation to the perturbations. Although the transition to a radiation dominated universe takes place within less than $1/4$ of e-fold, perturbations in very small scales are very much affected by the finite width of the transition, as was discovered by Inomata et al. \cite{Inomata:2019zqy,Inomata:2019ivs,Inomata:2020lmk}. The main reason is clear if we neglect the expansion of the universe, which is justified as we are interested in very subhorizon scales. As PBHs evaporate their energy density decays as \cite{Inomata:2020lmk}
\begin{align}
\rho_{\rm PBH}\propto  M_{\rm PBH}\propto  \left(1-\frac{t}{t_{\rm eva}}\right)^{1/3}\,.
\end{align}
Now, we are interested in any effect that this decay may have on $\Phi$. The Newtonian potential is related to fluctuations by the Poisson equation which in Fourier space reads
\begin{align}\label{eq:Poisson}
2\frac{k^2}{a^2}\Phi\approx \rho_{\rm PBH}{\delta_{\rm PBH}}+\rho_r \delta_r\,,
\end{align}
where $\delta_Q\equiv \delta\rho_Q/\rho_Q$ is the density contrast. We will do a big step here that can be easily confirmed by looking at the formulas in the appendix of Ref.~\cite{Domenech:2020ssp} or in Ref.~\cite{Inomata:2020lmk}. If we completely neglect the expansion of the universe we find that $\delta_{\rm PBH}\approx {\rm constant}$ is a solution to the equations of motion of PBH density fluctuations. With this solution, we can estimate the size of the fluctuations in the radiation fluid sourced by the evaporation. With the Green's method to solve the equations of motion of radiation fluctuations, we find that 
\begin{align}\label{eq:deltar}
\delta_r\sim\frac{\rho_{\rm PBH}}{\rho_r}  \frac{a^2\Gamma^2}{k^2}\delta_{\rm PBH}\,,
\end{align}
where $\Gamma$ is the evaporation rate of PBHs and it is given by
\begin{align}
\Gamma\equiv -\frac{d\ln M_{\rm PBH}}{dt}\sim \frac{1}{t-t_{\rm eva}}\,.
\end{align}
What \eqref{eq:deltar} tell us is that fluctuations in the radiation fluid $\delta_r$  with $k/a\gg \Gamma$ are very suppressed with respect to $\delta_{\rm PBH}$ due to radiation pressure. This means that for such modes the dominant contribution to $\Phi$ is given by $\delta_{\rm PBH}$, even when PBH do not dominate the universe. Then by Eq.~\eqref{eq:Poisson} we see that $\Phi\propto \rho_{\rm PBH}$ until $\Gamma\sim k/a$, at which point the fluctuations in radiation dominate and perturbations behave as in a radiation dominated universe. The temporal decay of $\Phi$ due to the decay of $\rho_{\rm PBH}$ yields a relative suppression given by \cite{Inomata:2020lmk}
\begin{align}
S_\Phi(k/k_{\rm eva})\equiv \frac{\Phi}{\Phi_{\rm instant}}\approx \left(\sqrt{\frac{2}{3}}\frac{k}{k_{\rm eva}}\right)^{-1/3}\,,
\end{align}
where $\Phi_{\rm instant}$ refers to the value of $\Phi$ if the transition is instantaneous. 

Now, we can use our estimate for an instantaneous transition \eqref{eq:Phgaussiandust3} but taking into account the suppression of $\Phi$ until the transition to radiation domination is completely achieved. This yields a spectrum of curvature fluctuations at the start of the lRD given by
\begin{align}\label{eq:PpbhlRD}
{\cal P}_{\Phi,{\rm lRD}}&=S^2_\Phi(k/k_{\rm eva}){\cal P}_{\Phi,{\rm PBH}}\nonumber\\&=\frac{3}{8\pi}\left({\frac{2}{3}}\right)^{-1/3}\left(\frac{k_{\rm UV}}{k_{\rm eva}}\right)^{-2/3}\left(\frac{k_{\rm eq}}{k_{\rm UV}}\right)^4\left(\frac{k}{k_{\rm UV}}\right)^{-5/3}\,\Theta(k_{\rm UV}-k)\Theta(k-k_{\rm eq})\,.
\end{align}
Comparing \eqref{eq:PpbhlRD} with \eqref{eq:pphidust} we see that $n=5/3$ and
\begin{align}
{\cal A}_\Phi=\frac{3}{8\pi}\left({\frac{2}{3}}\right)^{-1/3}\left(\frac{k_{\rm UV}}{k_{\rm eva}}\right)^{-2/3}\left(\frac{k_{\rm eq}}{k_{\rm UV}}\right)^4\,.
\end{align}

\begin{figure}
\centering
\includegraphics[width=0.55\columnwidth]{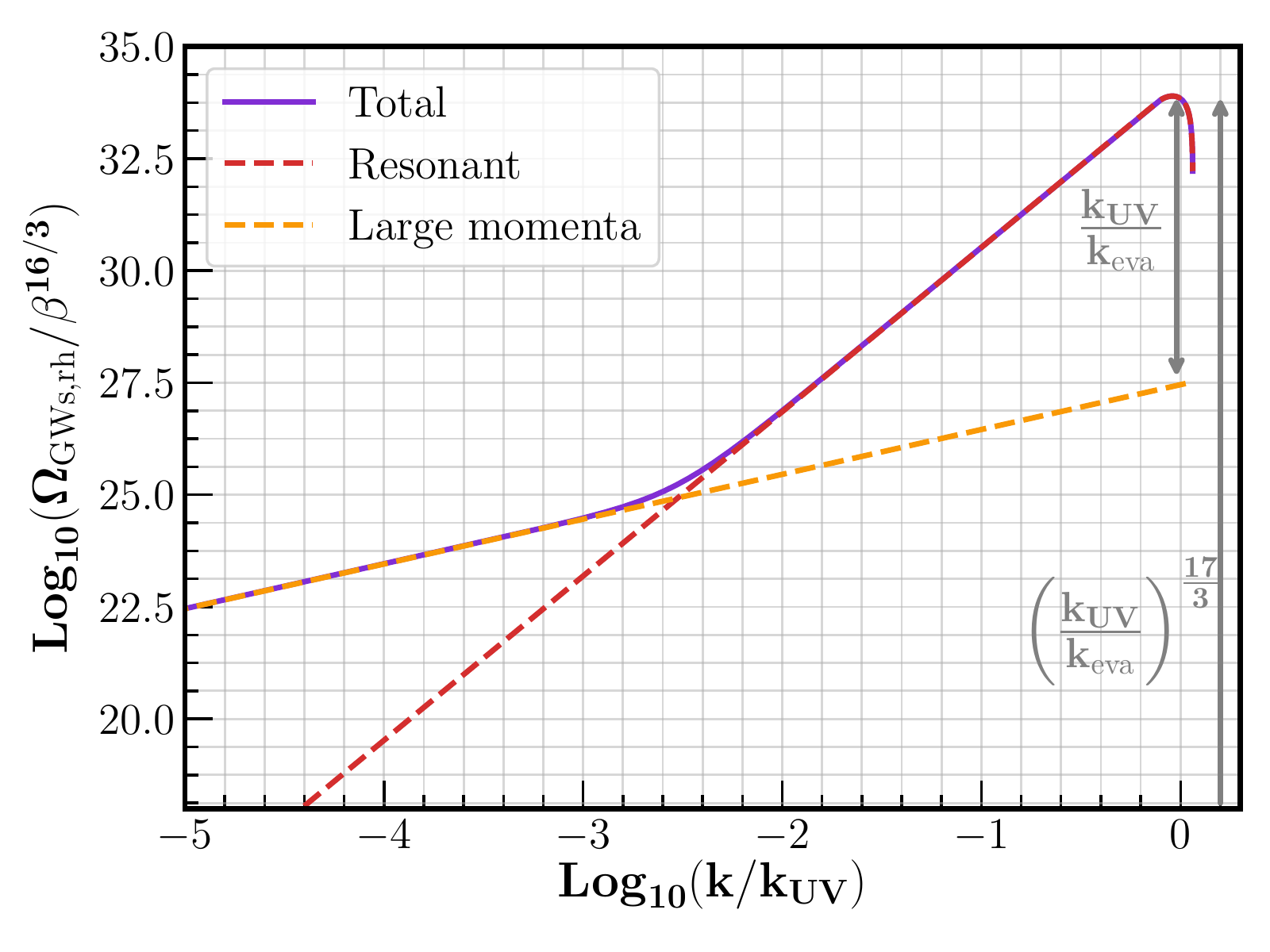}
\caption{Induced GW spectral density after PBH evaporation normalized by $\beta^{16/3}$ and in terms of comoving wavenumber $k/k_{\rm UV}$. The dashed red line shows the resonant contribution from Eq.~\eqref{eq:Phgaussiandust2}. The dashed orange line is the large momentum contribution which can be found in \cite{Domenech:2020ssp}. The purple line is the total GW spectrum. Due to the almost sudden transition the GW spectrum is amplified. This implies a very small initial fraction $\beta$ of PBH not to overproduce induced GWs. \label{fig:plotPBH}}
\end{figure}

The final ingredient to derive the induced GW spectrum is the hierarchy of the scales involved. First, note that the ratio between the cut-off $k_{\rm UV}$ and $k_{\rm eva}$ is independent of $\beta$ and it is given by \cite{Domenech:2020ssp}
\begin{align}\label{eq:relatiosk2}
\frac{k_{\rm UV}}{k_{\rm eva}}\approx {2.3}\times 10^6
 \left(\frac{g_H(T_{\rm PBH})}{108}\right)^{-1/3}\left(\frac{M_{{\rm PBH},i}}{10^4{\rm g}}\right)^{2/3}\,.
\end{align}
The $\beta$ independence of the ratio is due to the fact that $H_{\rm eva}$ only depends on the PBH mass and $a_{\rm eva}/a_i\propto n_{{\rm PBH},i}^{1/3}$ because it is a dust dominated universe. The hierarchy is closed with the second and third relations respectively given by \cite{Domenech:2020ssp}
\begin{align}\label{eq:relatiosk}
\frac{k_{\rm eq}}{k_{\rm UV}}={\sqrt{2}\gamma^{1/3}\beta^{2/3}}\quad{\rm and}\quad\frac{k_{\rm eq}}{k_p}={\sqrt{2}\beta}\,.
\end{align}
Now, inserting Eqs.~\eqref{eq:relatiosk2}, \eqref{eq:relatiosk} and \eqref{eq:PpbhlRD} into \eqref{eq:Phgaussiandust4} yields an amplitude of induced GWs at the peak of approximately
\begin{align}\label{eq:PhgaussiandustPBH}
\Omega^{\rm peak}_{\rm GWs,eva}\approx 1.5\times10^{-2}\left(\frac{\beta}{10^{-6}}\right)^{16/3}\left(\frac{\gamma}{0.2}\right)^{8/3}\left(\frac{g_H(T_{\rm PBH})}{108}\right)^{-17/9}
\left(\frac{M_{{\rm PBH},f}}{10^4{\rm g}}\right)^{34/9}\,.
\end{align}
Note that the estimate Eq.~\eqref{eq:PhgaussiandustPBH} is the amplitude of induced GWs right after evaporation. It should also be noted that \eqref{eq:PhgaussiandustPBH} is valid for very sharp PBH mass functions. As the mass function widens, the induced GW counterpart gets suppressed \cite{Inomata:2020lmk}. On top of that, we have that scales $k>k_{NL}$ \eqref{eq:kNL}, specially for $k\sim k_{UV}$, PBH density fluctuations $\delta_{\rm PBH}$ become $O(1)$ or larger. However, our estimate \eqref{eq:PhgaussiandustPBH} has been derived in the linear regime and might receive corrections due to non-linear effects. Nevertheless, we expect it to be a crude order of magnitude estimate since the main source of induced GWs, the curvature perturbation, remains always smaller than unity,\footnote{For example, this is also the case of a matter inhomogeneity in the universe, such as a galaxy, where the matter density is clearly larger than the mean density of the universe but the gravitational potential can be considered as a perturbation. In this perturbative expansion one recovers Newtonian gravity which is very accurate in galactic scales.} i.e., $\Phi\ll1$. Numerical relativity simulations are needed to derive a preciser estimate of the induced GWs generated in the non-linear regime. We show the resulting induced GW spectrum in Fig.~\ref{fig:plotPBH}. Using \eqref{eq:keva} and \eqref{eq:relatiosk2} we find that the peak of induced GWs today lies at a frequency given by
\begin{align}
f_{\rm UV}\approx 1.7\times 
10^{3}{\rm Hz}\left(\frac{g_H(T_{\rm PBH})}{108}\right)^{1/6}\left(\frac{g_*(T_{\rm eva})}{106.75}\right)^{1/4}
\left(\frac{g_{*,s}(T_{\rm eva})}{106.75}\right)^{-1/3}\left(\frac{M_{{\rm PBH},i}}{10^4{\rm g}}\right)^{-5/6}\,.
\end{align}
Quite surprisingly, the frequency of the peak enters the observational range of LISA, DECIGO and LIGO for $5\times 10^8\,{\rm g}>M_{\rm PBH,i}>2\times 10^4\,{\rm g}$. Thus, this scenario is testable in the future.

To compute the amount of induced GWs measured today we must consider the redshift of GWs until today. We did this in Sec.~\ref{sec:typical} with Eq.~\eqref{eq:spectraldensitytoday2}. However, if we want to evaluate the amplitude of induced GWs at BBN, we only need to know the fraction of GWs that compose the total radiation after reheating. This can be done by considering the degrees of freedom of the standard model, which leads us to
\begin{align}\label{eq:spectraldensityBBN}
\Omega_{\rm GW,BBN}&=0.39\left(\frac{g_*(T_{\rm eva})}{106.75}\right)\left(\frac{g_{*s}(T_{\rm eva})}{106.75}\right)^{-4/3}\Omega_{\rm GW,\rm eva}\,.
\end{align}
Since these induced GWs act as additional radiation, they contribute to the effective number of relativistic species $N_{\rm eff}$ at BBN. The contribution of GWs to $N_{\rm eff}$ is by convention\footnote{It is parametrised as the effective number of neutrinos below the electron-positron annihilation temperature. In this way, the factor ${7}/{8}$ is the relative factor of the energy density of fermions with respect to bosons. The factor ${4}/{11}$ is the relative factor of the entropy of neutrinos and photons. Since $T\propto s^{1/3}$ and $\rho\propto T^4$ we get a power of $4/3$. This is well explained in Baumann's lecture notes \url{http://cosmology.amsterdam/education/cosmology/}.} written as \cite{Caprini:2018mtu}
\begin{align}\label{eq:BBNconstraint}
\Omega_{\rm GW,BBN}=\frac{7}{8}\left(\frac{4}{11}\right)^{4/3}\Delta N_{\rm eff}< 0.05\,,
\end{align}
where in the last step we used that BBN \cite{Cyburt:2004yc,Fields:2019pfx} sets an upper bound $\Delta N_{\rm eff}<0.2$.\footnote{Note that one obtains similar bounds from studies of the CMB \cite{Sendra:2012wh,Pagano:2015hma}. However, these CMB constraints consider gravitational waves as a dark radiation component and depend on the initial conditions of such dark radiation fluctuations. As an order of magnitude estimate we limit ourselves to the BBN constraints \cite{Cyburt:2004yc,Fields:2019pfx} on the fraction of extra relativistic particles. We thank an anonymous referee for clarifying this point.} In this way, we can constrain the amplitude of GWs from existing BBN bounds. Note that to be precise the BBN bound is a constrain on the total energy density of relativistic particles. Thus, we should have actually integrated our induced GW spectrum over $\ln k$ and then compare it with the bound \eqref{eq:BBNconstraint}. However, since the GW spectrum is very peaked and we are already working with order of magnitude estimates, it is justified to just look at the peak of the GW spectrum. If we use the current constraints from BBN, then we obtain that
\begin{align}
\beta< 1.5\times10^{-6}\left(\frac{M_{{\rm PBH},i}}{10^4{\rm g}}\right)^{-17/24}\,.
\end{align}
This is a new constraint on early epochs of PBH dominations which cannot be obtained by any other means.

Interestingly, PBH evaporation also emits ``gravitons'', i.e., very high frequency GWs. While these gravitons are not observable by GW detectors, they might be seen as effective relativistic particles at BBN \cite{Dong:2015yjs,Arbey:2019jmj,Kuhnel:2019zbc,Arbey:2021ysg,Masina:2021zpu}. Future CMB probes such as CMB-S4 \cite{Abazajian:2016yjj} will be able to improve current bounds down to $\Delta N_{\rm eff}\sim 0.02$. Unfortunately, the graviton contribution of non-spinning PBHs to $N_{\rm eff}$ is out of reach for future CMB probes. Nevertheless, if PBH have a large spin (and dominate the universe at some point), the graviton contribution to $N_{\rm eff}$ will be accessible to future experiments \cite{Arbey:2021ysg,Masina:2021zpu}. This signature from graviton emission together with induced GWs will be an interesting probe of PBH dominated eras. While the graviton contribution to $N_{\rm eff}$ tells us about PBH spin, the induced GW counterpart has information on the formation mechanism \cite{Domenech:2021wkk}. As a clarification, note that the graviton emission and induced GWs are two very different effects. While the former is emitted by Hawking evaporation, the latter is generated by the time dependence of density fluctuations. The composition of such density fluctuations is not important, e.g., if a fraction are gravitons, as long as it behaves as fluctuations of a radiation fluid.

Before ending this section, we should note that we also have the induced GWs generated in the PBH dominated stage by using the Kernel \eqref{eq:IeMD}. This is studied in detail in Ref.~\cite{Papanikolaou:2020qtd}. However, as we discussed at the beginning of this section this contribution is sensitive to the detailed transition to radiation domination. In the present case, where the transition is almost instantaneous, we can continue the tensor mode amplitude build up during eMD to the lRD. Nevertheless, this contribution is always subdominant compared to the one after reheating \cite{Inomata:2019ivs}. Thus, our estimate \eqref{eq:PhgaussiandustPBH} provides the dominant peak of the induced GW spectrum.


\section{The gauge issue \label{sec:gauge}}

Despite the large amount of research on induced GWs, there is still the open question of what we really observe. This was made explicit in 2017, when Hwang, Jeong and Noh \cite{Hwang:2017oxa} showed that the induced GW spectrum generated in a dust dominated universe depends very much (by orders of magnitude) on the gauge chosen. Although the gauge issue was mentioned earlier by Matarrese, Mollerach and Bruni \cite{Matarrese:1997ay}, it was not until \cite{Hwang:2017oxa} that it received more attention. From then and on, there has been a big activity to understand the problem \cite{Domenech:2017ems,Gong:2019mui,Tomikawa:2019tvi,DeLuca:2019ufz,Inomata:2019yww,Yuan:2019fwv,Chang:2020iji,Chang:2020tji,Lu:2020diy,Ali:2020sfw,Chang:2020mky,Domenech:2020xin,Gurian:2021rfv}. For example in Refs.~\cite{Gong:2019mui,Tomikawa:2019tvi} other cosmological backgrounds rather than dust were studied. They found that the gauge dependence is in general present. Later studies \cite{DeLuca:2019ufz,Inomata:2019yww,Yuan:2019fwv,Chang:2020iji,Chang:2020tji,Lu:2020diy,Ali:2020sfw,Chang:2020mky,Domenech:2020xin,Gurian:2021rfv}, proposed partial solutions and investigated in which situations the spectrum of induced GW might be well defined. We dedicate sec.~\ref{sec:origingauge} to explain the origin of the gauge issue. Then, we discuss these approaches in detail in Sec.~\ref{sec:gaugesol}. \\

\noindent\faBook\,\,\textit{Main references:} The discussion in sec.~\ref{sec:origingauge} is mainly based on \cite{Domenech:2017ems,Domenech:2020xin}. Then, in Sec.~\ref{sec:gaugesol} we revisit the main arguments of \cite{DeLuca:2019ufz,Inomata:2019yww} and explain the improved results of \cite{Domenech:2020xin}.

\subsection{The origin of the issue \label{sec:origingauge}}

To understand the source of problem, we must distinguish between what we call tensor modes and GWs. On the theoretical side, we define tensor modes in a homogeneous and isotropic universe as the transverse-traceless fluctuations of the spatial metric.\footnote{Without the symmetries of an FLRW background the definition of tensor modes is more subtle. For example, in anisotropic inflation they are constructed from the 2D scalar and vector perturbations \cite{Watanabe:2010fh,Soda:2012zm}.} However, this definition must depend on how one chooses the spatial hypersurface. Although tensor modes are gauge invariant at linear order, they fail to be so at second order in cosmological perturbation theory. This is reflected by the fact that tensor, vector and scalar modes mix at second order \cite{Matarrese:1997ay}. This may pose a problem when we want to relate the tensor modes in a given slicing to the observational effect, which are the GWs. Note that this is not necessarily an ``issue''. If the quantity related to observation is clear and calculations are done properly (perhaps in a reasonable coordinate system), one should get proper results. For example, in studies of the CMB the temperature anisotropies and the polarization patterns of the CMB are well defined at second order. The interested reader is referred to Refs.~\cite{Naruko:2013aaa,Saito:2014bxa,Namikawa:2021obu} and references therein. 

The trouble with GW observations is that we only know the GW detector response to linear GWs. Only at first order the relation between the tensor modes and GWs in the linearised theory is clear, since tensor modes are gauge invariant at that order. There is a subtlety, though. A tensor mode whose wavelength is larger than the horizon is frozen, it does not evolve. So it acts as a constant anisotropy or shear and, as such, it has nothing to do with actual GWs. Furthermore, since the wavelength of the tensor modes is larger than the curvature scale, the averaging procedure used to derive the (pseudo) energy momentum tensor of GWs in Sec.~\ref{sec:induced1} does not apply. For these reasons, we will be only concerned about tensor modes on subhorizon scales. These tensor modes have a wavelength smaller than the horizon and behave as a wave, which are the GWs we could observe.

In the absence of the GW detector response at second order, we face the problem that we do not know the relation between tensor modes and the true GWs. In most situations though this should not really be a problem, unless one is choosing a very strange coordinate system to describe the observations. For instance, to describe the GWs measured by a GW detector today we usually do not need to consider second order terms as the amplitude of the GWs is already tiny enough. However, this issue is particularly relevant for induced GWs due to their secondary nature. Now, there are two important aspects of the problem. First, we have that the derived tensor spectrum depends on the gauge choice. Second, we expect that once the source term in \eqref{eq:eominducedGW2} is absent, i.e., its amplitude decayed enough, such previously induced tensor modes behave as a freely propagating GWs. From that moment on we should be able to relate the induced tensor modes to GWs. Thus, we must reconcile our physical intuition with the fact that tensor modes are gauge dependent. To arrive to that, there are two points that we would like to make as clear as possible. The first one concerns the tensor modes $h_{ij}$ themselves and the second one to the actual observable, which in Sec.~\ref{sec:induced1} we decided it would be the energy density of induced GWs \eqref{eq:spectumdensity}. We present them in the following subsections and we later discuss the current state of the issue in Sec.~\ref{sec:gaugesol}.

\subsubsection{The definition of tensor modes is gauge dependent} When working at second order in cosmological perturbation theory, it is very important to be clear about our convention. Here we will use the conformal decomposition from Eqs.~\eqref{eq:confdecom} and \eqref{eq:exph} which in a general gauge reads
\begin{align}\label{eq:confdecom2}
ds^2&=a^2(\tau)\left(-N^2d\tau^2+{\rm e}^{2\phi}\Upsilon_{ij}\left(dx^i+N^id\tau\right)\left(dx^j+N^jd\tau\right)\right)\,,
\end{align}
where $N$ and $N^i$ respectively are the lapse and shift vector. We then perturb the variables as follows
\begin{align}\label{eq:confdecom3}
N&=1+\alpha\quad,\quad N_i=\partial_i\beta\\
\left[\ln\Upsilon\right]_{ij}&=h_{ij}+2\left(\partial_{i}\partial_{j}-\tfrac{1}{3}\delta_{ij}\Delta\right)E\label{eq:confdecomh}\,,
\end{align}
where we only focused on the tensor and scalar components and $\phi$ is already a perturbation. Our convention differs from the common expansions, which expands the metric linearly in the perturbation variables instead of exponentially. Nevertheless, our choice simplifies considerably calculations. To relate the two different conventions we refer the reader to the appendix of Ref.~\cite{Domenech:2020xin}. We also use the exponential mapping for a gauge transformation (e.g., see the review by Malik and Wands \cite{Malik:2008im}) which naturally follows from the Hamiltonian flow (see e.g.,~\cite{Domenech:2017ems} for the application to second order cosmological perturbations). To be more concrete, let us do an infinitesimal coordinate transformation by
\begin{align}
\tilde x^\mu=x^\mu-\xi^{\mu} \quad{\rm with}\quad \xi^\mu=(T,\partial^i L)\,,
\end{align}
where we focused only on the scalar component of the spatial vector $\xi^i$.
Then a quantity $Q$ in one gauge is related to the same quantity evaluated in another gauge $\tilde Q$ by the exponential mapping
\begin{align}\label{eq:gaugetransf}
\tilde Q=e^{\raisebox{0.1ex}{{-}}\hspace{-1.5mm}{\cal L}_\xi}Q\,,
\end{align}
where $\raisebox{0.2ex}{{--}}\hspace{-2mm}{\cal L}_\xi$ is the Lie derivative along $\xi$. With this convention, we find that the tensor modes at second order change according to
\begin{align}\label{eq:2ordergauge}
\tilde h_{ij}=h_{ij}-\widehat{TT}_{ij}\,^{ab}\Big\{&-2\partial_a\left(\beta-E'\right)\partial_b T+2\partial_a\partial_kL\partial_k\partial_{b}E+\partial_aT\partial_bT+\partial_a\partial_kL\partial_b\partial_kL\Big\}\,.
\end{align}
Eq.~\eqref{eq:2ordergauge} is the first cause of the gauge issue for induced GWs. Since the source to induced tensor modes acts at second order in perturbation theory, the actual gauge in which we do the calculations might change drastically the form of the solutions for $h_{ij}$ \cite{Hwang:2017oxa}. The reader interested in the gauge transformations particularly related to induced GWs might go to Ref.~\cite{Domenech:2020xin}. General reviews of cosmological perturbation theory are listed at the \hyperref[sec:structure]{beginning of this review}.

\subsubsection{The definition of GW energy density is gauge dependent} Once we showed that the tensor modes $h_{ij}$ are gauge dependent, let us look at the definition of the ``observable'' energy density of GWs. Using \eqref{eq:pseudotensor} we have that
\begin{equation}\label{eq:pseudotensor2}
\rho_{\rm GW}=t_{00}^{\rm GW}=\frac{M_{pl}^2}{8a^2}\bigg\langle h'^{ij} h'_{ij}+\partial_k h^{ij}\partial^k h_{ij}\bigg\rangle\,.
\end{equation}
Now, looking at \eqref{eq:2ordergauge} and \eqref{eq:pseudotensor2} the first question that comes to mind is: $h_{ij}$ in \eqref{eq:pseudotensor2} is evaluated in which gauge? Indeed, if we naively apply the gauge transformation \eqref{eq:2ordergauge} to \eqref{eq:pseudotensor2} we see that
\begin{equation}\label{eq:pseudotensor3}
\tilde\rho_{\rm GW}\sim\langle \tilde h'^{ij} \tilde h'_{ij}\rangle=\rho_{\rm GW}+\langle\partial_iT\partial_jT\partial^iT\partial^jT\rangle + ...\,,
\end{equation}
where we neglected other terms to illustrate the main point, which is that the term quartic in $T$ in \eqref{eq:pseudotensor3} does not vanish in general. This shows something that we were expecting in Sec.~\ref{sec:estimates}: the energy density of GWs \eqref{eq:pseudotensor2} in analogy of the Ricci flat result \eqref{eq:pseudotensor} is strictly speaking not well-defined on a cosmological background at second order in perturbation theory. Furthermore, it does not matter when you evaluate the energy density, e.g., today. In principle, we can find a coordinate transformation in which $\rho_{\rm GW}$ is very different from the usual estimate. For example, if we go to a frame where the detector is wildly oscillating, Eq.~\eqref{eq:pseudotensor2} might not yield sensible results. This is the second and the most important cause for the gauge dependence of induced GWs.

\subsection{Current ``solutions''\label{sec:gaugesol}}

In view of the current lack of a description at second order in cosmological perturbation theory of the detector response to passing GWs, all we can do is to try to argue when Eq.~\eqref{eq:pseudotensor2} yields sensible results. Before we start discussing the details, it is important to realize that a gauge invariant formulation of the equations of motion of induced GWs \eqref{eq:eominducedGW2} does not help at all. There are infinitely many ways to define gauge invariant variables, each definition corresponding to a particular gauge choice. Thus, the same question remains, although with a slightly different language: which gauge invariant definition of the tensor modes goes into \eqref{eq:pseudotensor2}? Similarly, if one tries to find a gauge invariant formulation of \eqref{eq:pseudotensor2}, which implies going to fourth order in cosmological perturbation theory (and it is clearly not the point), it is not clear what is the relation with the real observable. So, what can we do?

An interesting direction, first explored by \cite{DeLuca:2019ufz} and later corrected by \cite{Inomata:2019yww}, is to argue what is the gauge (or coordinate frame) which best describes the GW detection. According to Ref.~\cite{DeLuca:2019ufz}, based on the analogy with asymptotically flat spacetimes, the best gauge is the so-called transverse-traceless (TT) gauge where the metric reads
\begin{align}\label{eq:TTgauge}
ds^2=-dt^2+(\delta_{ij}+h_{ij})dx^i dx^j\,.
\end{align}
In Minkowski spacetime, there are only two degrees of freedom, the ones in $h_{ij}$, since there is no source to the energy momentum tensor. In this gauge and in the linearised theory, the GW detector is at rest, even for passing GWs, which makes the physical interpretation (and calculations) much easier \cite{Misner:1974qy,Maggiore:1900zz}. The closest gauge to the TT gauge in a cosmological background is the synchronous gauge which is a coordinate system spanned by a family of geodesics. In our convention the metric in the synchronous gauge is given by
\begin{align}\label{eq:confdecomsyncrhonous}
ds^2&=a^2(\tau)\left(-d\tau^2+{\rm e}^{2\phi}\Upsilon_{ij}dx^idx^j\right)\,,
\end{align}
where we set $\alpha=\beta=0$ in \eqref{eq:confdecom2}. Let us remind the reader that
\begin{align}
\left[\ln\Upsilon\right]_{ij}&=h_{ij}+2\left(\partial_{i}\partial_{j}-\tfrac{1}{3}\delta_{ij}\Delta\right)E\label{eq:confdecomh2}\,,
\end{align}
and therefore we have two more variables, $\phi$ and $E$, than in \eqref{eq:TTgauge}. In the end, only one of them is dynamical. Thus, although at linear order \eqref{eq:confdecomh2} resembles the TT gauge \eqref{eq:TTgauge} it is certainly not the same. At linear order and if we neglect the drift due to the expansion $a$, the metric \eqref{eq:confdecomh2} does describe a fixed detector. However, at second order the detector will not be fixed and will feel the influence of $E$. Another ``problem'' of the synchronous gauge is that it is not a completely fixed gauge. There is a residual gauge ambiguity which corresponds to a change of the reference geodesics. This gauge ambiguity turns out to be very important in the final spectrum of induced tensor modes. If one does not properly fix the synchronous gauge, then induced tensor modes could be very different than, e.g., in the Newtonian gauge \cite{Lu:2020diy}. This is because if we start with a family of geodesics defined in the very early universe and we use the same geodesics to describe the GW detection, it is very possible that these geodesics do not match the rest frame of the detector. The resulting induced tensor spectrum might be singular if those geodesics happen to have focusing singularities near the detector \cite{Domenech:2020xin}. Nevertheless, by properly fixing the gauge one finds that the induced GW spectrum in the synchronous gauge and in the Newtonian gauge yield the same prediction in a radiation dominated universe \cite{DeLuca:2019ufz,Inomata:2019yww,Yuan:2019fwv,Lu:2020diy}. This also holds for any cosmological background with a constant equation of state \cite{Domenech:2020xin} if it is not dust, i.e., $c^2_s\neq0$. This result provides evidence that perhaps the energy density of GWs \eqref{eq:pseudotensor2} computed in the Newtonian gauge, which we do for simplicity of the calculations, yields meaningful results. However, it does not explain why these two predictions coincide. Furthermore, the original issue raised in \cite{Hwang:2017oxa} in a dust dominated universe persist.

Here is where the direction explored in Ref.~\cite{Domenech:2020xin} becomes relevant: when and why the energy density of GWs \eqref{eq:pseudotensor2} yields the same predictions and for which gauges? The physical intuition is that the analogy with Ricci flat spacetimes should work well if: 
\begin{enumerate}[label=\textit{(\roman*)}]
\item we look at really subhorizon scales ($k\gg {\cal H}$), where cosmology should be less relevant,
\item there is (barely) no source of induced GWs \eqref{eq:eominducedGW2}, so that induced tensor modes are freely propagating GWs,
\item we use a coordinate system suitable for small distances calculations, so that we are not confused by strange coordinate artefacts.
\end{enumerate}
As shown in \cite{Domenech:2020xin}, if points $(i)-(iii)$ are satisfied then \eqref{eq:pseudotensor2} is gauge independent up to corrections of $O({\cal H}^2/k^2)$. Note that conditions $(ii)$ and $(iii)$ yield some restrictions respectively on the type of cosmological background and on the class of gauges. In order to show such approximate gauge independence, Ref.~\cite{Domenech:2020xin} argued that a good gauge choice to start with is the Newtonian gauge. The physical reason for that choice is that on small distances the Newtonian gauge reduces to the well-known Newtonian gravity. The practical reason is that calculations are simpler. Nevertheless, the argument provided in \cite{Domenech:2020xin} does not strongly depend on the initial choice provided that it is suitable for subhorizon calculations. For example, one could have started with the synchronous gauge or the uniform Hubble gauge. 

Starting from the Newtonian gauge we can relate the curvature perturbation $\Phi$ in any gauge, say $\Phi^G$, with that of the Newtonian gauge, say $\Phi^N$, by \eqref{eq:gaugetransf} which yields
\begin{align}\label{eq:phiG}
\Phi_G&= \Phi_N+{\cal H}T_G+\tfrac{1}{3}\Delta L_G\,.
\end{align}
Then, the class of suitable gauges for subhorizon physics is defined as those gauges in which
\begin{align}\label{eq:phiGcondition}
\Phi_G(k\gg{\cal H})=O(\Phi_N(k\gg{\cal H}))\,.
\end{align}
This means that the curvature perturbation need not be the same but have the same qualitative behaviour. We shall see that although this is a requirement on the scalar variable $\Phi$, it yields the gauge independence of $h_{ij}$. Using \eqref{eq:phiGcondition} into \eqref{eq:phiG} we find conditions on the gauge parameters $T_G$ and $L_G$. For example, the gauge time parameter must be
\begin{align}\label{eq:TG}
T_G(k\gg{\cal H})&=O({\cal H}^{-1}\Phi_N(k\gg{\cal H}))\quad{\rm or\,\,higher\,\,order}.
\end{align}
Note that this abstract condition \eqref{eq:phiGcondition} actually includes most of the commonly used gauges such as the synchronous, the flat and the constant Hubble gauges \cite{Domenech:2020xin}. However, it excludes the comoving slicing gauge \cite{Domenech:2020xin}. The main reason for such exclusion is that the comoving slicing gauge is a choice where the spatial hypersurface is orthogonal to the flow lines of the fluid. This is a good choice for superhorizon scales where the flow lines are slowly changing with time, but it is a poor choice for subhorizon scales where fluid velocities oscillate. To convince ourselves of the gauge independence of \eqref{eq:pseudotensor2}, take the solution to $\Phi^N$ \eqref{eq:phisolsol} and $h_{ij}^N$ \eqref{eq:kernelaverage} for constant $w$, which read
\begin{align}
\Phi_k^N\propto \left(\frac{k}{{\cal H}}\right)^{-2-b}\quad{\rm and}\quad h_k^N\propto \left(\frac{k}{{\cal H}}\right)^{-1-b}\,.
\end{align}
For such solutions, the condition \eqref{eq:TG} translates into
\begin{align}\label{eq:TGNew}
T_G=O(h^N_k)\quad{\rm or\,\, higher\,\, order}\,.
\end{align}
Then, using \eqref{eq:2ordergauge} with $L_G=0$ and in Fourier space we have
\begin{align}\label{eq:2ordergauge3}
h^G_{k}=h^N_{k}-\int \frac{d^3q}{(2\pi)^3}e^{ij}q_iq_jT_G(q)T_G(|\mathbf{k}-\mathbf{q}|)\,,
\end{align}
where we used that in the Newtonian gauge we have $E=\beta=0$. So when substituting \eqref{eq:TGNew} in \eqref{eq:2ordergauge3} we conclude that for all gauges which satisfy \eqref{eq:phiGcondition} we have that
\begin{align}
h^G_{k}(k\gg{\cal H})=h^N_{k}(k\gg{\cal H})+O({\cal H}^2/k^2)\,,
\end{align}
which proves the approximate gauge independence of $\rho_{\rm GW}$ \eqref{eq:pseudotensor2}. It should be noted that while the work of \cite{Domenech:2020xin} is not a solution \textit{per se}, it shows when and why the predictions for induced GWs are well-defined and gives strong evidence that the calculations in this class of gauges are meaningful. In any case, the final word would be to find the GW detector response at second order.

Before ending this section, let us place some more attention to point $(ii)$, i.e., that the source term to induced GWs is not active. This means that the amplitude of scalar fluctuations has decayed enough so that they are no longer a significant source of induced GWs. This is often implied by taking the subhorizon limit $k\gg {\cal H}$ since induced GWs are mainly generated at horizon crossing, i.e., $k\sim {\cal H}$, and not afterwards. However, there is one exception: the dust dominated universe studied in Sec.~\ref{sec:dust}. In a dust dominated universe, the Newtonian potential $\Phi$ is constant on all scales and, therefore, is constantly sourcing induced GWs. This is the reason why Ref.~\cite{Hwang:2017oxa} finds such large gauge dependence of induced GWs in a dust dominated universe. Yet, we learned in Sec.~\ref{sec:dust} that the final spectrum of induced GWs is very sensitive to the transition from dust to radiation, where $\Phi$ changes from a constant to an oscillating function. Thus, one must follow the induced tensor modes generated in the dust dominated universe until they are deep inside the horizon in the radiation dominated era. It is from that moment on that the conditions $(i)-(iii)$ of \cite{Domenech:2020xin} apply. Then we know that the spectrum of induced GWs is approximately gauge invariant.


\section{Other GW sources related to PBH formation \label{sec:othersources}}

So far, we have focused on the GW induced by large primordial fluctuations. These large fluctuations collapse to form PBH if their rms amplitude is large enough. Once PBH form they may source other GWs independent of the induced GWs. Let us anticipate that the other GW counterparts are from PBHs mergers and graviton emission by Hawking evaporation. We briefly comment on these possibilities below.

On one hand, PBHs will form binaries in the early universe and eventually merge. The formation of such binaries is mainly due to the three body interaction with the nearest PBH \cite{Nakamura:1997sm}. These PBH binaries can be detected by the observation of resolved or unresolved merger events. In the former we would see the full GW waveform, e.g., as in the LIGO detections \cite{TheLIGOScientific:2017qsa}, and the latter will appear as a stochastic GW background. These makes two additional possible GW counterparts to PBHs. Note that if PBHs are lighter than $10^{15}\,{\rm g}$ they evaporated by today and, therefore, we probably cannot observe any resolved merger. PBHs heavier than $10^{15}\,{\rm g}$ have not yet evaporated and their binaries could be merging in the nearby universe. The GW spectrum from the unresolved mergers is studied in Refs.~\cite{Mandic:2016lcn,Wang:2016ana,Wang:2019kaf} while the estimates for the merger rates can be found in Refs.~\cite{Nakamura:1997sm,Sasaki:2016jop,Bird:2016dcv,Ali-Haimoud:2017rtz,Sasaki:2018dmp,Garriga:2019vqu}. A very rough order of magnitude estimate of the frequency at which the GWs from the mergers of PBH binaries will show up is given by the frequency of the GWs emitted at the Innermost Stable Circular Orbit \cite{Maggiore:1900zz}
\begin{align}\label{eq:fmax}
f_{\rm GW, max}\approx 2f_{\rm ISCO}\approx 4.4\,{\rm kHz}\left(\frac{M_{\odot}}{M}\right)\,,
\end{align}
where $M$ is the total mass of the binary and $M_\odot$ is a solar mass. For a monochromatic PBH mass function then $M=2M_{\rm PBH,i}$. This may change by a factor of a few depending on the redshift at the time of merger, so that $f_{\rm GW, max,0}=f_{\rm GW, max}/(1+z)$. For PBHs with $M_{\rm PBH,i}\gg 10^{15}\,{\rm g}$ we can assume that most of the mergers occur in the nearby universe and so the frequency \eqref{eq:fmax} already gives a good intuition of the position of the peak in the SGWB from mergers and the chirp frequency of the GW waveform, see for example Ref.~\cite{Wang:2019kaf}. However, PBHs with $M_{\rm PBH,i}<5\times 10^{8}\,{\rm g}$ have merged and evaporated long before BBN. In this case, we can assume that most of the mergers occur close to the evaporation time and, if PBH dominated the universe as in Sec.~\ref{sec:PBHdom}, we can take into account the redshift of the frequency \eqref{eq:fmax} until today. A quick estimate yields
\begin{align}
f^{\rm merger}_{\rm peak,0}\approx 10^{15}{\rm Hz}\left(\frac{M_{\rm PBH,i}}{10^{4}\,{\rm g}}\right)^{1/2}\left(\frac{g_*(T_{\rm eva})}{106.75}\right)^{1/4}\left(\frac{g_{s*}(T_{\rm eva})}{106.75}\right)^{-1/3} \left(\frac{g_H(T_{\rm PBH})}{108}\right)^{-1/2}\,,
\end{align}
which gives the right order of magnitude estimate for the position of the peak. For the detailed spectral shape in this case see Ref.~\cite{Inomata:2020lmk}. The peak frequency of this SGWB is unfortunately too high to be observed by current laser interferometers. Note that the PBH mass range $10^{9}\,{\rm g}<M_{\rm PBH,i}<10^{14}\,{\rm g}$ is subject to tight constraints from entropy production, changes in the abundance of light elements, the extragalactic photon background and damping of CMB temperature anisotropies. For a detailed review see Ref.~\cite{Carr:2020gox}.

On the other hand, when PBHs evaporate they also emit ``gravitons'' by Hawking radiation. The typical frequency of such gravitons can be estimated by assuming that Hawking radiation follows a black body spectrum. Then, the peak frequency is roughly at $f_{\rm peak}\approx 2.82\, T_{\rm PBH,i}$ where $T_{\rm PBH,i}$ is given in Eq.~\eqref{eq:TPBH} in terms of the initial PBH mass $M_{\rm PBH,i}$. Note that the frequency $f_{\rm peak}$ is the peak frequency of the spectrum when PBH evaporate. Thus, we have to take into account that frequencies redshift proportional to $a^{-1}$ to calculate the frequency of these gravitons today. To do that, we will use the calculations of Sec.~\ref{sec:PBHdom} which considers the case that PBH dominate the universe. Then, we can compute the evaporation temperature $T_{\rm eva}$ in terms of $M_{\rm PBH,i}$ by Eq.~\eqref{eq:Teva}. Inserting all numerical coefficients, we find that the graviton peak frequency today is given by
\begin{align}
f^{\rm graviton}_{\rm peak,0}\approx 10^{16}\,{\rm Hz}\left(\frac{M_{\rm PBH,i}}{10^4 {\rm g}}\right)^{1/2}\left(\frac{g_*(T_{\rm eva})}{106.75}\right)^{1/4}\left(\frac{g_{s*}(T_{\rm eva})}{106.75}\right)^{-1/3} \left(\frac{g_H(T_{PBH})}{108}\right)^{-1/2}\,.
\end{align}
A more detailed derivation of the graviton spectrum and the frequency range can be found in Ref.~\cite{Inomata:2020lmk}. Unfortunately, these are very high frequency GWs which are not observable by direct detection with the current capabilities. Nevertheless, they might be within range of future CMB experiments by looking at the contribution of these gravitons to the effective number of relativistic species at BBN \cite{Dong:2015yjs,Arbey:2019jmj,Kuhnel:2019zbc,Arbey:2021ysg,Masina:2021zpu}. Very interestingly if PBHs have a large spin, their imprint on the effective number is within range of the CMB-S4 experiment \cite{Arbey:2021ysg,Masina:2021zpu}. 

It is also important to notice that in some inflationary models involving axions coupled to gauge fields the enhancement in the primordial curvature perturbation is accompanied by an enhancement in the primordial tensor spectrum \cite{Peloso:2016gqs,Garcia-Bellido:2016dkw,Garcia-Bellido:2017aan}. In this case the primordial GW signal is in general larger than the induced GW one. Thus, in this scenario the induced GW signal may be buried inside the primordial GWs. It is possible that a detailed observation of the spectrum and the PBH counterpart helps to distinguish such scenario.


\section{Current and future observational prospects \label{sec:observations}}

In this section, we briefly discuss the future observational prospects regarding induced GWs. We will not focus on the prospects for PBHs, although they are equally interesting. Nevertheless, let us mention the most interesting PBH windows as they provide good motivation to look for induced GWs in particular ranges.  The first one is the LIGO binary black hole mergers \cite{Nakamura:1997sm,Sasaki:2016jop,Bird:2016dcv,Ali-Haimoud:2017rtz}. Notably, although some models are already ruled out by current data \cite{Hall:2020daa} there seems to be evidence for a small fraction of PBHs in addition to astrophysical ones \cite{DeLuca:2021wjr,Franciolini:2021tla}. Interestingly, the corresponding induced GWs to the PBHs in the LIGO window fall in the PTA band.\footnote{That is assuming an early universe dominated by radiation and using the estimates of Sec.~\ref{sec:PBH}.} Coincidently, the NANOGrav collaboration reported a possible SGWB in such range \cite{Arzoumanian:2020vkk}, yet to be confirmed. We discuss more on NANOGrav a bit later. Another interesting window is the planet mass PBHs \cite{Niikura:2019kqi} where the OGLE collaboration reported few microlensing events of planet mass objects \cite{2017Natur.548..183M}. Future microlensing observations such as with Subaru/HSC \cite{Sugiyama:2020roc} will be able to confirm the results from OGLE. If we assume that radiation dominated early universe, planet mass PBHs have an induced GW counterpart that falls in between PTA and LISA. The last promising window is for PBHs in the asteroid mass range where there are practically no observational constraints yet \cite{Sasaki:2018dmp,Carr:2020gox} and therefore PBHs can make up for all dark matter. In this case, the induced GW counterpart falls right inside the LISA band. Thus, induced GW will be an essential probe of PBH as dark matter. For a review see Ref.~\cite{Yuan:2021qgz}. Also see Ref.~\cite{Wang:2019kaf} for a detailed discussion.

Let us now turn to the observational prospects of induced GWs. Although there are weak constraints from LIGO \cite{LIGOScientific:2019vic} on the induced SGWB \cite{Kapadia:2020pnr,Wang:2016ana,Romero-Rodriguez:2021aws}, the most interesting part is the future prospects. On one hand, the absence of induced GWs directly translates into bounds on the primordial spectrum \cite{Gow:2020bzo}. In this way, assuming radiation domination, we see from Ref.~\cite{Gow:2020bzo} that the bounds are roughly ${\cal P}_{\cal R}\sim 10^{-5}$ for $k\sim 10^6-10^8\,{\rm Mpc}^{-1}$ in the PTA band, ${\cal P}_{\cal R}\sim 10^{-4}$ for $k\sim 10^{11}-10^{14}\,{\rm Mpc}^{-1}$ in the LISA band and ${\cal P}_{\cal R}\sim 10^{-3}$ for $k\sim 10^{15}-10^{18}\,{\rm Mpc}^{-1}$ in the ET band. CMB spectral distortions will probe down to ${\cal P}_{\cal R}\sim 10^{-8}$ for $k\sim 10-10^{5}\,{\rm Mpc}^{-1}$ \cite{Kite:2020uix,Unal:2020mts}. However, one of the problems is that such future bounds depend on the shape of the induced GW spectrum. For example, if we only see part of the induced GW spectrum, e.g., the UV or IR tail but not the peak, then in order to extract information on the amplitude of primordial fluctuations we must extrapolate the results beyond the range of the GW detector by assuming some template \cite{Kuroyanagi:2018csn}. On top of that, we have the issue that some induced GW templates might degenerate with other sources. As a good example for these potential problems, we proceed to discuss the various induced GW explanations of the NANOGrav results \cite{Arzoumanian:2020vkk}.

\begin{figure}
\centering
\includegraphics[width=0.55\columnwidth]{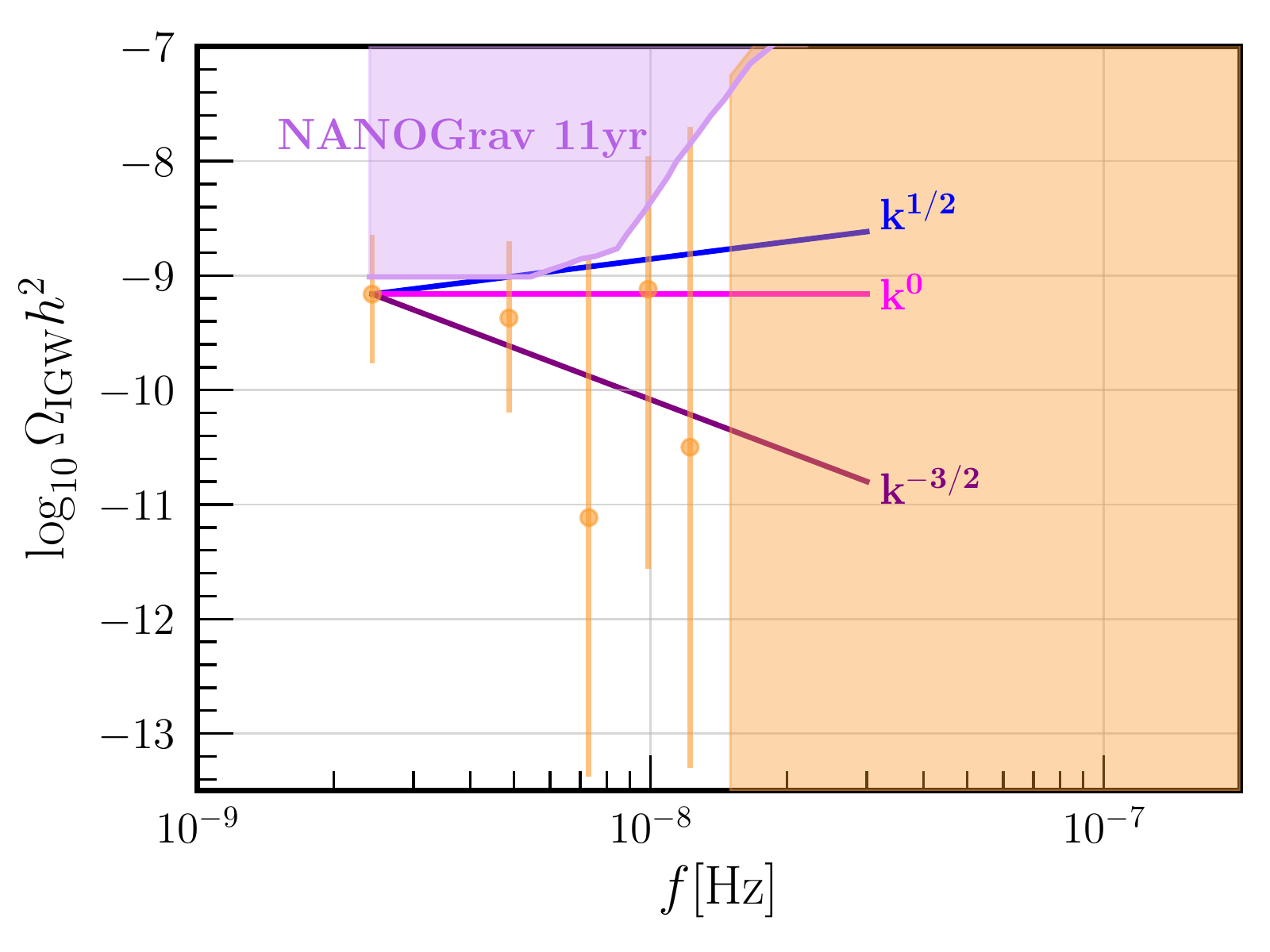}
\caption{NANOGrav results on the possible SGWB \cite{Arzoumanian:2020vkk} in terms of spectral density. The orange dots are the data points from the timing residuals with their error bars. We only show the first five data points which are the most relevant. The points in the orange shaded region were compatible with white noise and not considered for the fit. In light purple we also show the NANOGrav 11-yr sensitivity curve from \cite{NANOGRAV:2018hou}. In solid lines we show the allowed slopes within the 1-$\sigma$ contours. In blue and purple we respectively show the upper and lower limits on the slope, $k^{1/2}$ and $k^{-3/2}$. In magenta we give the approximate best fit for the data, which is a scale invariant spectrum.\label{fig:plotnanograv}}
\end{figure}

The NANOGrav collaboration reported a common process signal in the time residuals of the frequencies of pulsars \cite{Arzoumanian:2020vkk}. While no quadrupolar nature of the signal has been found, it is interesting to consider the possibility that the signal measured is in fact the SGWB of induced GWs. The main features of the reported spectrum \cite{Arzoumanian:2020vkk}, in cosmologists terms, are that it has an amplitude of $\Omega_{\rm GW,0}h^2\sim 10^{-9}$ at a frequency of $2\times 10^{-9}{\rm Hz}$ and a spectral index, that is  $\Omega_{\rm GW,0}\propto k^{n}$, that ranges from $n\sim 1/2$ to $n\sim -3/2$ within the $1$-$\sigma$ bounds. This means that the NANOGrav data seems to prefer a rather flat GW spectrum. This is illustrated in Fig.~\ref{fig:plotnanograv}. Using the very rough estimate of Sec.~\ref{sec:estimates}, that is $\Omega_{\rm GW,0}h^2\sim 10^{-6}{\cal P}_{\cal R}^2$, we find that if the peak of the induced GW spectrum lies around $f\sim2\times 10^{-9}{\rm Hz}$ then the amplitude of primordial fluctuations that sourced such induced GWs is about ${\cal P}_{\cal R}\sim O(10^{-2})$. Such amplitude of primordial fluctuations is right at the value when PBH formation is interesting, and their masses are in the LIGO window. 

The main complication is that the NANOGrav GW spectrum is rather flat and the induced GWs from a sharp peak are very peaked and decay as $k^3$ in the IR. This means that there are two possibilities: either the GWs were induced by a broad peak in the primordial spectrum (including a flat spectrum) \cite{Vaskonen:2020lbd,DeLuca:2020agl,Kohri:2020qqd,Bhattacharya:2020lhc,Yi:2021lxc} or the GWs were induced in non-radiation dominated universe with $w<1/3$ \cite{Domenech:2020ers}. Below we list the main results of these works: 
\begin{itemize}
\item[--] Ref.~\cite{Vaskonen:2020lbd} found a negligible amount of PBH in LIGO region but the PBH could be the seeds of supermassive black holes. \item[--] Ref.~\cite{Kohri:2020qqd} found a larger amount of PBH in the LIGO band than \cite{Vaskonen:2020lbd} and also pointed out that the SGWB from unresolved mergers could detectable in the future by LISA and ET. 
\item[--] Ref.~\cite{DeLuca:2020agl} showed that the induced GWs from a flat primordial spectrum could explain all dark matter.
\item[--] Ref.~\cite{Inomata:2020xad} proposed a particular inflationary model with an axion-like curvaton which could explain the NANOGrav result as well as the LIGO events due to primordial non-gaussianities. Similar claims appear in Ref.~\cite{Yi:2021lxc} with a Higgs-like inflation with non-canonical kinetic terms. 
\item[--] In a different direction, Refs.~\cite{Domenech:2020ers} and \cite{Bhattacharya:2020lhc} studied the induced GWs from a peaked primordial spectrum in a non-radiation dominated universe. They both find that a soft equation of state seems to be preferred. In particular, Ref.~\cite{Domenech:2020ers} use the IR tail of the induced GW spectrum from a very peaked primordial spectrum (see Sec.~\ref{sec:typical}) to fit the NANOGrav results. They found that the 1$\sigma$ contours on the spectral tilt translate to bounds for the equation state parameter as $-0.1<w<0.05$. 
\item[--] Ref.~\cite{Atal:2021jyo} uses the UV tail of the induced GW spectrum from a broken power-law primordial spectrum. They concluded that the NANOGrav results imply a small non-gaussianity parameter $f_{NL}<0.7$.
\end{itemize}

It is important to note that the induced GWs are not the only explanations of the NANOGrav results. For example, among many others, possible candidates are cosmic strings \cite{Ellis:2020ena,Blasi:2020mfx,Buchmuller:2020lbh,Samanta:2020cdk} and first order phase transitions \cite{Nakai:2020oit,Addazi:2020zcj,Neronov:2020qrl,Ratzinger:2020koh,Bian:2020urb,Li:2020cjj,Liu:2020mru,Paul:2020wbz}. Within the next decades, we expect to have more information from PTA experiments. We also expect the GW detector LISA to launch. In order to distinguish these candidates to the NANOGrav data, we will need better resolution for a better Bayesian analysis \cite{Bian:2020urb} but crucially we will need other discriminators of cosmic GWs. In the case of induced GWs, if some of the LIGO binary black hole mergers are found to be PBHs, the NANOGrav signal would have a strong case to be induced GWs.

\section{Summary of main formulas \label{sec:summary}}

We dedicate this section to recollect the main formulas scattered around the review and present them so that they are ready for use. We first give the GW spectrum measured today $\Omega_{\rm GW,0}$ in terms of the spectral density of GWs at a pivot scale $\Omega_{\rm GW,rh}$. The subscript ``rh'' stands for reheating and refers to the time of instantaneous transition to the standard radiation dominated era. In the case where GWs are induced during radiation domination, the pivot time ``rh'' refers to the time when induced GWs of given wavenumber $k$ are sufficiently inside the cosmological horizon to be treated as a radiation fluid in an expanding universe. With these clarifications, we can use the results of Secs.~\ref{sec:analytical} and \ref{sec:typical} to write
\begin{align}\label{eq:spectraldensitytoday3}
\Omega_{\rm GW,0}h^2=1.62\times 10^{-5}\left(\frac{\Omega_{r,0}h^2}{4.18\times 10^{-5}}\right)\left(\frac{g_*(T_{\rm rh})}{106.75}\right)\left(\frac{g_{*s}(T_{\rm rh})}{106.75}\right)^{-4/3}\Omega_{\rm GW,rh}\,.
\end{align}
We give the detailed expression of $\Omega_{\rm GW,rh}$ for a general equation of state in Sec.~\ref{sec:generaleos} and for the particular case of a dust universe with an instantaneous transition to radiation domination in Sec.~\ref{sec:dustestimate}.

\subsection{General equation of state \label{sec:generaleos}}

We studied in the detail the derivation of $\Omega_{\rm GW,rh}$ for general expansion histories in Sec.~\ref{sec:analytical}. The main approximations used are the following: $(i)$ there is an epoch of constant $w$ and $c_s$ after inflation, $(ii)$ the peak (or plateau) in the primordial spectrum enters the horizon during such era and $(iii)$ the transition to radiation domination is instantaneous. Note that in the case of standard radiation domination these assumptions are unnecessary. The presence of a transition to radiation domination introduces another relevant scale $k_{\rm rh}$, which is the last comoving scale that entered the horizon at the transition. The GW spectrum for scales $k\gg k_{\rm rh}$ takes a general form given by
\begin{equation}\label{eq:Phgaussianfinal}
\Omega_{\rm GW,rh}=\left(\frac{k}{k_{\rm rh}}\right)^{-2b}\int_0^\infty dv\int_{|1-v|}^{1+v}du\,{\cal T}(u,v,b,c_s){{\cal P}_{\cal R}(ku)}{{\cal P}_{\cal R}(kv)}\,,
\end{equation}
where we defined the transfer function as
\begin{align}\label{eq:kernelaveragefinal}
{\cal T}(u,v,b,c_s)=& {\cal N}(b,c_s)\left(\frac{4v^2-(1-u^2+v^2)^2}{4u^2v^2}\right)^2{|1-y^2|^{b}}\nonumber\\&
\times\Bigg\{\left(\mathsf{P}^{-b}_{b}(y)+\frac{b+2}{b+1}\mathsf{P}^{-b}_{b+2}(y)\right)^2\Theta[c_s(u+v)-1]\nonumber\\&
+\frac{4}{\pi^2}\left(\mathsf{Q}^{-b}_{b}(y)+\frac{b+2}{b+1}\mathsf{Q}^{-b}_{b+2}(y)\right)^2\Theta[c_s(u+v)-1]\nonumber\\&
+\frac{4}{\pi^2}\left({\cal Q}^{-b}_{b}(-y)+2\frac{b+2}{b+1}\mathsf{\cal Q}^{-b}_{b+2}(-y)\right)^2\Theta[1-c_s(u+v)]\Bigg\}\,,
\end{align}
and the numerical coefficient is given by
\begin{align}
{\cal N}(b,c_s)\equiv \frac{4^{2b}}{3c_s^4}\Gamma^4\left[b+\tfrac{3}{2}\right]\left(\frac{b+2}{2b+3}\right)^2\left(1+b\right)^{-2(1+b)}\,.
\end{align}
In Eq.~\eqref{eq:kernelaveragefinal} we have introduced the following notation for convenience:
\begin{equation}\label{eq:y2}
b=\frac{1-3w}{1+3w}\quad{\rm and}\quad y=1-\frac{1-c_s^2(u-v)^2}{2c_s^2 uv}\,.
\end{equation}
These equations are general enough for $k\gg k_{\rm rh}$ that can be used for any shape of the primordial spectrum as long as the main feature, whether it is a peak or a plateau, enters during the $w={\rm constant}$ epoch. If the transition to radiation domination is gradual, then we only expect a change in the IR tail of the spectrum for modes which entered during the relevant stages of the transition. Following with the instantaneous transition, it is important to note that for $b>1$ the IR tail has a red tilt and the spectrum for $k\ll k_{\rm rh}$ becomes important. Nevertheless, a good approximation is to stop the spectrum at $k\sim k_{\rm rh}$ and match it to the typical IR scaling in radiation domination, i.e., $\Omega_{\rm GW,rh}\propto k^3$. Details on the spectrum for $k\ll k_{\rm rh}$ can be found in Sec.~\ref{sec:analytical}. It should be noted that \eqref{eq:kernelaveragefinal} is not valid for $c_s=0$. The case of $c_s=0$ is studied in detail in Sec.~\ref{sec:dust} and a quick estimate is presented below in Sec.~\ref{sec:dustestimate}.

The transfer function \eqref{eq:kernelaveragefinal} is expressed in terms of Associated Legendre functions which are provided in App.~\ref{app:legendre} in terms of Hypergeometric functions. This general form, while very useful, might be a hindrance to a quick implementation. For this reason, we present below particular examples of the transfer function \eqref{eq:kernelaveragefinal}. We choose the cases where $b\in \mathbb{Z}$ and $b\in \mathbb{Z}+\tfrac{1}{2}$ since then the associated Legendre functions (and the Hypergeometric functions) reduce to polynomials. Note that only when $b\in \mathbb{Z}$ logarithmic terms appear. We will first check the standard case of radiation domination ($b=0$, $w=1/3$) and we will later turn to other situations. We give the transfer function for a stiff fluid $b=-1/2$ ($w=1$), a soft fluid $b=1/2$ ($w=1/9$), a pressureless fluid $b=1$ ($w=0$) and negative equation of state fluid $b=2$ ($w=-1/9$). We keep the sound speed $c_s$ arbitrary. For an adiabatic perfect fluid, we have $c_s^2=w$ and for a canonical scalar field $c^2_s=1$.

\subsubsection{Radiation domination}
The transfer function \eqref{eq:kernelaveragefinal} for $b=0$ ($w=1/3$) reads
\begin{align}\label{eq:w13}
{\cal T}_{RD}(u,v,c_s,w=1/3)=&\frac{y^2}{3c_s^4}\left(\frac{4v^2-(1-u^2+v^2)^2}{4u^2v^2}\right)^2\nonumber\\&\times
\left\{\frac{\pi^2}{4}y^2\Theta[c_s(u+v)-1]
+\left(1-\frac{1}{2}y \ln\left|\frac{1+y}{1-y}\right|\right)^2\right\}\,,
\end{align}
where we kept the variable $y$ as it simplifies considerably the expressions. This recovers the result of Kohri and Terada \cite{Kohri:2018awv} after taking into account all numerical factors in \eqref{eq:spectraldensity}.

\subsubsection{Stiff fluid (kinetic) domination}

Another notable case of interest is a stiff fluid with $b=-1/2$ ($w=1$). This is for example typical of quintessential inflation scenarios \cite{Spokoiny:1993kt,Peebles:1998qn,Brax:2005uf,Hossain:2014xha}. For a kinetic dominated period, the transfer function \eqref{eq:kernelaveragefinal} is given by
\begin{align}\label{eq:w1}
&{\cal T}_{KD}(u,v,c_s,w=1)=\frac{4}{3\pi c_s^4 |1-y^2|}\left(\frac{4v^2-(1-u^2+v^2)^2}{4u^2v^2}\right)^2\nonumber\\&\times
\Bigg\{\left(1+3y^2\right)\Theta[c_s(u+v)-1]
+\left(1-3y^2+3y\sqrt{|1-y^2|}\right)^2\Theta[1-c_s(u+v)]\Bigg\}\,.
\end{align}
This coincides with the formulas derived in Ref.~\cite{Domenech:2019quo,Domenech:2020kqm}. It has also been studied numerically in Ref.~\cite{Dalianis:2020cla}.

\subsubsection{Soft fluid domination}

The next case is that of a soft fluid with $b=1/2$ ($w=1/9$). This could be due to a scalar field rolling down an exponential potential. After some algebra, we find that  
\begin{align}\label{eq:w12}
{\cal T}(u,v,c_s,w=1/9)&=\frac{2^8}{3^8\pi c_s^4  }\left(\frac{4v^2-(1-u^2+v^2)^2}{4u^2v^2}\right)^2\nonumber\\&\times
\Bigg\{\left(4+45y^2\right)\Theta[c_s(u+v)-1]\nonumber\\&
+\left(y(3-10y^2)+(2+10y^2)\sqrt{|1-y^2|}\right)^2\Theta[1-c_s(u+v)]\Bigg\}\,.
\end{align}

\subsubsection{Pressure-less fluid domination}
We turn now to the case of a pressure-less fluid with $b=1$ ($w=0$). Note that this is not dust in the sense that $c_s\neq0$. So that even though the fluid has no pressure, perturbations still propagate at a finite speed. This is the case of a scalar field rolling down an exponential potential and might also be close to the coherent oscillations of a scalar field around the bottom of the potential \cite{Martin:2020fgl}. The kernel for $b=1$ is given by
\begin{align}\label{eq:w0}
{\cal T}(u,v,c_s,w=0)=&\frac{3^3 \,5^2}{2^{14} c_s^4}\left(\frac{4v^2-(1-u^2+v^2)^2}{4u^2v^2}\right)^2\nonumber\\&\times
\Bigg\{\frac{\pi^2}{4}(1-y^2)^2(1+3y^2)^2\Theta[c_s(u+v)-1]
\nonumber\\&+\left(y(1-3y^2)-\frac{1}{2}(1+2y^2-3y^4) \ln\left|\frac{1+y}{1-y}\right|\right)^2\Bigg\}\,.
\end{align}

\subsubsection{Negative EoS fluid domination}

The last case we are interested in is a negative EoS parameter. When $-1/3<w<0$ the universe is decelerating very slowly which yields to interesting slopes in the IR tail. For convenience we present the kernel for the case $b=2$ ($w=-1/9$), which reads
\begin{align}\label{eq:wm19}
{\cal T}(u,v,c_s,w=-1/9)=&\frac{5^2\,7^2}{2^8\,3^9 c_s^4 }\left(\frac{4v^2-(1-u^2+v^2)^2}{4u^2v^2}\right)^2\nonumber\\&\times
\Bigg\{\frac{225\pi^2}{4}(1-y^2)^2(1+y^2-2y^4)^2\Theta[c_s(u+v)-1]
\nonumber\\&+\left(y(9+35y^2-30y^4)+\frac{15}{2}(1-3y^4+2y^6) \ln\left|\frac{1+y}{1-y}\right|\right)^2\Bigg\}\,.
\end{align}

\subsection{Dust domination\label{sec:dustestimate}}

Here we provide a useful order of magnitude estimate for the following situation: $(i)$ the universe is dominated by dust $w=c_s^2=0$ when induced GWs are generated, $(ii)$ the transition to radiation domination is instantaneous and $(iii)$ the spectrum of fluctuations for the Newtonian potential $\Phi$ is given by 
\begin{align}\label{eq:pphidustfinal}
{\cal P}_{\Phi}={\cal A}_\Phi\left(\frac{k}{k_{\rm UV}}\right)^{-n}\Theta(k_{UV}-k)\,,
\end{align}
where $n<2$. This power spectrum includes a power-law generated during inflation and the case of PBH density fluctuations. The details of the derivation can be found in Sec.~\ref{sec:dust}. With these assumptions, the spectrum of induced GWs is approximately given by
\begin{align}\label{eq:Phgaussiandust4final}
\Omega_{\rm GWs,rh}(k\sim k_{\rm UV})\approx\Omega^{\rm peak}_{\rm GWs}\left(\frac{k_{\rm UV}}{k_{\rm rh}}\right)^{7-2n}\Theta(k_{\rm UV}-k)\,,
\end{align}
where the amplitude of the peak reads
\begin{align}\label{eq:Phgaussiandust3final}
\Omega^{\rm peak}_{\rm GWs}\approx\frac{\pi}{9216 \sqrt{3}}\left(\frac{4}{3}\right)^{n}\left(\frac{k_{\rm UV}}{k_{\rm rh}}\right)^7{\cal A}^2_{\Phi}\,.
\end{align}
We can then use Eq.~\eqref{eq:Phgaussiandust4} together with \eqref{eq:spectraldensitytoday3} to estimate the induced GW spectrum today. Note that if the UV cut-off $k_{\rm UV}$ is at very small scales, i.e., $k_{\rm UV}\gg k_{\rm rh}$, the amplitude of induced GW is very much enhanced and strong constraints apply. It is important to clarify that if $k_{\rm UV}$ is very large (larger than $k_{NL}$ in Eq.~\eqref{eq:kNL}), some density fluctuations would enter the non-linear regime. Thus, the above result must be considered as a rough estimate. For detailed discussion we refer the reader to Sec.~\ref{sec:dust}.

\section{Conclusions \label{sec:conclusions}}

Gravitational waves induced by primordial fluctuations, so-called induced GWs, have been shown to be a very promising tool to complete our picture of inflation and the early universe. Although this effect was first pointed out fifty years ago, induced GWs together with primordial black holes are receiving serious attention since the first GW detection by LIGO \cite{Abbott:2016nmj}. They are even more interesting after the NANOGrav collaboration recently reported a possible stochastic GW signal \cite{Arzoumanian:2020vkk}. Within the next couple of decades, LISA will launch, and we will have a new data on stochastic backgrounds of gravitational waves in a wide frequency range. Even in the worst-case scenario, where no conclusive induced GW signal is found, we will obtain new constraints on the physics of inflation and the early universe in completely unexplored regimes.

The field of cosmology with induced GWs is relatively new and still developing. In this review we have focused on the current analytical techniques and estimates to compute the induced GW spectrum. We believe they will be most useful in analysing future SGWB data.  We have not dwelled into the details of PBH formation nor on the observational forecasts for future GW detectors, both of which deserve a separate review. In the next years there will be more studies on induced GWs; new and better analytical and numerical results might be developed. Until then, particularly interesting and important directions are: $(i)$ the impact of primordial non-gaussianities, not necessarily in the local shape, $(ii)$ possible discriminators of induced GWs to distinguish them from other sources, $(iii)$ the GW detector response at second order and $(iv)$ GWs from the non-linear regime in an early matter dominated stage. We hope this review will be useful to the next advancements in the cosmology of induced GWs.


\vspace{6pt} 



\section*{Funding} 
G.D. as a Fellini fellow was supported by the European Union’s Horizon 2020 research and innovation programme under the Marie Sk{\l}odowska-Curie grant 
agreement No 754496.

\section*{Acknowledgments} 
I have benefited from many helpful and insightful discussions and collaborations over the years while carrying out the works in which this review is based. I would like to thank Misao Sasaki for his constant support and encouragement. I thank Jinn-Ouk Gong and Shi Pi for awaking my interest on induced GWs some years ago. I also thank Sabino Matarrese for insightful discussions and for pointing out the first papers on induced GWs. I have also learned a lot from discussions with Vicente Atal, Nicola Bartolo, Albert Escriv{\'a}, Joseph Fedrow, Jacopo Fumagalli, Jaume Garriga, Cristiano Germani, Marc Kamionkowski, David Langlois, Chunshan Lin, Atsushi Naruko, Sébastien Renaux-Petel, Javier Rubio, Volodymyr Takhistov, Vicent Vennin and Lukas T. Witkowski. I acknowledge very helpful correspondence with Jai-chan Hwang, Keisuke Inomata, Kazunori Kohri, Caner Unal and Zach Weiner. Last but not least, I would like to thank D.~Rojas and A.~D.~Rojas for their constant support.


\appendix
\section{Useful formulas and numerical values \label{app:formulasuseful}}
In this appendix we present the numerical values and formulas used to derive the parameters in the main text. First, we used the following values from the Planck results \cite{Aghanim:2018eyx}: $k_{\rm eq}\approx 0.0104 \,{\rm Mpc}^{-1}$, $z_{\rm eq}\approx 3400$, $T_0=2.7255\,{\rm K}$. We also used the value for the reduced Planck mass $M_{\rm pl}\approx 4.235\times 10^{18}\,{\rm GeV}\approx 4.34\times 10^{-6}\,{\rm g}$ and the solar mass $M_\odot\approx 2\times 10^{33}\,{\rm g}$. A useful relation between units (in the Planck units) is the following:
\begin{align}
1\,{\rm K}\approx 8.62\times 10^{-5}\,{\rm eV}\approx4.34\,{\rm cm}^{-1}\approx1.31\times 10^{11}\,{\rm Hz}
\end{align}

On the effective degrees of freedom, we used that for radiation at temperature $T$ the energy density is given by
\begin{align}
\rho=\frac{\pi^2}{30}g_*(T)T^4\quad{\rm with}\quad g_*(T)=\sum g_b+\frac{7}{8}\sum g_f\,,
\end{align}
where $g_b$ and $g_f$ are respectively the degrees of freedom of bosons and fermions. The entropy can then be calculated by
\begin{align}
s=\frac{2\pi^2}{45}g_{*s}(T)T^3\,.
\end{align}
The different values of $g_{*}$ and $g_{*s}$ in the standard model are review in Ref.~\cite{Husdal:2016haj}. At very high temperatures, i.e., $T\gg 100\,{\rm GeV}$ one has $g_{*}\approx g_{*s}\approx 106.75$. At the time of matter-radiation equality and at present we took $g_*(T_{\rm eq})\approx g_{*}(T_{0})\approx 3.38$ and $g_{*s}(T_{\rm eq})\approx g_{*s}(T_{0})\approx 3.94$.

The Hubble parameter in the radiation dominated universe can be related by the Friedmann equations to the temperature as
\begin{align}
H=\frac{\pi}{3\sqrt{10}M_{\rm pl}}g_*^{1/2}(T)T^2\,,
\end{align}
Another important relation is that since the entropy is conserved, which in an expanding universe means $s\propto a^{-3}$, there is a direct relation between the scale factor and the temperature. We have that 
\begin{align}
\frac{a}{a_\star}=\frac{T_\star}{T}\left(\frac{g_{*s}(T_\star)}{g_{*s}(T)}\right)^{1/3}\,.
\end{align}
More detailed explanations can be found, e.g., in Baumann's lecture notes \url{http://cosmology.amsterdam/education/cosmology/}

\section{Green's function method \label{sec:green}}

In this appendix we present the Green's method to find particular solutions to a differential equation where the homogeneous solutions are known. In the main text we found that the induced GWs have the following equations of motion:
\begin{equation}\label{eq:eominducedGWapp}
h''_{\mathbf{k},\lambda}+2{\cal H} h'_{\mathbf{k},\lambda}+k^2 h_{\mathbf{k},\lambda}={\cal S}_{\mathbf{k},\lambda}\,.
\end{equation}
Assuming that we know the two homogeneous solutions, say $h_1$ and $h_2$, we can check that the following Green's function,
\begin{equation}\label{eq:G}
G(\tau,\tilde \tau)=\frac{1}{W(h_1,h_2,\tilde \tau)}\left(h_1(\tau)h_2(\tilde\tau)-h_1(\tilde\tau)h_2(\tau)\right)\,,
\end{equation}
is a solution to
\begin{equation}\label{eq:eominducedGWapp2}
G''_{\mathbf{k},\lambda}(\tau,\tilde\tau)+2{\cal H} G'_{\mathbf{k},\lambda}(\tau,\tilde\tau)+k^2 G_{\mathbf{k},\lambda}(\tau,\tilde\tau)=\delta(\tau-\tilde\tau)\,.
\end{equation}
In Eq.~\eqref{eq:G} we have defined the Wronskian as
\begin{equation}
W(h_1,h_2,\tilde \tau)=h_1'(\tilde\tau)h_2(\tilde\tau)-h_1(\tilde\tau)h'_2(\tilde\tau)\,.
\end{equation}
Then with the Green's function \eqref{eq:G} we have that the particular solution to $h_{\mathbf{k},\lambda}$ with initial conditions $h_{\mathbf{k},\lambda}(\tau_i)=h'_{\mathbf{k},\lambda}(\tau_i)=0$ is given by
\begin{equation}
h_{\mathbf{k},\lambda}=\int_{\tau_i}^{\tau}d\tilde\tau G(\tau,\tilde \tau){\cal S}_{\mathbf{k},\lambda}(\tilde\tau)\,.
\end{equation}

\section{ADM formalism\label{app:ADM}}
In this appendix we briefly give the ADM or (3+1)-decomposition of the Einstein-Hilbert action, that is
\begin{align}
S=\int d^4x \sqrt{-g} \frac{M_{pl}^{2}}{2}R\,.
\end{align}
More details can be found in Ref.~\cite{Poisson:2009pwt}. First, the line element is parametrised by
\begin{align}
ds^2=-N^2dt^2+h_{ij}\left(dx^i+N^idt\right)\left(dx^j+N^jdt\right)\,,
\end{align}
where $N$ is the lapse and $N^i$ is the shift vector and describe the foliation of our spacetime. In this decomposition the Einstein-Hilbert action, after integrating out boundary terms, is given by
\begin{align}
\begin{split}
S&=\int d^3x dt N \sqrt{h}\,\frac{M_{pl}^{2}}{2}\left(R^{(3)}[h]+K_{ij}K^{ij}-K^2\right)\,,
\end{split}
\end{align}
where 
\begin{align}
K_{ij}\equiv\frac{1}{2N}\left(\dot{h}_{ij}-D_{i}N_{j}-D_{j}N_{i}\right)\,,
\end{align}
is the extrinsic curvature, $K=h^{ij}K_{ij}$ and $D_i$ is the covariant derivative of $h_{ij}$. Explicit expressions of the metric, its inverse and all the Christoffel symbols can be found in Appendix A of Ref.~\cite{Deffayet:2015qwa}.

\section{Fourier conventions and polarization tensors\label{app:polarization}}

In this appendix we specify the conventions used for Fourier transforms and the polarization tensors. First, we define the Fourier transform of a scalar mode as
\begin{align}
\Phi(\tau,\mathbf{k})=\int\frac{d^3 k}{(2\pi)^3}\,\Phi_{\mathbf{k}}(\tau)e^{i\mathbf{k}\mathbf{x}}\,,
\end{align}
and of a tensor mode as
\begin{align}
h_{ij}(\tau,\mathbf{k})=\sum_\lambda\int\frac{d^3 k}{(2\pi)^3}\,e^\lambda_{ij}(\mathbf{k})\,h_{\mathbf{k},\lambda}(\tau)e^{i\mathbf{k}\mathbf{x}}\,,
\end{align}
where $\lambda$ are the polarization, e.g., $+$ and $\times$, and $e^\lambda_{ij}(\mathbf{k})$ are the polarization tensors. Such expansion is valid for real polarization tensors. Then, reality conditions\footnote{In case of doubt about conventions and normalisation conditions, it is advisable to treat $h_{ij}$ as a field operator and express the Fourier expansion in the Fock representation. If the polarization tensors are complex, such as in circular polarization, the general normalization conditions are $e^{\lambda}_{ij}(\mathbf{k})\left(e^{\lambda ij}(\mathbf{k})\right)^*=\delta^{\lambda,\lambda'}$, where an asterisk refers to complex conjugate. We thank M.~Sasaki for explaining this point.} of $h_{ij}$ imply \cite{Caprini:2018mtu} $e^\lambda_{ij}(\mathbf{k})=e^\lambda_{ij}(-\mathbf{k})$ and $h^*_{\mathbf{k},\lambda}=h_{-\mathbf{k},\lambda}$. We define the polarization tensors in terms of polarization vectors as follows:
\begin{align}
e^{+}_{ij}(\mathbf{k})&=\frac{1}{\sqrt{2}}\left[e_{i}(\mathbf{k})e_{j}(\mathbf{k})-\bar e_{i}(\mathbf{k})\bar e_{j}(\mathbf{k})\right]\,,\\
e^{\times}_{ij}(\mathbf{k})&=\frac{1}{\sqrt{2}}\left[e_{i}(\mathbf{k})\bar e_{j}(\mathbf{k})+\bar e_{i}(\mathbf{k}) e_{j}(\mathbf{k})\right]\,.
\end{align}
Then, using that $e_{i}(\mathbf{k})k^i=\bar e_{i}(\mathbf{k})k^i=0$ and $e_{i}(\mathbf{k})e^{i}(\mathbf{k})=\bar e_{i}(\mathbf{k})\bar e^{i}(\mathbf{k})=1$, we arrive at the usual properties of the polarization tensors, namely
\begin{align}
e^{+}_{ij}(\mathbf{k})e^{+ ij}(\mathbf{k})&=1\quad,\quad e^{\times}_{ij}(\mathbf{k})e^{\times ij}(\mathbf{k})=1\quad,\quad e^{+}_{ij}(\mathbf{k})e^{\times ij}(\mathbf{k})=0\,,\nonumber\\
\delta_{ij}e^{+ij}(\mathbf{k})&=\delta_{ij}e^{\times ij}(\mathbf{k})=k_ie^{+ij}(\mathbf{k})=k_ie^{\times ij}(\mathbf{k})=0\,.
\end{align}

\subsection{Spherical parametrisation}
When calculating the projection of polarization tensors with the scalar momenta we choose spherical coordinates for convenience. Then, one can write in general that the wavenumber of a tensor mode is given by
\begin{align}
\mathbf{k}=k(\sin\theta_k\cos\varphi_k,\sin\theta_k\sin\varphi_k,\cos\theta_k)\,,
\end{align}
where $\theta_k$ and $\varphi_k$ respectively are the polar and azimuthal angles. With this choice, a pair of orthonormal polarization vectors are
\begin{align}
\mathbf{e}(\mathbf{k})&=(\cos\theta_k\cos\varphi_k,\cos\theta_k\sin\varphi_k,-\sin\theta_k)\,,\\
\bar{\mathbf{e}}(\mathbf{k})&=(-\sin\varphi_k,\cos\varphi_k,0)\,.
\end{align}
We parametrize two scalar momenta $\mathbf{q}$ and $\mathbf{l}$ as
\begin{align}
\mathbf{q}=q(\sin\theta_{q}\cos\varphi_{q},\sin\theta_{q}\sin\varphi_{q},\cos\theta_{q})\,,
\end{align}
and
\begin{align}
\mathbf{l}=l(\sin\theta_{l}\cos\varphi_{l},\sin\theta_{l}\sin\varphi_{l},\cos\theta_{l})\,.
\end{align}

In the calculations of the tensor spectrum of Sec.~\ref{sec:derivation} we considered general local primordial non-gaussianity. In this case, due to the symmetry between $\mathbf{l}$ and $\mathbf{q}$, we find most convenient to choose $\mathbf{k}$ in the z-axis, namely we set
\begin{align}
\theta_k=0\quad,\quad \varphi_k=0\,.
\end{align}
Then one has that
\begin{align}
|\mathbf{k}-\mathbf{q}|^2&=k^2+q^2-2kq\cos\theta_k\,,\\
|\mathbf{k}-\mathbf{l}|^2&=k^2+l^2-2kl\cos\theta_l\,,\\
|\mathbf{q}-\mathbf{l}|^2&=q^2+l^2-2lq\left(\cos\theta_{q}\cos\theta_{l}+\cos(\varphi_q-\varphi_l)\sin\theta_{q}\sin\theta_{l}\right)\,.
\end{align}
Note that the azimuthal dependence is only in $|\mathbf{q}-\mathbf{l}|$. The projections of $\mathbf{q}$ and $\mathbf{l}$ with the polarization tensors then read
\begin{align}
e^{+}_{ij}(\mathbf{k})q^iq^j&=\frac{1}{\sqrt{2}}q^2\sin^2\theta_{k}\cos(2\varphi_q)\quad,\quad
e^{\times}_{ij}(\mathbf{k})q^iq^j=\frac{1}{\sqrt{2}}q^2\sin^2\theta_{k}\sin(2\varphi_q)\\
e^{+}_{ij}(\mathbf{k})l^il^j&=\frac{1}{\sqrt{2}}l^2\sin^2\theta_{l}\cos(2\varphi_l)\,,\quad,\quad
e^{\times}_{ij}(\mathbf{k})l^il^j=\frac{1}{\sqrt{2}}l^2\sin^2\theta_{l}\sin(2\varphi_l)\,.
\end{align}
Lastly, what we need to derive the induced GW spectrum are the following results:
\begin{align}
\left(e^{+}_{ij}(\mathbf{k})q^iq^j\right)^2+\left(e^{\times}_{ij}(\mathbf{k})q^iq^j\right)^2=\frac{q^4}{2}\sin^4\theta_{k}\,,
\end{align}
and
\begin{align}
e^{+}_{ij}(\mathbf{k})q^iq^je^{+}_{ij}(\mathbf{k})l^il^j+e^{\times}_{ij}(\mathbf{k})q^iq^je^{\times}_{ij}(\mathbf{k})l^il^j=\frac{q^2l^2}{2}\sin^2\theta_{k}\sin^2\theta_{l}\cos(2(\varphi_q-\varphi_l))\,.
\end{align}

\section{Formulas in a general gauge \label{app:generalgauge}}

In this appendix we present the explicit expressions for Einstein equations in a perturbed flat FLRW universe filled with a perfect fluid. We present the results in a general gauge. The convention for the perturbed metric is given in Eq.~\eqref{eq:confdecom2} and the energy momentum tensor in Eq.~\eqref{eq:Tmunu}.

\subsection{Background}
At zeroth order in perturbation theory, we have the two Friedmann equations and energy conservation equations. These are respectively given by
\begin{align}\label{eq:friedman}
3M_{\rm pl}^2{\cal H}^2=a^2\rho\,,\\
M_{\rm pl}^2\left({\cal H}^2+2{\cal H}'\right)=-a^2P\,,
\end{align}
and
\begin{align}
\rho'+3{\cal H}(\rho+P)=0\,.
\end{align}

\subsection{First order}
At first order we find that the Hamiltonian and momentum constraints yield \cite{Malik:2008im}
\begin{align}
6{\cal H}(\phi'-{\cal H}\alpha)+2\Delta(\phi+{\cal H}\sigma)=M_{\rm pl}^2a^2\delta\rho\,,\\
2\phi'-2{\cal H}\alpha=M_{\rm pl}^2a^2(\rho+P)(v+\beta)\,.
\end{align}
The remaining Einstein equations are
\begin{align}
2\phi''+4{\cal H}\phi'-2{\cal H}\alpha'-(4{\cal H}'+2{\cal H}^2)\alpha=-M_{\rm pl}^2a^2\delta P\,,\\
\sigma'+2{\cal H} \sigma+\phi+\alpha=0\,.
\end{align}
We assumed that there is no anisotropic stress for matter, and we defined
\begin{align}
\sigma\equiv\beta-E'\,,
\end{align}
which corresponds to the scalar shear of the metric \cite{Malik:2008im}. The energy and momentum conservation equations can be derived from the Einstein equations.

\subsection{Second order}
Here we only present the second order equations for the tensor modes with a scalar squared source. In the source terms there is in general vector and tensor components as well although they are often subdominant \cite{Gong:2019mui}. After a long algebra, the equations of motion for tensor modes are given by \cite{Domenech:2017ems}
\begin{align}\label{eq:heom}
(\hat D_\tau-\Delta) h_{ij}=\widehat{TT}^{ab}_{ij}S_{ab}\,,
\end{align}
where $\hat D_\tau\equiv a^{-2}\partial_\tau(a^2\partial_\tau)$, $\Delta\equiv\delta^{ij}\partial_i\partial_j$ is the flat 3-dimensional Laplacian. $\widehat{TT}^{ab}_{ij}$ is the transverse-traceless projector which is given by
\begin{align}\label{eq:tt}
   \widehat{TT}_{ij}\,^{ab}\equiv&
   \left(\delta_{i}^{(a}-\partial_i\partial^{(a}\Delta^{-1}\right)\left(\delta_j^{b)}-\partial_j\partial^{b)}\Delta^{-1}\right)-\frac{1}{2}\left(\delta_{ij}-\partial_i\partial_j\Delta^{-1}\right)\left(\delta^{ab}-\partial^a\partial^b\Delta^{-1}\right)\,,
\end{align}
where the parenthesis in the indexes denote normalised symmetrisation. The source term in a general gauge can be written as
\begin{align}\label{eq:heomsource}
S_{ab}=4&\partial_a\Phi\partial_b\Phi+2a^2\left(\rho+p\right)\partial_aV\partial_bV-(\hat D_\tau-\Delta)\left[\partial_a\sigma\partial_b\sigma+\partial_a\partial^kE\partial_k\partial_bE\right]\,,
\end{align}
where we defined for convenience
\begin{align}
\Phi\equiv\phi-\frac{1}{3}\Delta E+\sigma\quad,\quad
V\equiv v+E'\,.
\end{align}
The source term \eqref{eq:heomsource} reduces to the result of Sec.~\ref{sec:derivation} in the shear-free or Newtonian gauge, that is $\sigma=E=0$.

\section{Bessel functions \label{app:bessel}}
In this appendix  we write useful formulas and relations of the Bessel functions. First, the asymptotic expansion for small argument is given by
\begin{align}
{J_\nu(x\ll1)}\approx x^\nu\frac{2^{-\nu}}{\Gamma[1+\nu]}+O(x^{\nu+1})
\quad,\quad
{Y_\nu(x\ll1)}\approx-\frac{2^{\nu}}{\pi}{\Gamma[\nu]}x^{-\nu}+O(x^{-\nu+1})\,.
\end{align}
 For large arguments we have that the Bessel functions oscillate periodically as
\begin{align}
{J_\nu(x\gg1)}\approx\sqrt{\frac{2}{\pi x}}\cos\left(x-\frac{\nu\pi}{2}-\frac{\pi}{4}\right)+O(x^{-1})\,,
\end{align}
and
\begin{align}
{Y_\nu(x\gg1)}\approx\sqrt{\frac{2}{\pi x}}\sin\left(x-\frac{\nu\pi}{2}-\frac{\pi}{4}\right)+O(x^{-1})\,.
\end{align}
Lastly, useful relations between derivative and Bessel functions of similar order are respectively given by
\begin{align}
\partial_x{J}_{\nu}\left(x\right)={J}_{\nu-1}\left(x\right)-(\nu/x)%
{J}_{\nu}\left(x\right)\,,
\end{align}
and
\begin{align}
{J}_{\nu-1}\left(x\right)+{J}_{\nu+1}\left(x\right)=(2\nu/x)%
{J}_{\nu}\left(x\right)\,.
\end{align}

\section{Integrals with two and three Bessel functions \label{app:integralbessel}}

In this appendix we give the expression for the integrals used in Sec.~\ref{sec:analytical}. Most of the formulae can be found in \cite{threebesselI} and \cite{NIST:DLMF}.

\subsection{Superhorizon approximation}

The integrals \ref{eq:Isimpledef} in the kernel \ref{eq:Isimple} for the superhorizon approximation ($x\ll1$) of Sec.~\ref{sec:superhorizonkernel} involve an integral with two Bessel functions of the same order. Then, the relevant integrals necessary for Sec.~\ref{sec:superhorizonkernel} are
\begin{align}
\int dx x J^2_{\mu}\left(ax\right)=\tfrac{%
1}{2}x^{2}\left(J^2_{\mu}\left(ax\right)-%
J_{\mu-1}\left(ax\right)J_{\mu+1}(ax)\right)\,,
\end{align}
and
\begin{align}
\int dx x^{-2\mu+1}J^2_{\mu}\left(ax\right)%
=\frac{x^{-2\mu+2}}{2(1-2\mu)}\*\left(J^2_{\mu}\left(%
ax\right)+J^2_{\mu-1}\left(ax\right)\right)\,.
\end{align}
Using the above formulas it is straightforward to show that the integrals \ref{eq:Isimpledef} reduce after integration to
\begin{align}
{\cal I}_J&\approx \frac{3+2b}{1+b}\frac{2^{-b-1/2}}{\Gamma[b+3/2]}\frac{x}{\pi v}\,,\\
{\cal I}_Y&\approx \frac{2b+3}{b(1+b)c_s\pi}\left({2^{b-1/2}\Gamma[b+1/2]}\frac{x^{-2b}}{\pi v}-2^{-b-3/2}\frac{c_s^{2b}}{\Gamma[b+3/2]}\frac{1+b+b^2}{1+b}\right)\,.
\end{align}

\subsection{Subhorizon approximation}

The integrals \ref{eq:Isimpledef} in the kernel \ref{eq:Isimple} for the subhorizon approximation ($x\gg1$) of Sec.~\ref{sec:subhorizonkernel} involve definite integral of three Bessel functions. The relevant integrals necessary for Sec.~\ref{sec:subhorizonkernel} can be found in Ref.~\cite{threebesselI}. Here we review their main results. On one hand, they find that for $|a-b|<c<a+b$ the integrals \ref{eq:Isimpledef} read
\begin{align}
{\cal I}_{J/Y}&=\int_0^{\infty} d\tilde x \tilde x^{1-\rho}
\left\{ 
\begin{aligned}
    J_\rho(c\tilde x)\\
    Y_\rho(c\tilde x)
\end{aligned}
\right\}
J_{\nu}(a\tilde x)J_{\nu}(b\tilde x)\nonumber\\&=\frac{1}{\pi}\sqrt{\frac{2}{\pi}}\frac{(ab)^{\rho-1}}{c^\rho}\left(\sin\varphi\right)^{\rho-1/2}\left\{
\begin{aligned}
    \frac{\pi}{2}\mathsf{P}^{-\rho+1/2}_{\nu-1/2}(\cos\varphi)\\
    -\mathsf{Q}^{-\rho+1/2}_{\nu-1/2}(\cos\varphi)
\end{aligned}
\right\}\,,
\end{align}
where
\begin{align}
16\Delta^2\equiv\left(c^2-(a-b)^2\right)\left((a+b)^2-c^2\right)\,,
\end{align}
and
\begin{align}
\cos\varphi=\frac{a^2+b^2-c^2}{2ab}\quad,\quad\sin\varphi=\frac{2\Delta}{ab}\,.
\end{align}
On the other hand, if $c>a+b$ the integrals are instead given by
\begin{align}
{\cal I}_{J/Y}&=\int_0^{\infty} d\tilde x \tilde x^{1-\rho}
\left\{    
\begin{aligned}
    J_\rho(c\tilde x)\\
    Y_\rho(c\tilde x)
\end{aligned}
\right\}
J_{\nu}(a\tilde x)J_{\nu}(b\tilde x)\nonumber\\&
\nonumber\\&=\frac{1}{\pi}\sqrt{\frac{2}{\pi}}\frac{(ab)^{\rho-1}}{c^\rho}\left(\sinh\phi\right)^{\rho-1/2}\Gamma[\nu-\rho+1]{\cal Q}^{-\rho+1/2}_{\nu-1/2}(\cosh\phi)\left\{  
\begin{aligned}
    -\sin\left[(\nu-\rho)\pi\right]\\
    \cos\left[(\nu-\rho)\pi\right]
\end{aligned}
\right\}\,,
\end{align}
where
\begin{align}
16\tilde\Delta^2\equiv\left(c^2-(a-b)^2\right)\left(c^2-(a+b)^2\right)\,,
\end{align}
and
\begin{align}
\cosh\phi=\frac{c^2-(a^2+b^2)}{2ab}\quad,\quad\sinh\phi=\frac{2\tilde\Delta}{ab}\,.
\end{align}

To use these formulas in the main text, we identify
\begin{align}
c=1\quad,\quad a=c_sv\quad,\quad b=c_su\,.
\end{align}
Then, the range $|1-v|<u<1+v$ can be split into $1>c_s(u+v)$ ($c>a+b$) and $1<c_s(u+v)$ ($c<a+b$). In the latter case we always have that $c_s|u-v|<1$ by momentum conservation and so the formulas presented above cover all ranges of interest.

\section{Associated Legendre functions \label{app:legendre}}
In this appendix, we present useful formulae for the associated Legendre functions that can be found in Ref.~\cite{NIST:DLMF}. We begin with the definitions of the associated Legendre functions of the first and second kind in terms of hypergeometric functions which read
\begin{align}
\mathsf{P}^{\mu}_{\nu}\left(x\right)=\left(\frac{1+x}{1-x}\right)^{\mu/2}%
\mathbf{F}\left(\nu+1,-\nu;1-\mu;\tfrac{1}{2}-\tfrac{1}{2}x\right)\,,
\end{align}
\begin{align}
\mathsf{Q}^{\mu}_{\nu}\left(x\right)=&\frac{\pi}{2\sin\left(\mu\pi\right)}\Bigg%
(\cos\left(\mu\pi\right)\left(\frac{1+x}{1-x}\right)^{\mu/2}\mathbf{F}\left(%
\nu+1,-\nu;1-\mu;\tfrac{1}{2}-\tfrac{1}{2}x\right)\nonumber\\&\qquad\qquad\qquad-\frac{\Gamma\left(\nu+\mu+1%
\right)}{\Gamma\left(\nu-\mu+1\right)}\left(\frac{1-x}{1+x}\right)^{\mu/2}%
\mathbf{F}\left(\nu+1,-\nu;1+\mu;\tfrac{1}{2}-\tfrac{1}{2}x\right)\Bigg)\,,
\end{align}
and
\begin{align}
{\cal Q}^{\mu}_{\nu}\left(x\right)=&%
\frac{\pi}{2\sin\left(\mu\pi\right)\Gamma\left(\nu+\mu+1\right)}\Bigg(\left(\frac{x+1}{x-1}\right)^{\mu/2}\mathbf{F}\left%
(\nu+1,-\nu;1-\mu;\tfrac{1}{2}-\tfrac{1}{2}x\right)\nonumber\\&\qquad\qquad\qquad-\frac{\Gamma\left(\nu+\mu+1\right)}{%
\Gamma\left(\nu-\mu+1\right)}\left(\frac{x-1}{x+1}\right)^{\mu/2}\mathbf{F}\left(\nu+1,-\nu;1+\mu;%
\tfrac{1}{2}-\tfrac{1}{2}x\right)\Bigg)\,.
\end{align}
Note that $\mathsf{P}^{\mu}_{\nu}(x)$ and  $\mathsf{Q}^{\mu}_{\nu}(x)$ are also called the Ferrer's functions and are valid for $|x|<1$. Then, ${\cal Q}^{\mu}_{\nu}(x)$ is also known as Olver's function which is real for $|x|>1$. In the above definitions we have used the compact notation
\begin{align}
\mathbf{F}\left(a,b;c;x\right)=\frac{1}{\Gamma[c]}{F}\left(a,b;c;x\right)\,,
\end{align}
with ${F}\left(a,b;c;x\right)$ being the Gauss's Hypergeometric function.

\subsection{Asymptotic limits \label{sec:asymptotics}}
We write down the asymptotic behaviour of the Associated Legendre functions near the singular points $x\sim 1$ and $x\to \infty$. It should be noted that the formulas below are valid for $\mu\notin \mathbb{Z}$. Relevant cases with $\mu\notin \mathbb{Z}$ are given in Sec.~\ref{sec:summary}.

\textit{The limit $x\to 1^{-}$:} First, the Ferrer's function of the first kind asymptotically goes as
\begin{align}
\mathsf{P}^{\mu}_{\nu}\left(x\right)\sim\frac{1}{\Gamma\left(1-\mu\right)}%
\left(\frac{2}{1-x}\right)^{\mu/2}\,.
\end{align}
Second, the Ferrer's function of the second kind goes for $\mu>0$ as
\begin{align}
\mathsf{Q}^{-\mu}_{\nu}\left(x\right)\sim\frac{\Gamma\left(\mu\right)\Gamma%
\left(\nu-\mu+1\right)}{2\Gamma\left(\nu+\mu+1\right)}\left(\frac{2}{1-x}%
\right)^{\mu/2},
\end{align}
while for $\mu<0$ behaves as 
\begin{align}
\mathsf{Q}^{\mu}_{\nu}\left(x\right)\sim\frac{1}{2}\cos\left(\mu\pi\right)%
\Gamma\left(\mu\right)\left(\frac{2}{1-x}\right)^{\mu/2}\qquad {\mu\neq 1/2}\,.
\end{align}

\textit{The limit $x\to 1^{+}$:} Since $|x|>1$, this case only concerns the Olver's function of the second kind. In the limit $x\to 1^{+}$ we have that
\begin{align}
{\cal Q}^{\mu}_{\nu}\left(x\right)\sim\frac{\Gamma\left(\mu\right)}{2%
\Gamma\left(\nu+\mu+1\right)}\left(\frac{2}{x-1}\right)^{\mu/2}\,.
\end{align}

\textit{The limit $x\to \infty$:} This case also applies only to the Olver's function of the second kind. The asymptotic behaviour is then given by
\begin{align}
{\cal Q}^{\mu}_{\nu}\left(x\right)\sim\frac{\pi^{1/2}}{\Gamma\left(\nu+%
\frac{3}{2}\right)(2x)^{\nu+1}}\,.
\end{align}

\subsubsection{Useful relations}

In order to apply the asymptotic limits of Sec.~\ref{sec:asymptotics} we have to evaluate the functions for negative argument and negative order. This can be done by the useful relations below:
\begin{align}
\mathsf{P}^{-b}_{b}(-x)=\mathsf{P}^{-b}_{b}(x)\quad,\quad \mathsf{P}^{-b}_{b+2}(-x)=\mathsf{P}^{-b}_{b+2}(x)\,,
\end{align}
\begin{align}
\mathsf{Q}^{-b}_{b}(-x)=-\mathsf{Q}^{-b}_{b}(x)\quad,\quad \mathsf{Q}^{-b}_{b+2}(-x)=-\mathsf{Q}^{-b}_{b+2}(x)\,,
\end{align}
and
\begin{align}
{\cal Q}^{-b}_{b}(x)={\cal Q}^{b}_{b}(x)\quad,\quad {\cal Q}^{-b}_{b+2}(x)={\cal Q}^{b}_{b+2}(x)\,.
\end{align}

\subsubsection{The resonance limit}
Using the formulas in the Sec.~\ref{sec:asymptotics} we can compute the terms that appear in Sec.~\ref{sec:subhorizonkernel} in Eq.~\ref{eq:kernelaverage} near the resonant point $y\sim1$. In this way, we have that on one hand for $y\to -1^+$
\begin{align}
\mathsf{P}^{-b}_{b}(y)+\frac{b+2}{b+1}\mathsf{P}^{-b}_{b+2}(y)\sim\frac{3+2b}{1+b}\frac{1}{\Gamma[1+b]}\left(\frac{2}{1+y}\right)^{-b/2}\,,
\end{align}
and
\begin{align}
\mathsf{Q}^{-b}_{b}(y)+\frac{b+2}{b+1}\mathsf{Q}^{-b}_{b+2}(y)\sim-\frac{3+2b}{1+b}\left\{
\begin{aligned}
&(1+b+b^2)\frac{\Gamma[b]}{\Gamma[2b+3]}\left(\frac{2}{1+y}\right)^{b/2}\quad &b>0\\
&\frac{1}{2}\cos(b\pi)\Gamma[-b]\left(\frac{2}{1+y}\right)^{-b/2}\quad &b<0
\end{aligned}
\right.\,.
\end{align}
On the other hand, for $y\to 1^+$ we find
\begin{align}
{\cal Q}^{-b}_{b}(y)+2\frac{b+2}{b+1}{\cal Q}^{-b}_{b+2}(y)\sim\frac{3+2b}{1+b}\left\{
\begin{aligned}
&(1+b+b^2)\frac{\Gamma[b]}{\Gamma[2b+3]}\left(\frac{2}{1-y}\right)^{b/2}\quad &b>0\\
&\frac{1}{2}\Gamma[-b]\left(\frac{2}{1-y}\right)^{-b/2}\quad &b<0
\end{aligned}
\right.\,.
\end{align}

\bibliographystyle{JHEP}
\bibliography{bibliographyreview}

\end{document}